\documentclass{sn-jnl}

\usepackage{graphicx}%
\usepackage{multirow}%
\usepackage{amsmath,amssymb,amsfonts}%
\usepackage{mathrsfs}%
\usepackage[title]{appendix}%
\usepackage{xcolor}%
\usepackage{textcomp}%
\usepackage{manyfoot}%
\usepackage{booktabs}%
\usepackage{algorithm}%
\usepackage{algorithmicx}%
\usepackage{algpseudocode}%
\usepackage{listings}%
\usepackage{mathdots}%
\usepackage{adjustbox}%
\usepackage[inkscape]{svg}%
\usepackage{listings}

\newcommand{\operator}[1]{\hat{#1}}
\newcommand{\ansatz}[1]{\operator{#1}}
\newcommand{\crt}[1]{\operator{a}^\dagger_{#1}}
\newcommand{\dst}[1]{\operator{a}^{\phantom{\dagger}}_{#1}}
\newcommand{\pauli}[1]{\hat{\mathsf{#1}}}
\newcommand{\software}[1]{$\mathsf{#1}$}
\newcommand{\device}[1]{$\mathsf{ibm\_#1}$}
\newcommand{\myarray}[1]{\mathbf{#1}}

\newcommand{\hartree}{E_{\mathrm{h}}}
\newcommand{\method}[1]{\mathrm{#1}}
\newcommand{\code}[1]{\lstinline{#1}}

\usepackage[
  backend=biber,
  bibstyle=numeric,        
  citestyle=numeric-comp,  
  sorting=none,   
  sortcites=true          
]{biblatex}
\usepackage{subcaption}

\usepackage{tikz}
\usetikzlibrary{quantikz2}

\BiblatexSplitbibDefernumbersWarningOff

\begin{document}

\title[Article Title]{From Promise to Practice: Benchmarking Quantum Chemistry on Quantum Hardware}

\author[1,3]{\fnm{Osama M.} \sur{Raisuddin}}\email{raisuo2@rpi.edu}

\author[4]{\fnm{Haimeng} \sur{Zhang}}\email{Haimeng.Zhang@ibm.com}

\author[4]{\fnm{Mario} \sur{Motta}}\email{Mario.Motta@ibm.com}

\author*[2]{\fnm{Fabian M.} \sur{Faulstich}}\email{faulsf@rpi.edu}

\affil[1]{
\orgdiv{Scientific Computation Research Center}, 
\orgname{Rensselaer Polytechnic Institute}, \orgaddress{\street{8th Street}, \city{Troy}, \postcode{12180}, \state{NY}, \country{USA}}
}

\affil[2]{
\orgdiv{Department of Mathematical Sciences},
\orgname{Rensselaer Polytechnic Institute}
}

\affil[3]{
\orgdiv{Future of Computing Institute},
\orgname{Rensselaer Polytechnic Institute}
}

\affil[4]{
\orgdiv{IBM Quantum}, 
\orgname{IBM T.J. Watson Research Center},
\orgaddress{\street{1101 Kitchawan Road}, \city{Yorktown Heights}, \postcode{10598}, \state{NY}, \country{USA}}
}

\abstract{
We provide a systematic evaluation of the sample-based quantum diagonalization (SQD) method for electronic structure based on the W4-11 thermochemistry dataset, comprising  124 total atomization, 83 bond dissociation, 20 isomerization, 505 heavy-atom transfer, and 13 nucleophilic substitution processes, covering diverse bonding situations and reaction mechanisms. This is the largest study assessing the accuracy and precision of a quantum-hybrid algorithm on a digital quantum device across a variety of molecular systems and chemical reactions, using 16.85 hours on the superconducting quantum processor \device{rensselaer} and 724.22 node hours on the supercomputer \texttt{AiMOS}. To ensure a fair comparison, our study employs commensurate resource allocation for both classical and quantum simulations. Although SQD exhibits large statistical deviations from ground-state reference energies, energy extrapolations yield CCSD-level accuracy. While bond-breaking reactions show a systematic improvement as computational resources increase, nucleophilic substitution or heavy atom transfer reactions do not. The limitations quantified in this manuscript indicate opportunities for improvement in SQD-based algorithms. This work provides a benchmark and community resource for exploring new quantum algorithms and devices, supported by an online benchmark challenge and an open-source Python library for direct comparison.}

\maketitle

\begin{refsegment}
\section*{Introduction}

The accurate computational treatment of interacting electronic systems is a major challenge in contemporary quantum chemistry, materials science, and physics. The time-independent Schr\"{o}dinger equation is the theoretical foundation for these calculations~\cite{dirac1928quantum}. However, the search for numerically exact solutions of the Schr\"{o}dinger equation is challenging due to the combinatorial size of the Hilbert space associated with the distribution of $N$ electrons among $M$ spatial orbitals. As a result, numerically exact solutions are currently available for $N$ and $M$ around 20~\cite{vogiatzis2017pushing, gao2024distributed} and the Schr\"{o}dinger equation is almost exclusively solved by approximate methods, each offering different levels of accuracy, precision, and computational cost. Since precision and accuracy cannot be easily quantified or predicted, systematic knowledge of the approximations underlying computational methods is commonly developed through detailed benchmark studies~\cite{leblanc2015solutions, zheng2017stripe, motta2017towards, eriksen2020ground, williams2020direct}.

In recent years, progress in hardware manufacturing has produced quantum devices capable of performing computations of limited scale, and integrated within high-performance computing systems~\cite{alexeev2024quantum}. At the same time, the original suggestion of using quantum devices as simulators for other quantum systems~\cite{feynman1982simulating} has evolved into specific methods to approximately solve the Schr\"{o}dinger equation~\cite{georgescu2014quantum,cao2019quantum,cerezo2020variational,mcardle2020quantum,bauer2020quantum}, some of which are designed for pre-fault-tolerant devices. Such methods are shaped by the tension between two competing objectives: on the one hand, maximizing accuracy and precision; on the other hand, ensuring compatibility with resource constraints dictated by coherence times and error rates of quantum devices. As a result, a robust assessment of their performance and limitations relies on systematic benchmarking in lieu of theoretical accuracy guarantees.
However, detailed benchmark studies of such methods have been relatively rare~\cite{eisert2020quantum,proctor2025benchmarking}, in part because they require extensive access to quantum devices, which is less common than to classical devices due to their less widespread availability and more specialized infrastructure.

In this study, we use the W4-11 dataset developed by Karton, Daon, and Martin~\cite{karton2011w4,karton2017w4} to benchmark quantum computing methods for near-term devices. The W4-11 dataset is a collection of 745 thermochemical reactions of diverse nature (atomization, bond dissociation, isomerization, nucleophilic substitution, and heavy-atom transfer) involving 152 unique chemical species.
Because the electronic structure of W4-11 species in their equilibrium geometries is predominantly dynamic in character, quantum chemistry methods for classical computers -- particularly coupled cluster singles, doubles, and perturbative triples, CCSD(T)~\cite{raghavachari1989fifth} -- can deliver reference results to evaluate the accuracy and precision of a quantum computing method over the entire database.
Therefore, determining the electronic ground-state energies of W4-11 species at any given level of theory, along with the corresponding thermochemical energy differences, is (i) an ambitious test for a quantum computing method targeting near-term devices with limited coherence times and error rates, (ii) an assessment of the extent to which a method can resolve a typical molecular system (medium-sized with no indications of strong electron correlation) or a thermochemical reaction connecting multiple species, and (iii) an occasion to identify algorithmic limitations and guide research towards improvements.

\section*{Methods}

\begin{figure*}[ht!]
\centering
\includegraphics[width=\textwidth]{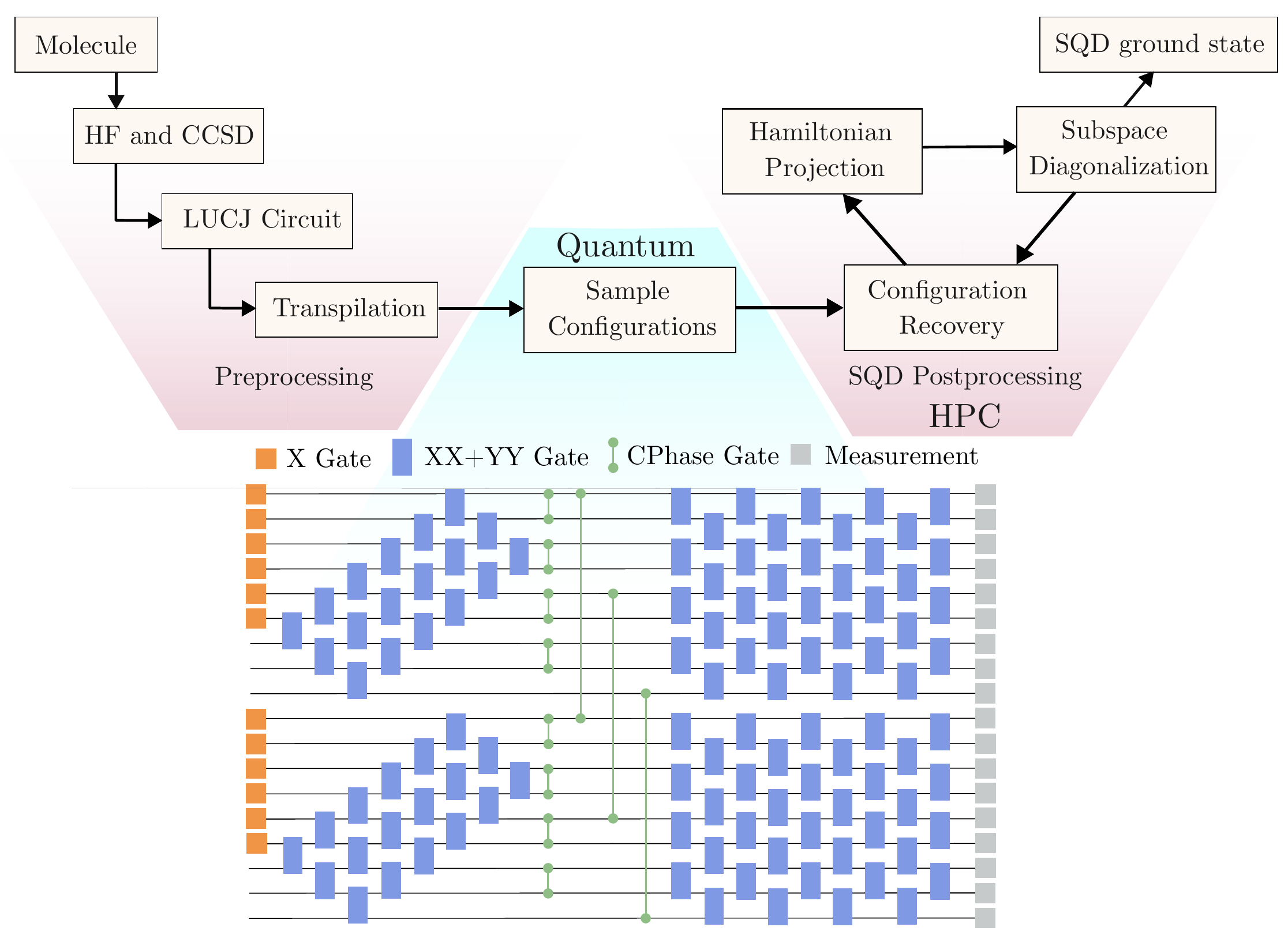}
\caption{Computational workflow of SQD used in this work: The preprocessing stage (left, pink) comprises classical electronic-structure calculations, the construction of the corresponding quantum circuits (illustrated in the lower panel for a closed-shell configuration with $N=12$ electrons in M=9 spatial orbitals), and their subsequent transpilation. The transpiled circuits are then executed on a quantum device (middle, blue), and the resulting measurement data are analyzed during the postprocessing stage (right, pink).}
\label{fig:fig_1}
\end{figure*}

In this work, we use the sample-based quantum diagonalization (SQD) method~\cite{robledo2025chemistry} to compute the electronic ground-state energies and thermochemical reaction energies of W4-11 species, at STO-6G level of theory~\cite{hehre1969self} with the frozen-core approximation. Figure~\ref{fig:fig_1} provides an overview of our computational workflow, which is elaborated upon in the subsequent discussion. For brevity, the technical details of SQD and the auxiliary techniques are deferred to the Appendix; here, we present only the information necessary for a self-contained exposition.

SQD is related to classical selected configuration interaction~\cite{tubman2016deterministic,holmes2016heat,schriber2016communication} and quantum selected configuration interaction (QSCI)~\cite{kanno2023quantum}. It diagonalizes the electronic structure Hamiltonian in a subspace spanned by electronic configurations, i.e., Slater determinants, sampled by repeatedly executing a quantum circuit on a quantum device. In its current formulation, SQD uses quantum circuits derived from the local unitary cluster Jastrow ansatz (LUCJ)~\cite{motta2023bridging} to sample configurations, and a configuration recovery technique to mitigate errors arising from decoherence and imperfect realization of quantum gates.
Because SQD has been used in several recent studies~\cite{kaliakin2025accurate,shajan2025toward,kaliakin2025implicit,bazayeva2025quantum,danilov2025enhancing,liepuoniute2025quantum,duriez2025computing,barroca2025surface,smith2025quantum,barison2025quantum}, it is timely and useful to investigate its performance robustly and systematically, over a broad and diverse family of use cases.

\begin{figure*}[ht!]
    \centering
    \includegraphics[width=0.8\textwidth]{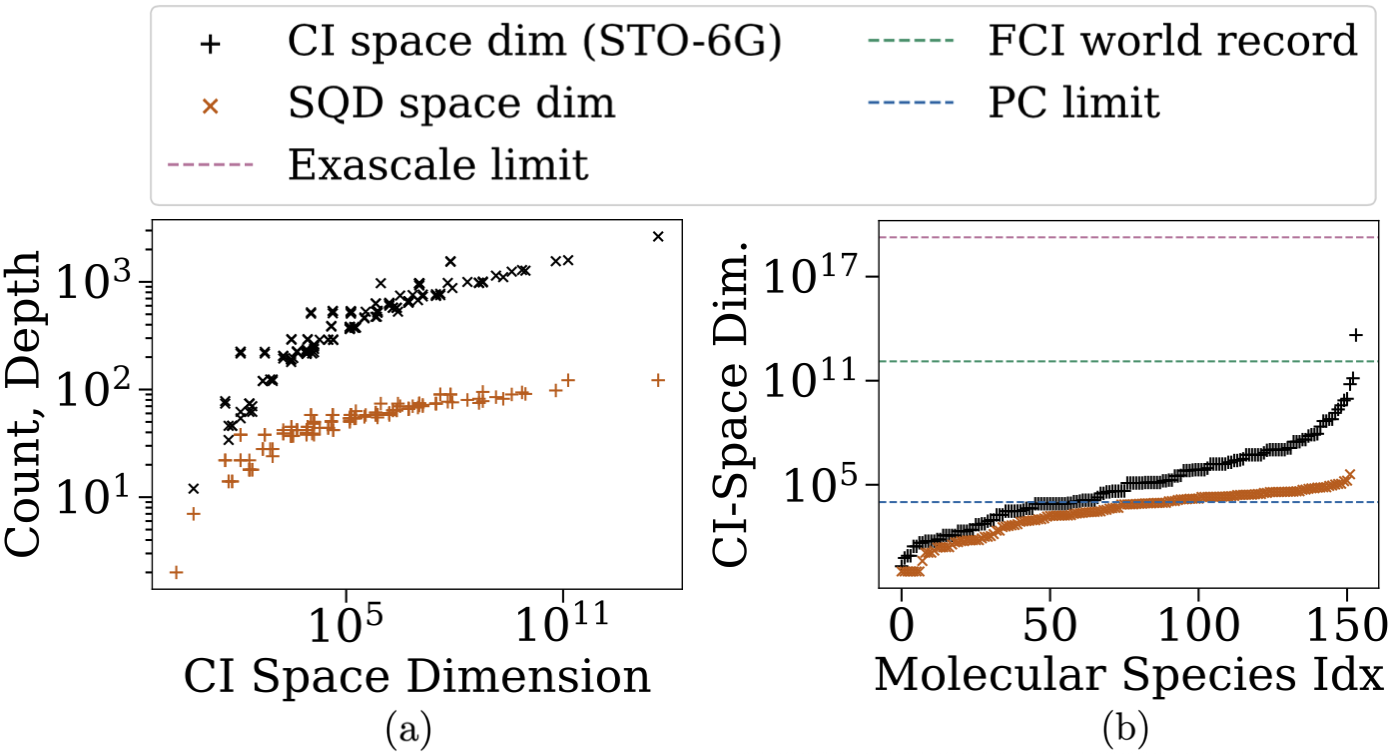}
\caption{(a) two-qubit gate count (black ``x'') and depth (orange ``+'') of LUCJ circuits used in SQD calculations as functions of the configuration space dimension, for the molecules in the W4-11 database at STO-6G level of theory. (b) dimension of the configuration space (black ``x''), dimension of the largest computed SQD wavefunctions (orange ``+'') for the molecules in the W4-11 database at STO-6G level of theory, compared with $10^{4}$ (blue dashed line) taken to represent a 1-minute timeout on a personal computer, the FCI limit at the time of writing ($1.3\cdot 10^{12}$, green dashed line, Ref.~\cite{gao2024distributed}), and $2^{64} \simeq 10^{19}$ (purple dashed line) taken to represent the exascale limit.
}
\label{fig:2}
\end{figure*}

As a quantum-hybrid algorithm, the SQD workflow comprises three distinct phases~\cite{robledo2025chemistry}: (1) the classical preprocessing phase, (2) the quantum computing phase, and (3) the classical postprocessing phase. In the classical preprocessing phase, we perform, for each molecule in the W4-11 dataset, Hartree-Fock (HF), M{\o}ller-Plesset second-order perturbation theory (MP2), configuration interaction singles and doubles (CISD), CCSD, and CCSD(T). 
These computations provide both the classical reference data and the inputs required for the subsequent quantum simulations. The latter involves the construction of LUCJ circuits, illustrated in the lower portion of Fig.~\ref{fig:fig_1}. In the quantum computing phase, the LUCJ circuits are executed on a quantum device to obtain measurement outcomes, i.e., bitstrings corresponding to Slater determinants. In the classical postprocessing phase, the resulting samples are processed on a classical high-performance computing (HPC) platform to obtain the final SQD result. This postprocessing serves two purposes: First, because contemporary quantum hardware is prone to noise, we apply a configuration recovery procedure that enforces conservation of electron number and total spin-$z$, thereby improving the quality of the sampled determinant configurations. Second, the final step of the SQD workflow constructs the Hamiltonian in the subspace spanned by the recovered configurations and then diagonalizes it classically. Although computationally demanding, this task is substantially less complex than solving the original full problem, as illustrated in Fig.~\ref{fig:2}b. This yields the SQD wavefunctions, i.e., sparse linear combinations of $d$ determinant configurations, and corresponding approximations to the electronic ground-state energies. The accuracy of these energies depends on both the number $d$ and the fidelity of the recovered configurations.

Although STO-6G calculations are not chemically realistic because they lack polarization and diffusion functions, they provide a challenging and valuable set of use cases for quantum algorithms running on present-day devices.
As qubits undergo decoherence and errors accumulate in the computation as quantum operations are applied, the signal emerging from circuits of increasing depth (number of layers of quantum gates) and size (total number of quantum operations) gradually weakens. This limits the size of use cases that an algorithm can tackle before the signal is corrupted by noise to an extent that precludes assessing the impact of algorithmic approximations. The depth and size of LUCJ circuits used in SQD, as illustrated in Fig.~\ref{fig:2}a, scale as $D \simeq 2.35 \, N_q$ and $n_g \simeq 0.98 \, N_q^2$ respectively, where $N_q = 2M$ is the number of qubits required to simulate electrons in $M$ spatial orbitals (for more details, see also Appendix~\ref{sec:LUCJ} and references therein). Such a scaling of quantum resources yields noisy quantum samples, which manifests in the accuracy of SQD energies along with algorithmic approximations. However, the strength of the noise -- attenuated by configuration recovery and diagonalization on classical HPC -- allows SQD to treat the entire W4-11 database subject to frozen core approximation at STO-6G level of theory.

Furthermore, the relatively modest size of the STO-6G basis allows for increasing the number $d$ of SQD configurations towards the dimension of the configuration space for many of the species in the database, as illustrated in Fig.~\ref{fig:2}b. As increasing the size of SQD wavefunctions leads to near-exact energy, precluding the assessment of the impact of quantum noise on SQD samples,
in this study, we choose to perform SQD calculations in subspaces of dimensions $d \leq \zeta N_{\method{CCSD}}$ where $\zeta = 25\%, 50\%, 100\%, 200\%, 400\%$ and $N_{\method{CCSD}} = \mathcal{O}(N^2 M^2)$ is the number of CCSD free parameters~\cite{scuseria1987closed}. As seen in Fig.~\ref{fig:2}b, this choice leads to $d$ much below the configuration space dimension, allowing us to assess the combined impact of quantum noise and algorithmic approximations in SQD.

In the case of SQD, we perform extrapolations toward an estimate of the ground-state energy using raw SQD energies $E(\Psi_{\method{SQD}})$ and variances $V(\Psi_{\method{SQD}})$ computed at various subspace dimensions, based on the linear relation $E(\Psi_{\method{SQD}}) \simeq m \, V(\Psi_{\method{SQD}}) + E_{\mathrm{gs}}$ \cite{mizusaki2003precise}, valid when $\Psi_{\method{SQD}}$ is sufficiently close to the ground-state wavefunction. As part of this study, we present and compare two extrapolation techniques -- one based on data clustering and labeled LMM, and the other based on subspace diagonalization and labeled GEV, both detailed in the Appendix -- and focus on their accuracy and precision, i.e., size of statistical uncertainties on the extrapolated $E_{\mathrm{gs}}$. To isolate the effect of the extrapolation from the raw SQD energies, we present both unextrapolated and extrapolated results.

\section*{Results}

The quantum computations presented here were conducted on \device{rensselaer}, a 127-qubit superconducting quantum processor based on the Eagle architecture. We sampled $10^6$ samples per species for a total of 16.85 hours of quantum wall-clock time. The classical pre-, peri, and post-processing was performed on \texttt{AiMOS} (an eight-petaflop IBM supercomputer) for a total of 724.22 node hours of computation on 20 cores of IBM Power9 processors clocked at 3.15~GHz (75.49 hours wall-clock time for the largest species).

\begin{figure*}[ht!]
\begin{subfigure}[t]{0.49\textwidth}
    \centering
    \includegraphics[width=\textwidth]{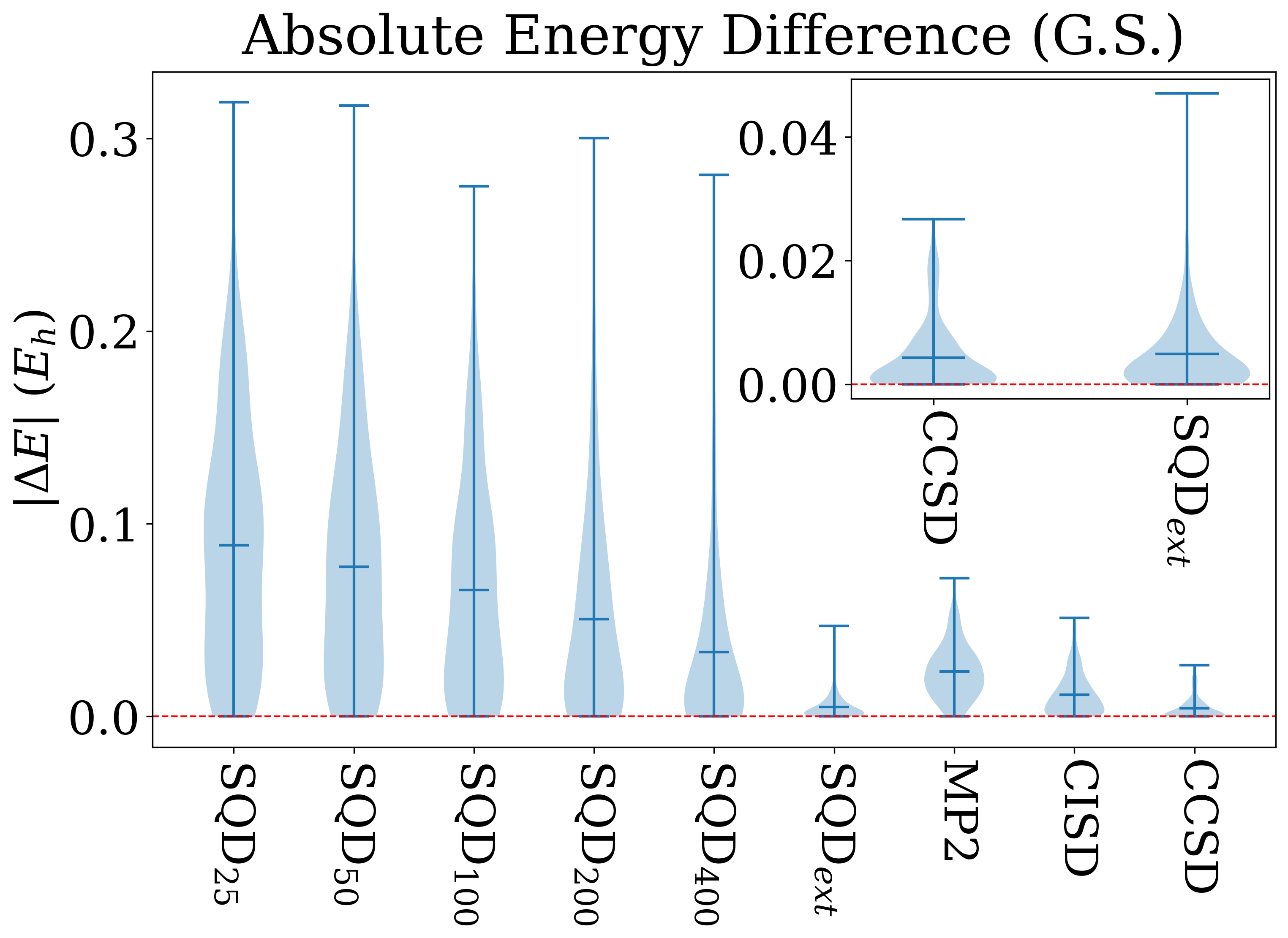}
    \caption{}
\end{subfigure}%
\hfill
\begin{subfigure}[t]{0.49\textwidth}
    \centering
    \includegraphics[width=\textwidth]{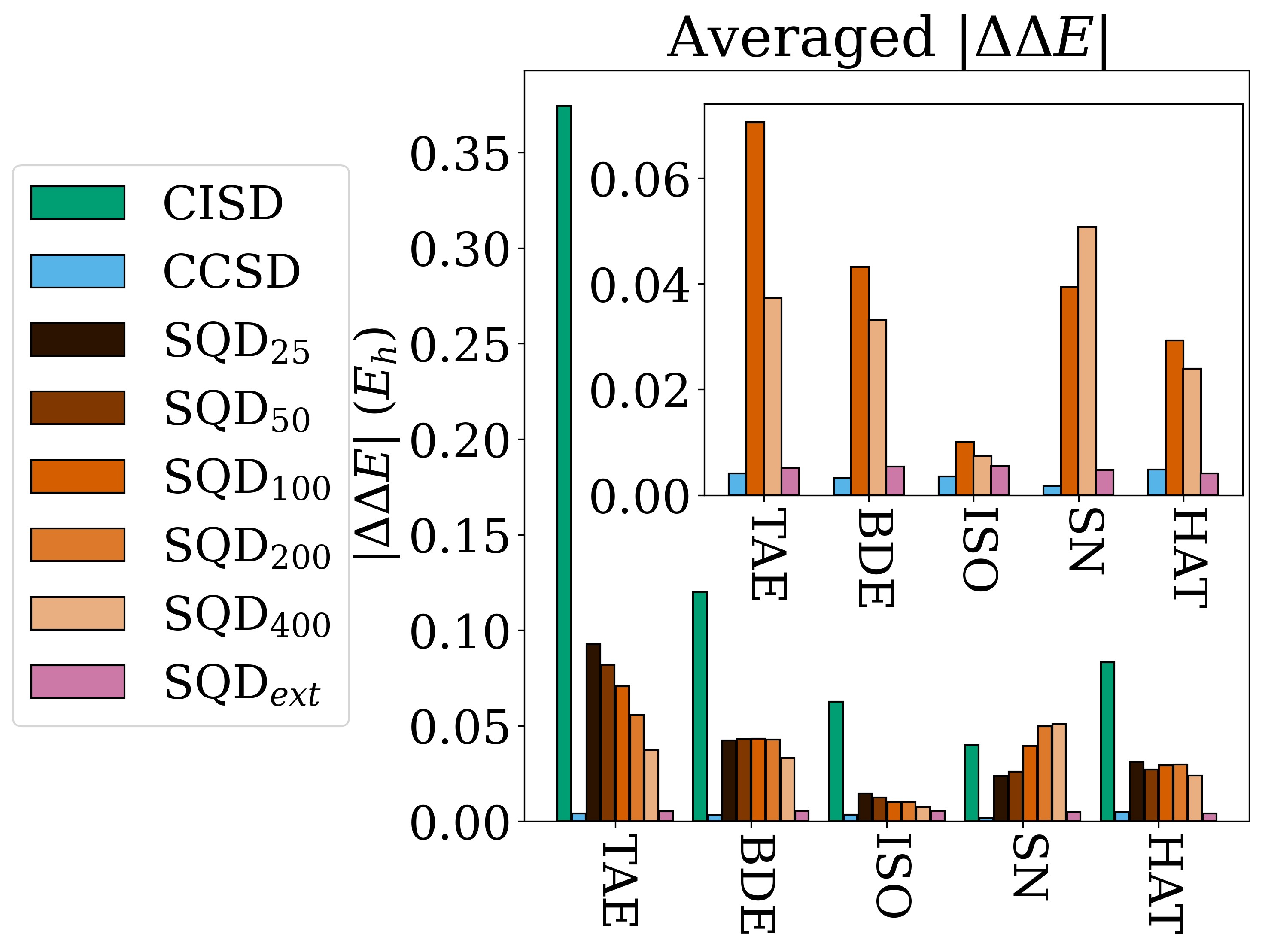}
    \caption{}
\end{subfigure}
\\
\begin{subfigure}[t]{0.49\textwidth}
    \centering
    \includegraphics[width=\textwidth]{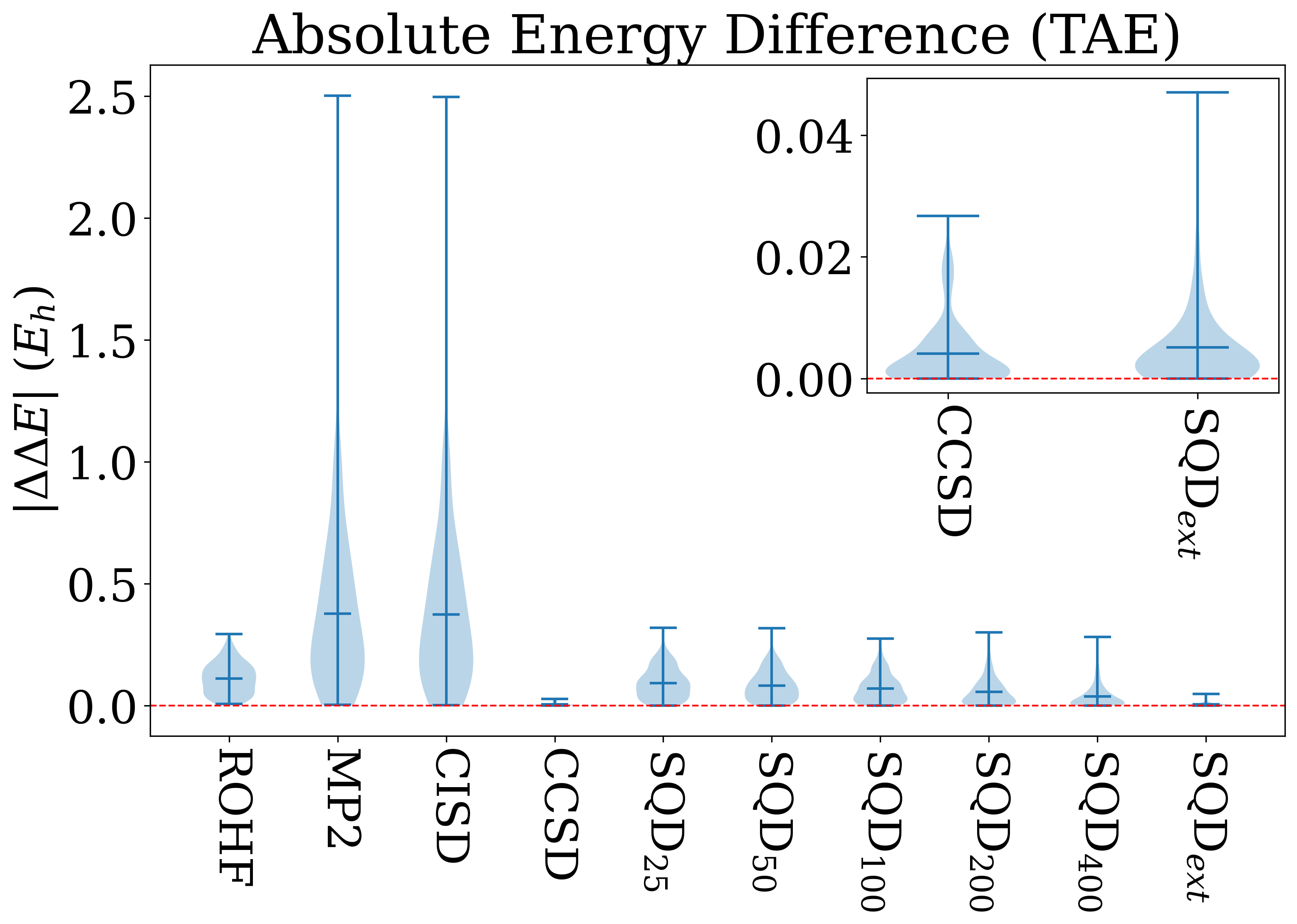}
    \caption{}
\end{subfigure}%
\hfill
\begin{subfigure}[t]{0.49\textwidth}
    \centering
    \includegraphics[width=\textwidth]{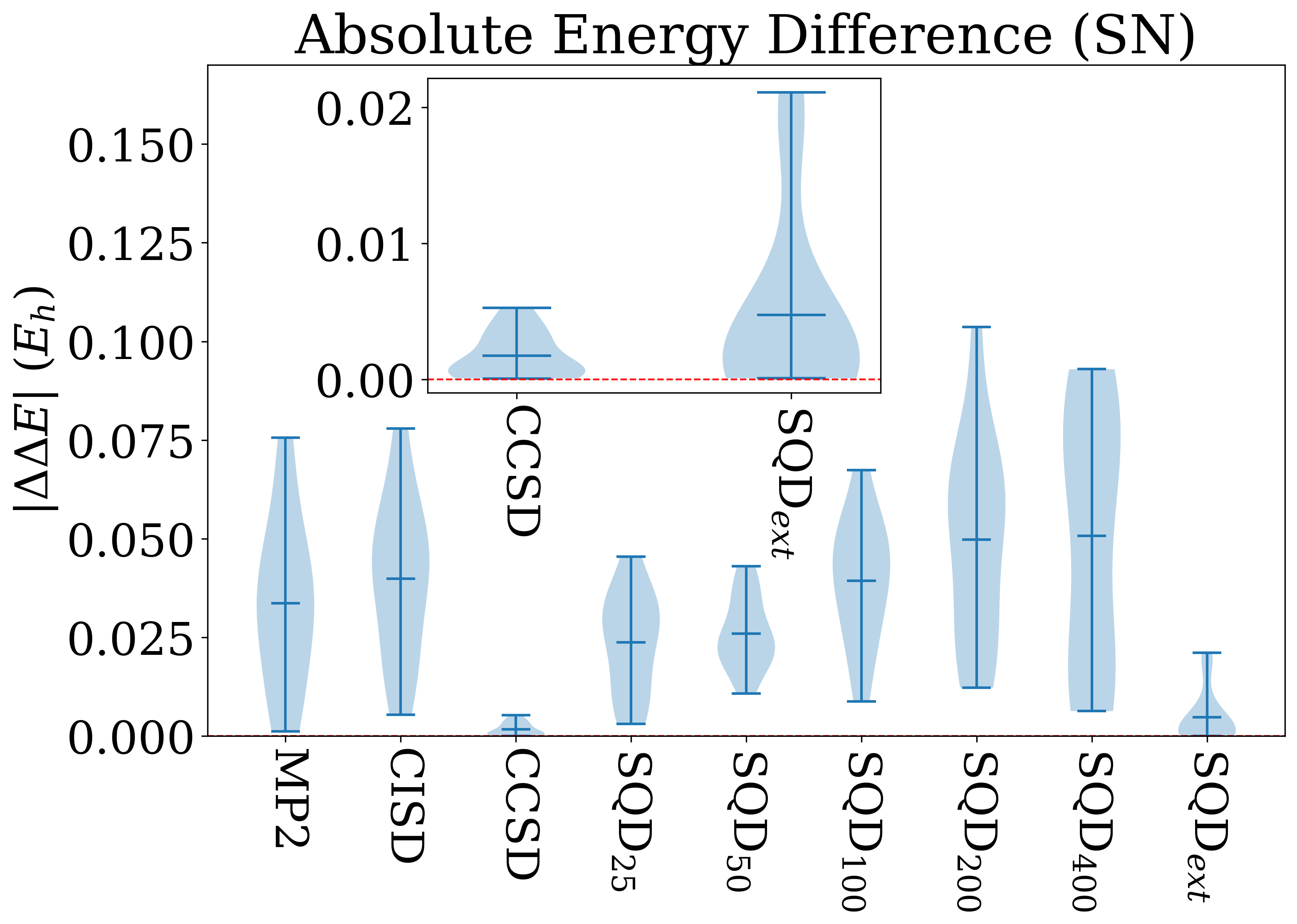}
    \caption{}
\end{subfigure}%
\caption{\label{fig:3} Comparison of MP2, CISD, CCSD, SQD with different subspace sizes (SQD$_\zeta$ with $\zeta = 25\%, 50\%, 10\%, 200\%, 400\%$) and SQD extrapolated with GEV (SQD$_\mathrm{ext}$) across the W4‑11 dataset. 
(a) Distribution of absolute ground-state energy errors ($|\Delta E|$) 
%
(b) Average absolute reaction energy error $|\Delta \Delta E|$ for each reaction class: total atomization (TAE), bond dissociation (BDE), isomerization (ISO), heavy‑atom transfer (HAT), and nucleophilic substitution (SN) 
%
(c) Distribution of absolute reaction energy errors for TAE processes 
%
(d) Distribution of absolute reaction energy errors for SN processes 
}
\end{figure*}

The key observations of this study are summarized in Fig.~\ref{fig:3}. We begin our analysis in panel (a) by assessing the accuracy of SQD for total energies. Specifically, we evaluate the absolute energy error,
\begin{equation}
\label{eq:EnergyDiviation}
|\Delta E_{s,\method{X}}| = |E_{s,\method{X}} - E_{s, \method{CCSD(T)}}| \;,
\end{equation}
where $s$ is one of the species in the W4-11 suite, $\method{X}$ labels MP2, CISD, CCSD, SQD$_\zeta$ (i.e., SQD with $d \leq \zeta N_{\method{CCSD}}$), or SQD$_\mathrm{ext}$ (i.e., SQD extrapolated with the GEV technique, see Section~\ref{sec:extrapolation} and Appendix~\ref{sec:GEV}), and CCSD(T) serves as the reference. As seen, SQD$_\zeta$ total energies are in worse agreement with CCSD(T) than the three classical methods reported. Notably, while the average $|\Delta E|$ by SQD$_\zeta$ decreases from ${\sim 0.10} \, \hartree$ to ${\sim 0.03 }\, \hartree$ as $\zeta$ increases, the distribution features outliers of up to $\sim 0.30 \, \hartree$ that are not present in classical methods. Furthermore, SQD$_{100}$ has a poorer accuracy than CISD, which uses a subspace of comparable size, indicating the sub-optimality of the SQD configurations. 
Upon extrapolation with GEV, SQD total energies are in reasonable agreement with those of CISD and CCSD, with an average error of $\sim 0.005 \, \hartree$ and outliers of up to $\sim 0.050 \, \hartree$. For further details and a comprehensive statistical analysis of the ground state approximations via SQD, we refer the reader to Appendix~\ref{sec:ResultsGS}.

Next, we assess the accuracy of SQD for energy differences corresponding to different thermochemical reactions. Specifically, we evaluate the absolute reaction energy error,
\begin{equation}
\label{eq:EnergyDifference}
\Big| \Delta \Delta E_{r\to p,\method{X}} \Big|
= 
\Big| \Delta E_{r,\method{X}} - \Delta E_{p,\method{X}} \Big|
\;,
\end{equation}
where $r \to p$ denotes a reaction in the W4-11 suite, $\method{X}$ labels MP2, CISD, CCSD, SQD$_\zeta$, or SQD$_\mathrm{ext}$, and $E_{r,\method{X}}$ ($E_{p,\method{X}}$) is computed as the sum of total energies of individual reactants in $r$ (products in $p$). Fig.~\ref{fig:3}b shows average absolute reaction energy errors over the families of reactions considered in the W4-11 suite, i.e., total atomization (TAE), bond dissociation (BDE), isomerization (ISO), nucleophilic substitution (SN), and heavy atom transfers (HAT). As seen, SQD lies between CISD and CCSD in terms of accuracy for all families of reactions. However, SQD errors depend on $\zeta$ in different ways for different families of reactions. For bond-breaking reactions (TAE, BDE, and HAT), SQD performs as naturally expected, i.e., for an increasing subspace size, reaction energies become more accurate (the average of $|\Delta \Delta E|$ decreases with $\zeta$). Similarly to the total energy calculations, we find that, without extrapolation, an accuracy comparable to CCSD is not reached. Remarkably, ISO can be accurately described even with comparably small subspaces, i.e., $\zeta = 100$-$200\%$. Notably, we observe a counterintuitive trend in SN, where an increase in the subspace size causes SQD to become less accurate (i.e., the average of $|\Delta \Delta E|$ increases with $\zeta$), and only upon extrapolation one recovers reaction energies comparable to CCSD (see inset of Fig.~\ref{fig:3}b). Fig.~\ref{fig:3}c and Fig.~\ref{fig:3}d report the distribution of $|\Delta \Delta E|$ for TAE and SN reactions, respectively (similar figures for BDE, ISO, and HAT are in the Appendix). For TAE, unextrapolated SQD performs better than MP2 and CISD, which feature outliers of up to $2.5 ~ \hartree$, however, its performance is comparable to ROHF (leftmost violin plot), suggesting this observation may not be general nor the result of a significant cancellation of errors. On the other hand, the average $|\Delta \Delta E|$ decreases appreciably with $\zeta$ along with the spread of the distribution, and upon extrapolation SQD has accuracy comparable with CCSD (see inset of Fig.~\ref{fig:3}c). The situation is different for SN, where the accuracy of unextrapolated SQD is comparable to MP2 and CISD (though not CCSD), however, the average $|\Delta \Delta E|$  increases with $\zeta$, and so does the spread of the distribution. Upon extrapolation, SQD is comparable with CCSD, although with a slightly broader distribution (see inset of Fig.~\ref{fig:3}d). Detailed investigations, including a comprehensive statistical analysis for the individual reaction processes, are performed in Appendix~\ref{sec:hermo-Chemical Reactions}. 

Beyond these averaged and distributional error metrics, Appendix \ref{sec:TAE}-\ref{sec:HAT} provide a detailed chemical-domain deficiency analysis that identifies regimes in which SQD fails to capture the requisite correlation effects, whereas CCSD reliably recovers the correct qualitative and quantitative chemical behavior. Across the different reaction families of the W4-11 dataset, SQD exhibits its largest deviations in electronically complex domains involving delocalized $\pi$-bonding, multi-center electronic structures, electronic reorganizations upon bond cleavage, strong charge separation, and multi-reference rearrangements. Such electronically flexible and correlation-sensitive systems, commonly found in conjugated organics, oxygenated and carbonyl species, and charge-transfer SN reactions, require a balanced description of electron correlation that SQD, particularly without extrapolation, does not reliably provide. In contrast, CCSD remains robust across these cases, with significant challenges arising only for a narrower class of strongly near-degenerate small heteroatomic fragments.

\section*{Extrapolation}
\label{sec:extrapolation}

\begin{figure*}[ht!]
\centering
\includegraphics[width=\textwidth]{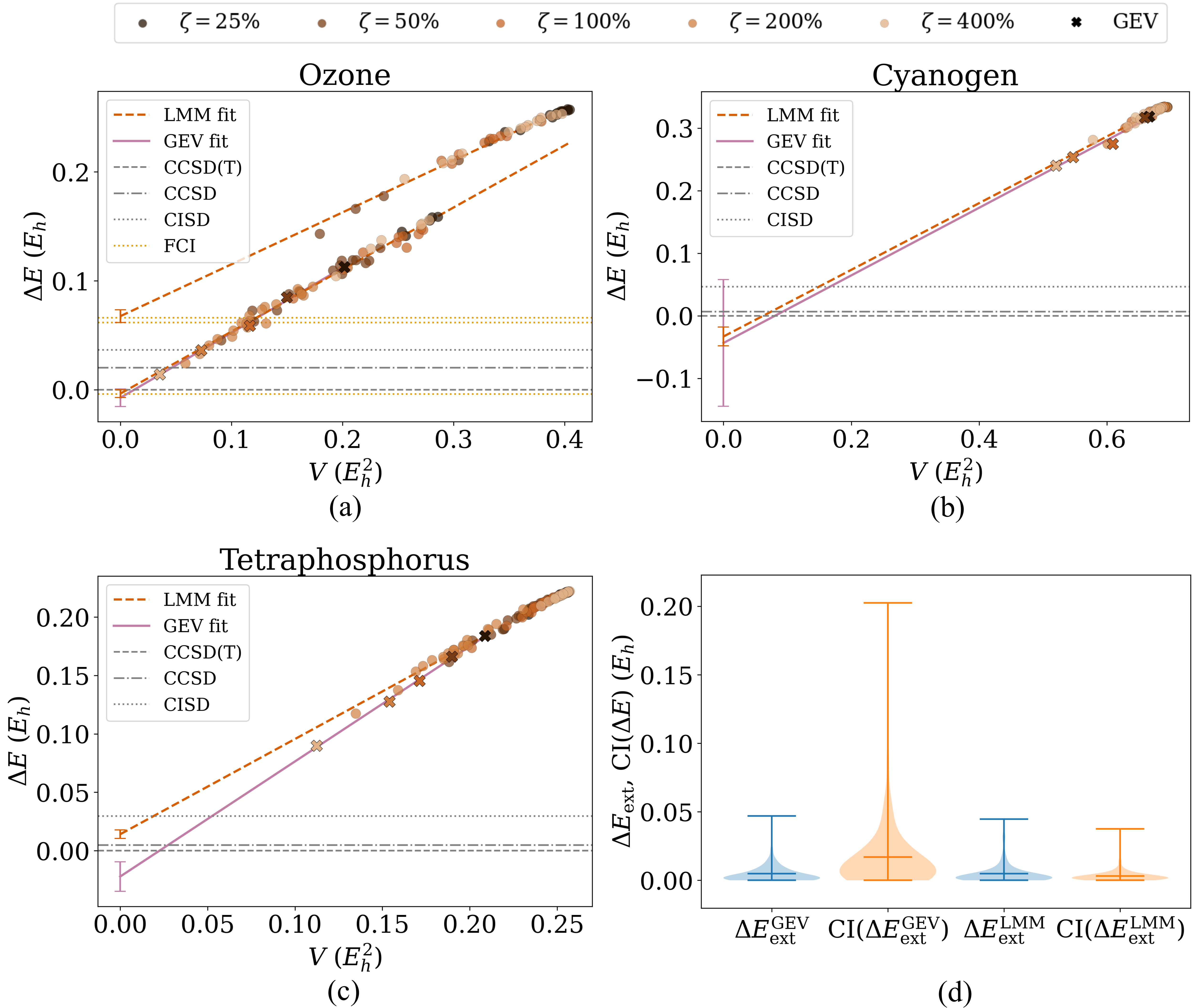}
\caption{(a-c) show energy-variance extrapolations for ozone, cyanogen, and tetraphosphorus. SQD and GEV variance-energy pairs are shown as brown circles and crosses, respectively. Horizontal lines indicate ground-state energies for CISD (dotted), CCSD (dot-dashed), CCSD(T) (dashed), as well as singlet excited-states from FCI (dotted orange lines).
(d) Absolute energy deviations $\Delta E$ between CCSD(T) and SQD extrapolated with LMM and GEV (blue), along with the corresponding statistical uncertainties $\method{CI}(\Delta E)$ across the W4-11 dataset.
}
\label{fig:4}
\end{figure*}

Our observations on extrapolation techniques for SQD energies are collected in Fig.~\ref{fig:3}. We see that raw SQD energies are less accurate than CCSD for total and reaction energies. Short of improvements in raw SQD energies, extrapolation is required to reach CCSD-level accuracy. As exemplified in Fig.~\ref{fig:4}, extrapolations are based on computing $(V_k,E_k) = \big( V(\Psi_{\method{SQD},k}), E(\Psi_{\method{SQD},k}) \big)$, where $V_k=\langle \Psi_{\method{SQD},k} |H^2| \Psi_{\method{SQD},k} \rangle - \langle\Psi_{\method{SQD},k} |H |\Psi_{\method{SQD},k} \rangle ^2$, for a collection of SQD wavefunctions $\Psi_{\method{SQD},k}$ from different values of $\zeta$ and fitting them with a linear regression. The intercept of the regression line provides an estimate of the ground-state energy.

While a reliable extrapolation requires the pairs $(V_k,E_k)$ to approximately lie on a single straight line, this condition is not always met, as illustrated in Fig.~\ref{fig:4}a. A potential remedy is the linear mixture model (LMM), in which the pairs $(V_k,E_k)$ are divided into clusters and each cluster is fit separately. As seen, clustering leads to more reliable extrapolations compared to ordinary linear regression on unclustered data points. One might expect that different clusters extrapolate to distinct eigenvalues. Although this behavior is observed in some instances (see Fig.~\ref{fig:4}a), it does not generally hold, see Appendix~\ref{Appendix:EnergyVarianceAnalysis}. Determining the precise conditions under which such correspondence arises remains an open question requiring further investigation. While LMM extrapolations improve total energy estimates, they may under- or overestimate the reference CCSD(T) values, see Fig.~\ref{fig:4}b and Fig.~\ref{fig:4}c, respectively. Moreover, clustering in the LMM is not automated, requiring manual labeling and case-by-case visual inspection. This introduces subjective elements into the definition and labeling of clusters, thereby limiting methodological consistency and potentially hindering reproducibility.

We propose a more robust and automated alternative to LMM, termed generalized eigenvalue extrapolation (GEV). The central idea is to reduce pollution in the estimated ground-state energies and variances by constructing lowest-energy linear combinations of SQD wavefunctions of the same $\zeta$ value. The energies associated with these linear combinations of states are subsequently extrapolated to obtain approximations to the ground-state energy, see Appendix~\ref{sec:GEV} for a more detailed exposition. As illustrated in Fig.~\ref{fig:4}a, GEV variance-energy pairs (illustrated by crosses) do not show clusters. The GEV extrapolation may incur larger statistical uncertainties in the intercept of the regression line compared to LMM, as illustrated in Fig.~\ref{fig:4}b and Fig.~\ref{fig:4}c, but these statistical uncertainties are observed to be more consistent with actual extrapolated errors. In Fig.~\ref{fig:4}d, we compare the performance of the LMM and GEV extrapolation techniques across the W4-11 dataset. The deviations from CCSD(T), shown in blue, are of similar magnitude for both methods. In contrast, the statistical uncertainties in the extrapolated total energies, shown in orange, are generally larger for GEV than for LMM. A comprehensive list of molecules for which the extrapolated energy using the LMM method or the GEV method is statistically incompatible with CCSD(T) is given in Table \ref{tab:incompatible_molecules}.

\section*{Conclusions}

In summary, we conducted a systematic study of the SQD method across the W4-11 thermochemistry suite. We sampled electronic configurations from quantum circuits using \device{rensselaer}, collecting $10^6$ samples per species for a total of 16.85 hours of computation, and performed pre-, peri-, and post-processing operations using \texttt{AiMOS}, for a total of 724.22 node hours of computation on 20 cores of IBM Power9 processors clocked at 3.15 GHz from the supercomputer \texttt{AiMOS} (75.49 hours wall-clock time for the largest species).

The result is an extensive assessment of SQD across a broad range of diverse chemical systems and reactions, allowing us to characterize its accuracy, precision, and computational cost relative to established classical reference methods. To highlight and quantify the approximations and limitations of SQD, we performed calculations with varying numbers of configurations. The unextrapolated SQD total energies are less accurate than those obtained from classical methods employing a comparable number of configurations (see, e.g., SQD$_{100}$ and CISD energies), indicating substantial opportunities for improvement. The extrapolated SQD total and reaction energies show reasonable agreement with CCSD; however, the extrapolations can be inaccurate or imprecise due to the relatively modest quality of the underlying unextrapolated SQD data, suggesting that improvements in the latter would directly enhance the reliability of the extrapolated results. 
Our study also reveals several noteworthy and unexpected behaviors, including a pronounced scatter in the SQD total energies, limited cancellation of errors in reaction energies, and a counterintuitive increase in deviation from CCSD(T) for SN reaction pathways. 

Our results may serve as a valuable reference for future benchmarks of quantum devices (e.g., across different qubit architectures or successive generations of a specific qubit architecture) and for guiding or calibrating future algorithmic advances in SQD and related quantum computational methods. These advances include, but are not limited to, different
(i) quantum circuits to sample configurations~\cite{yu2025quantum, piccinelli2025quantum}, 
(ii) parameterization of such quantum circuits~\cite{lin2025pushing}, 
(iii) orbital bases~\cite{moreno2023enhancing}, 
(iv) configuration recovery schemes, and 
(v) configuration carryover schemes~\cite{shirakawa2025closed}.
Indeed, by conducting controlled comparisons to the results presented here (i.e., all other factors being equal), researchers will be able to robustly and systematically establish cause-and-effect relationships between algorithmic modifications and the quality of total and reaction energies. Thus, this benchmark study not only provides a broader and deeper understanding of the performance, limitations, and development opportunities of SQD but also offers a foundation with significant potential to inform future methodological research and accelerate progress in quantum computational chemistry.

Beyond the present study, focused on SQD and the STO-6G level of theory, future extensions may include (i) the use of larger basis sets (such as the 6-31G Pople basis~\cite{hehre1972self}, which requires twice the number of qubits and four times the number of gates of this study) to substantiate projections for what accuracy and precision may be expected upon accessing more coherent quantum devices, and (ii) different quantum algorithms, upon conducting a rigorous estimate of the necessary quantum resources and of the impact of quantum noise on computed properties.

Finally, as the sophistication and range of applications of quantum algorithms and devices continue to grow, we conclude by encouraging the benchmarking of future methodological developments aimed at further improving their accuracy, precision, and computational cost against the data presented here. To support such efforts, a formatted version of the W4-11 dataset was created and made available as a Python package~\cite{website}. Moreover, the results presented here are accessible through the RPI W4 Challenge~\cite{w411Challenge} database. 

\section*{Acknowledgements}
This work was supported by IBM through the IBM-Rensselaer Future of Computing Research Collaboration, and HPC resources were made available by the Center for Computational Innovations. The authors would also like to acknowledge Kevin J. Sung for helpful discussions on the LUCJ ansatz and SQD calculations. H.Z.~additionally acknowledges Maria Joseph Lozano, Abdullah Ash Saki, and the qiskit-addon-sqd development team for their support.

\section*{Data Availability}
Within this effort, a formatted version of the W4-11 dataset was created and made available as a Python package~\cite{website} available via the Python package manager pip. The results presented here are accessible through the RPI W4 Challenge~\cite{w411Challenge} database. Code and computational data are available upon request.

\end{refsegment}

\printbibliography[segment=1,heading=bibliography,title={References}]

\pagebreak
\begin{refsegment}
\begin{appendices}
\begin{center}
\section*{From Promise to Practice: Benchmarking
Quantum Chemistry on Quantum Hardware
\\
-- Appendix -- }
\end{center}
\vspace{5mm}
\tableofcontents
\newpage
\section{Background}
\label{secBackground}

\subsection{The W4-11 dataset}

Thermochemistry is a central field in computational (quantum) chemistry. While interesting in itself, it is critical for broader applications across multiple chemical applications, such as green energy conversion, catalysis, and materials science. Thermochemistry is particularly challenging for classical simulations because of the vastly different quantum mechanical principles underlying the different reaction mechanisms. Notably, predicting chemical reaction mechanisms adequately requires highly accurate energy simulations. In our study, we employ the W4-11 dataset~\cite{karton2011w4}, a rigorous benchmark suite for quantum chemical methods that encompasses a diverse range of bonding environments, including single, double, and triple bonds with varying degrees of covalent and ionic character. It comprises 152 molecular species and 745 thermochemical reactions, spanning five key categories: 124 total atomization energies (TAEs), 83 bond dissociation energies (BDEs), 20 isomerization energies (ISO), 505 heavy atom transfer (HAT) reactions, and 13 nucleophilic substitution (SN) reactions. In the following, we briefly review and highlight the chemical and methodological significance of each reaction class and its relevance to broader applications across the physical and life sciences.\\

{\bf Total atomization energies} correspond to reactions in which a molecule is fully dissociated into free atoms, e.g.
\begin{equation}
\mathrm{H_2O \to 2H+O} \;.
\end{equation}
The complete loss of the bonding context makes atomization reactions particularly demanding benchmarks for electronic structure methods~\cite{karton2025highly}. Since all bonds are broken, TAEs typically exceed 1000 {kcal/mol} even for molecules of modest size~\cite{karton2017w4}. As a result, even a relative error of 0.1\% translates to more than 1 kcal/mol. Consequently, atomization reactions serve as a stringent test for evaluating the reliability of quantum chemical methods in thermochemistry~\cite{karton2017w4,karton2022quantum,feller2013improved,karton2011w4,goerigk2011thorough,mardirossian2017thirty,goerigk2017look}.\\

{\bf Bond dissociation energies} quantify the electronic energy required to homolytically cleave a specific bond within a molecule, yielding two radical species, e.g.,
\begin{equation}
\mathrm{CH_4 \to CH_3^\bullet+C^\bullet} \;,
\end{equation}
providing an approximation of the overall reaction enthalpy~\cite{benson1976thermochemical}. BDEs are widely employed in the prediction of reaction kinetics~\cite{gani2018understanding}, and offer valuable information on thermodynamically accessible reaction pathways, often serving as the first screening step to identify dominant mechanisms in diverse applications. This includes combustion~\cite{kim2019experimental}, polymer synthesis~\cite{lin2011linear} and thermal stability~\cite{giannetti2005thermal,bian2016thermal}, lignin depolymerization~\cite{kim2011computational}, drug metabolism~\cite{lienard2015predicting,drew2012impact,zhao2005assessment}, and design of energetic materials~\cite{harris1997ab,warr2014short,ahneman2018predicting,wilcox2018stable}.\\

\newpage
{\bf Isomerization energies} are energy differences between molecular isomers, i.e., molecules with the same chemical formula but different structural arrangements, for example, propyne ($\mathrm{CH_3C \equiv CH}$) and allene ($\mathrm{H_2C=C=CH_2}$). Because isomers often differ only in the arrangement of a few bonds or torsional angles, their total electronic energies lie close together (typically within 1–20~kcal/mol)~\cite{goerigk2017look}. This proximity leads to substantial cancellation of systematic errors in absolute energies, making isomerization energies an especially stringent test of a method’s ability to capture fine‑scale electron correlation and subtle changes in electronic structure~\cite{bak2000accuracy, ruzsinszky2015insight,lesiuk2022quintic,grimme2007compute,dasgupta2017standard,goerigk2017look}. Accurate isomerization energetics are crucial for predicting conformational equilibria in biomolecules~\cite{wedemeyer2002proline}, reaction selectivity in catalysis~\cite{laplaza2024overcoming,zahrt2019prediction}, and phase behavior in materials~\cite{yu2004alignment,legge1992photo}, hence, they serve as a sensitive probe of both the accuracy and transferability of quantum chemical methods.\\

{\bf Heavy-atom transfer} reactions involve migration of a non-hydrogen atom between fragments, e.g., 
\begin{equation}
\mathrm{CO+OH \to CO_2+H} \;.
\end{equation}
Such processes typically entail simultaneous changes in oxidation state~\cite{chen2015reactivity,holm1987metal}, bond order~\cite{masunov2016chemical}, and often spin multiplicity~\cite{harvey2007understanding,schroder2000two}, making them especially sensitive to both dynamic and static electron-correlation effects~\cite{zhao2005benchmark,cohen2012challenges}. Reaction energies in this class can span tens to hundreds of kilocalories per mole~\cite{semidalas2020canonical,goerigk2017look}, hence, small errors in electronic structure can lead to qualitatively incorrect predictions of thermodynamics or kinetics~\cite{dzib2024enhancing}. Heavy-atom transfer is fundamental to combustion chemistry~\cite{baulch2005evaluated}, atmospheric processes~\cite{atkinson2003atmospheric} (e.g. the formation and degradation of pollutants~\cite{saunders2003protocol}), and organometallic catalysis~\cite{wang2017oxygen}, where accurate energetics guide catalyst design and mechanistic understanding. Because these reactions probe diverse bonding environments and multireference character, they provide a rigorous test of a method’s ability to capture complex electronic rearrangements\\

{\bf Nucleophilic substitution} reactions replace a leaving group on a saturated center with a nucleophile, e.g., 
\begin{equation}
\mathrm{CH_3F+H \to F+CH_4} \;.
\end{equation}
Despite their apparent simplicity, accurate treatment requires robust handling of charge separation~\cite{hamlin2018nucleophilic}, polarization~\cite{kubelka2017activation}, and long-range correlation~\cite{hohenstein2012wavefunction}. Nucleophilic substitution reaction energies benchmark a method’s description of ionic and highly polar transition structures and intermediates~\cite{chabinyc1998gas}, with broad relevance to mechanism design in organic synthesis~\cite{houk2008computational} and to reactive trajectories in the gas phase~\cite{mikosch2008imaging}.

\pagebreak
\subsection{The Electronic Schr{\"o}diner Equation}

A central component of essentially all electronic-structure algorithms is the discretization of the electronic Hamiltonian into a tractable finite representation. To that end, the Hamiltonian that is originally defined over a continuous many-electron configuration space is mapped to a discrete space
suitable for numerical computation. The continuous electronic Hamiltonian in atomic units reads
\begin{equation}
H
= -\sum_i \frac{\nabla_{\mathbf{r}_i}^2}{2}
- \sum_{I,j} \frac{Z_I}{|\mathbf{R}_I - \mathbf{r}_j|}
+ \sum_{i<j} \frac{1}{|\mathbf{r}_i - \mathbf{r}_j|}
+ E_{\text{II}},
\end{equation}
where we have assumed the Born-Oppenheimer approximation so that the nuclear positions $\mathbf{R}_I$ are fixed, contributing only the constant nuclear-nuclear repulsion energy $E_{\text{II}}$. The variables $\mathbf{r}_i$ denote the electronic coordinates, and $Z_I$ denotes the nuclear charge. 

In a Galerkin discretization, one selects a set of (spatial) basis functions $\{\chi_p\}$ and imposes electronic antisymmetry at the operator level. The Hamiltonian can then be written in its standard second-quantized form. Using the usual chemists' notation for one- and two-electron integrals, the Hamiltonian in an arbitrary orthonormal basis becomes
\begin{equation}
\label{eq:elecHam}
  \hat{H}
  = \sum_{p r,\sigma} h_{pr}\,
      \hat{a}^\dagger_{p\sigma} \hat{a}_{r\sigma}
    + \frac{1}{2} \sum_{p r q s,\sigma\tau}
      (pr|qs)\,
      \hat{a}^\dagger_{p\sigma}
      \hat{a}^\dagger_{q\tau}
      \hat{a}_{s\tau}
      \hat{a}_{r\sigma},
\end{equation}
where the one-electron integrals are
\begin{equation}
  h_{pr}
  = \int d\mathbf{r}\;\overline{\chi}_p(\mathbf{r})
    \left[
      -\frac{\nabla_{\mathbf{r}}^2}{2}
      - \sum_I \frac{Z_I}{|\mathbf{R}_I - \mathbf{r}|}
    \right]
    \chi_r(\mathbf{r}),
\end{equation}
and the two-electron Coulomb integrals are
\begin{equation}
  (pr|qs)
  = \int d\mathbf{r}\,d\mathbf{r}'\;
    \frac{
      {\chi}^*_p(\mathbf{r})\chi_r(\mathbf{r})\,
      {\chi}^*_q(\mathbf{r}')\chi_s(\mathbf{r}')
    }{|\mathbf{r}-\mathbf{r}'|}.
\end{equation}

In this work, we employ the minimal STO-6G basis set. However, in the context of quantum simulations of electronic systems, there is a significant ongoing effort to design basis representations that compress the Hamiltonian while retaining systematic improvability~\cite{liu2022quantum}. Such compact bases aim to reduce qubit counts, gate complexity, and sparsity overheads, making electronic-structure simulation more tractable on near-term hardware. Recent approaches include real-space grid and finite-element bases \cite{berry2019qubitization,kassal2008polynomial,kivlichan2017bounding,chan2023grid}, multi-resolution or wavelet-based bases \cite{su2021fault,childs2022quantum,babbush2019quantum,berry2024quantum,georges2025quantum, mcclean2020discontinuous,faulstich2022discontinuous, feniou2025real}, and numerical atomic orbital and adaptive local bases \cite{barison2022quantum,motta2021low,kwon2023adaptive}. 

\pagebreak
\subsection{Overview of Quantum Algorithms}

A wide spectrum, in terms of resource requirements, of quantum algorithms for ground-state energy estimation have been proposed, given a procedure to prepare a quantum state approximating the ground state. Well-known algorithms include Quantum Phase Estimation (QPE) \cite{kitaev1995qpe} and the Hadamard Test (HT) and its variants \cite{Ding2023simultaneous, Ding2024quantummultiple} and the targets of this study, the Variational Quantum Eigensolver \cite{peruzzo2014vqe} and the Sample-based Quantum Diagonalization \cite{robledo2025chemistry}.

QPE estimates the ground-state energy by measuring the Hamiltonian operator over a trial state $\Psi_T$. The measurement is implemented by a quantum circuit comprising controlled powers of an approximation to the time evolution operator $\exp(-it \operator{H})$, which collapses the trial state onto an approximation of the ground state with probability $| \langle \Psi_T | \Psi_{\mathrm{gs}} \rangle|^2$. The cost to obtain the ground-state energy to precision $\varepsilon$ is~\cite{lee2023evaluating}
\begin{equation}
[C_{\Psi_T}+\mathrm{poly}(\mathrm{size}) \mathrm{poly}(1/\varepsilon)] \, \mathrm{poly}\left[ \frac{1}{| \langle \Psi_T | \Psi_{\mathrm{gs}} \rangle|} \right]
\end{equation}
which is optimal in the sense that it saturates the Heisenberg limit, with accuracy scaling as $1/\varepsilon$. However, QPE is not within reach of contemporary hardware due to prohibitive circuit depths and numbers of controlled operations, and practical QPE calculations need trial states having sufficiently high overlap with the ground state, which is a profound motivation for the development of complementary methods for ground-state approximation. Although the HT does not saturate the Heisenberg limit, recent algorithmic developments have achieved this limit through classical post-processing techniques. Nevertheless, the HT too is not within reach of contemporary hardware for the entire range of species considered in this study.

Before the availability of quantum error correction and fault-tolerant quantum computation, the development of devices based on physical qubits (i.e., subject to unwanted interaction with the environment and imperfect implementation of quantum operations) has motivated researchers to propose and validate a broad set of heuristic methods~\cite{bauer2020quantum,georgescu2014quantum,cao2019quantum,cerezo2020variational}. Among these, a well-known and intensely studied method is the Variational Quantum Eigensolver (VQE)~\cite{peruzzo2014vqe}. The VQE algorithm is a hybrid quantum-classical method, consisting of a parametrized quantum circuit serving as a variational ansatz to prepare trial states, and a classical optimization method to optimize the circuit parameters.

A parametrized quantum circuit $\ansatz{U}(\theta)$ operating on a reference state $| \psi _0 \rangle$ (typically chosen to be the Hartree-Fock state) prepares a second-quantized trial state,
\begin{equation}
\label{eq:ansatz}
| \psi (\theta) \rangle = \ansatz{U}(\theta) |\psi_0\rangle
\;,
\end{equation}
which is optimized as
\begin{equation}
\theta^* = \underset{\theta}{\mathrm{argmin}} \; C(\theta)
\;,
\end{equation}
where
\begin{equation} 
\label{eq:cost}
C(\theta) = \langle \psi(\theta)| \operator{H} | \psi (\theta) \rangle \;.
\end{equation}

\paragraph{Cost function evaluation}

The cost function Eq.~\eqref{eq:cost} may be computed by representing the Hamiltonian operator as a sum of measurable terms (e.g. Pauli string operators, although other choices are possible) and estimating the expectation value of each term separately,
\begin{equation}
\operator{H} = \sum_k c_k \operator{P}_k
\;,
\end{equation}
where $\operator{H}$ is the second-quantized electronic structure Hamiltonian c.f.~\eqref{eq:elecHam}. Resolving~\eqref{eq:cost} within precision $\varepsilon$ requires executing $\mathcal{O}(\|c\|_1^2/\varepsilon^2)$ circuits~\cite{mcclean2016theory}, and the large ratio $\|c\|_1/\varepsilon$ poses a substantial obstacle to large-scale VQE simulations of electronic structure~\cite{patel2025quantum}, motivating the development of alternative methods including SQD. 

\paragraph{Parameter optimization}

The accuracy of VQE depends in part on the ability to optimize the variational parameters $\theta$. A well-known phenomenon is the appearance of exponentially increasing numbers of local minima with respect to the number of parameters, due to which seeking a global optimum is NP-hard in general. As the number of parameters increases, barren plateaus and narrow gorges also emerge in the optimization landscape, which leads to difficulties in determining optimization directions~\cite{cerezo2021variational}. The presence of large statistical uncertainties in $C(\theta)$ further challenges the convergence of optimization operations. Remedying these challenges is an active area of research, and techniques like local cost functions~\cite{cerezo2021local} and stochastic approximations~\cite{spall2002multivariate} are typically suggested to improve convergence. 

\paragraph{Ansatz choice}

The accuracy of VQE depends on the choice of the Ansatz in Eq.~\eqref{eq:ansatz}. In Section~\ref{sec:vqe} we present a well-established VQE ansatz for electronic structure simulations on quantum computers and, in Section~\ref{sec:LUCJ}, we present a simplification of such an Ansatz used in this study.

\begin{figure*}[ht!]
    \centering
    \begin{subfigure}[t]{0.45\textwidth}
    \includegraphics[width=\textwidth]{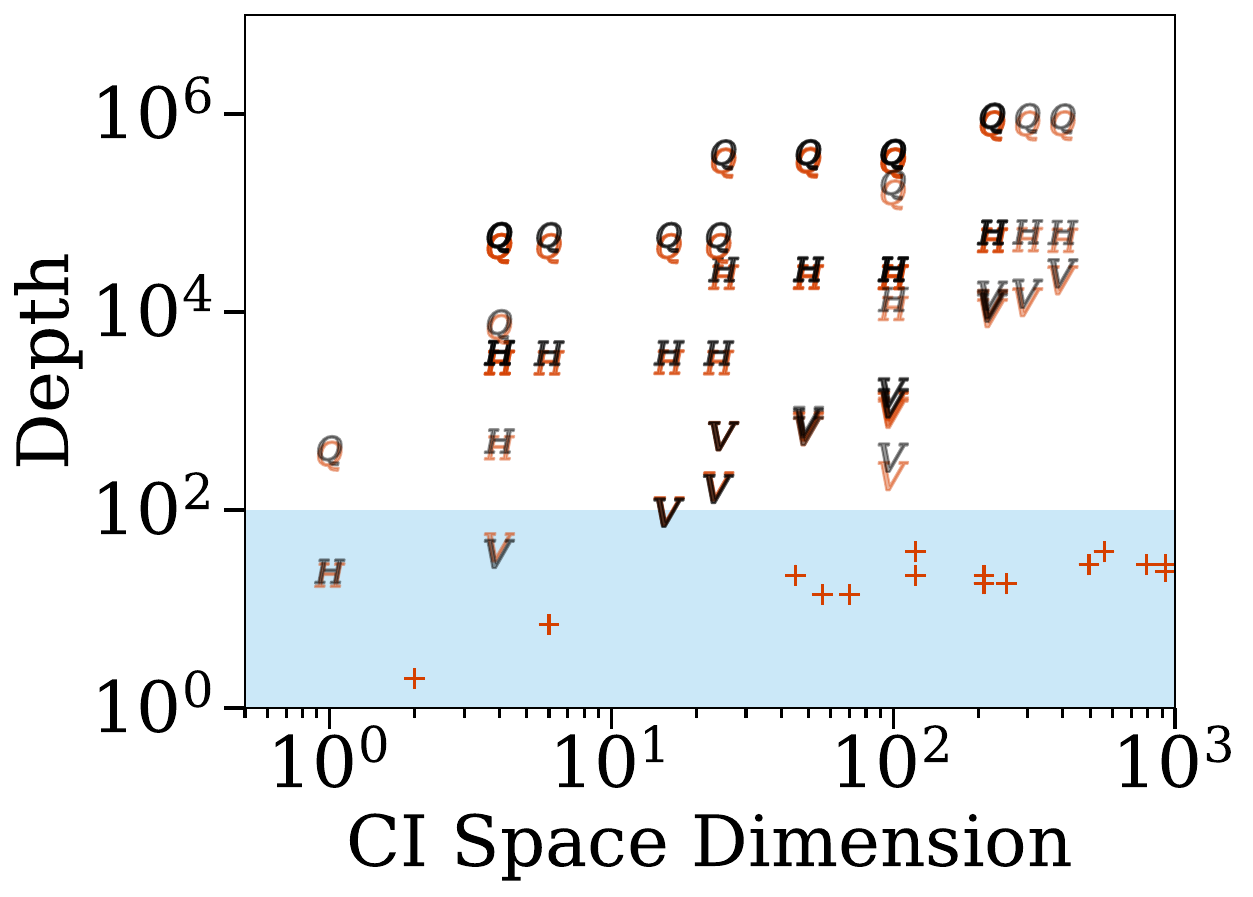}
    \caption{}
    \end{subfigure}
    \begin{subfigure}[t]{0.45\textwidth}
    \includegraphics[width=\textwidth]{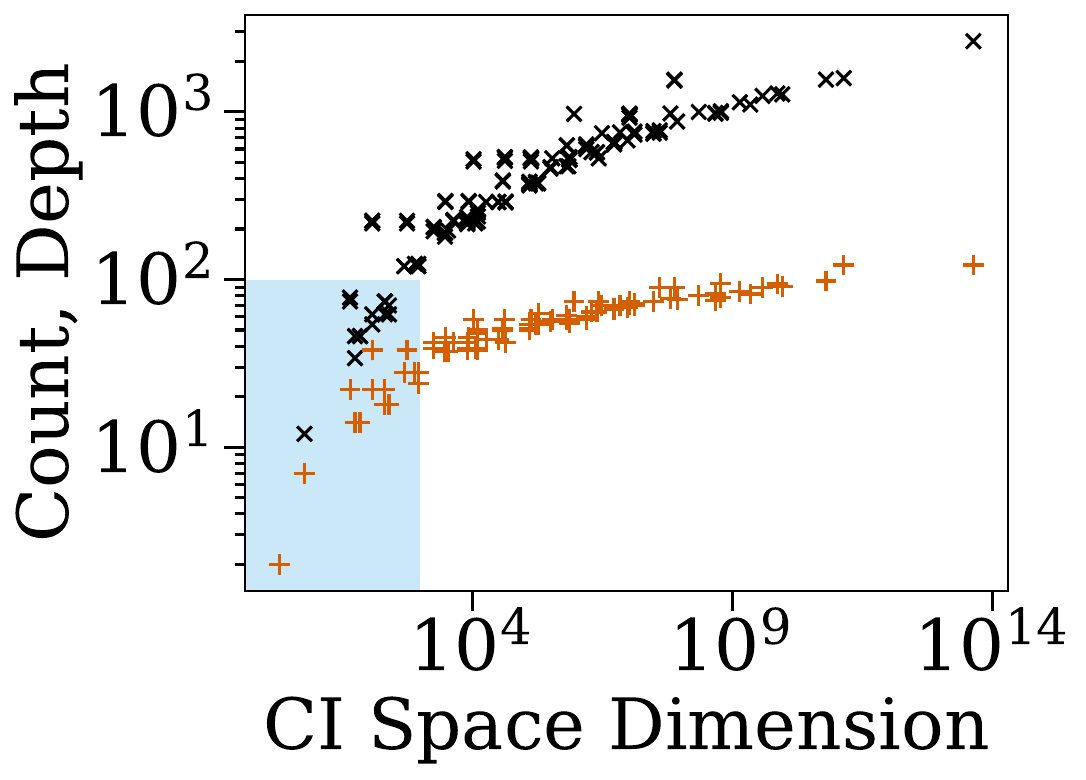}
    \caption{}
    \end{subfigure}
    \hfill
\caption{\label{fig:resource_estimates}
(a) Circuit depth for different quantum algorithms. We denote by ``V'', ``H'', ``Q'' and ``+'' the gate depth of VQE using USCC (converged to 1 m$E_h$ of reference CCSD(T) energy), HT, QPE, and LUJC ansatz, respectively. The orange and black colors describe the Jordan-Wigner and Bravyi-Kitaev mappings, respectively. (b) Gate count (black ``x'') circuit and depth (orange ``'+') of the LUCJ ansatz for all species in the dataset.}
\end{figure*}

\section{Methods}
\label{sec2}

\subsection{The Unitary Coupled Cluster Ansatz} \label{sec:vqe}

The unitary coupled cluster (UCC) Ansatz can be written as
\begin{equation}
\ansatz{U}_{\mathrm{UCC}}(\theta) = 
e^{ \operator{T} - \operator{T}^\dagger }
\end{equation}
where $\operator{T} = \sum_{i=1}^{n_e} \operator{T}_i$ is a linear combination of $n_e$-fold fermionic excitation operators. The included excitations, indexed by $i$, are typically truncated up to double excitations, resulting in the UCCSD (UCC singles and doubles) ansatz. The UCC ansatz can be approximated with the first-order Lie-Trotter formula
\begin{equation}
\label{eq:ucc}
e^{ \operator{T} - \operator{T}^\dagger }
\approx
\left[ \prod_{i=1}^{n_e} e^{ \frac{1}{r} \, (\operator{T}_i - \operator{T}_i ^\dagger) }
\right]^r
+ \mathcal{O}(1/r)
\end{equation}
where $r$ is the number of Trotter steps. One may use higher order formulae \cite{suzuki1976productformulae} or post-Trotter methods \cite{low2017qsphamsim,childs2012lcuhamsim} to approximate $\ansatz{U}_{\mathrm{UCC}}(\theta)$ with higher accuracy for a given number $r$ of Trotter steps.
We note that the ordering of terms in the Trotter sequence plays an important role in the approximation error~\cite{grimsley2020is,childs2021theory}.
In this study, we focus on a first-order Trotter approximation with $r=1$, motivated by hardware constraints on circuit depth for near-term quantum devices. This choice is theoretically justified by the proof that the UCC ansatze with single Trotter steps can exactly parametrize arbitrary fermionic wavefunctions \cite{evangelista2019exact}, and empirically supported by numerical studies demonstrating that they achieve ground state energies within chemical accuracy (1 kcal/mol) for some molecular systems~\cite{barkoutsos2018quantum}.

\paragraph{Cost of quantum circuits}

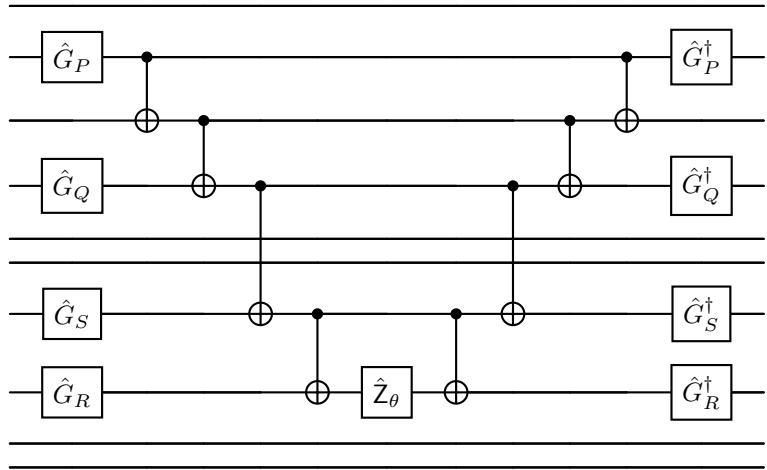
\begin{figure}
\centering
\begin{quantikz}[row sep=9, column sep=12]
\qw & \qw & \qw & \qw & \qw & \qw & \qw & \qw & \qw & \qw & \qw & \qw & \qw \\
\qw & \gate{\ansatz{G}_P} & \ctrl{1} & \qw & \qw & \qw & \qw & \qw & \qw & \qw & \ctrl{1} & \gate{\ansatz{G}^\dagger_P} & \qw \\
\qw & \qw & \targ{} & \ctrl{1} & \qw & \qw & \qw & \qw & \qw & \ctrl{1} & \targ{} & \qw & \qw \\
\qw & \gate{\ansatz{G}_Q} & \qw & \targ{} & \ctrl{3} & \qw & \qw & \qw & \ctrl{3} & \targ{} & \qw & \gate{\ansatz{G}^\dagger_Q} & \qw \\
\qw & \qw & \qw & \qw & \qw & \qw & \qw & \qw & \qw & \qw & \qw & \qw & \qw \\
\qw & \qw & \qw & \qw & \qw & \qw & \qw & \qw & \qw & \qw & \qw & \qw & \qw \\
\qw & \gate{\ansatz{G}_S} & \qw & \qw & \targ{} & \ctrl{1} & \qw & \ctrl{1} & \targ{} & \qw & \qw & \gate{\ansatz{G}^\dagger_S} & \qw \\
\qw & \gate{\ansatz{G}_R} & \qw & \qw & \qw & \targ{} & \gate{\pauli{Z}_\theta} & \targ{} & \qw & \qw & \qw & \gate{\ansatz{G}^\dagger_R} & \qw \\
\qw & \qw & \qw & \qw & \qw & \qw & \qw & \qw & \qw & \qw & \qw & \qw & \qw \\
\qw & \qw & \qw & \qw & \qw & \qw & \qw & \qw & \qw & \qw & \qw & \qw & \qw \\
\end{quantikz}
\caption{Quantum circuit to implement the exponential of the Pauli operator in Eq.~\eqref{eq:pauli_cnot_string}, with $n=10$ qubits and $R,S,Q,P = 3,4,7,9$. The single-qubit gates are $\ansatz{G} = H, SH$ for $\pauli{M} = \pauli{X}, \pauli{Y}$, and $\pauli{Z}_\theta$ denotes a single-qubit Z rotation of an angle $\theta$.}
\label{fig:cnot-ladder}
\end{figure}

The Ansatz~\eqref{eq:ucc} can be implemented by a quantum circuit by mapping fermionic excitation operators onto qubit operators. As fermionic operators obey canonical anticommutation relations, their qubit representations are non-local, to an extent determined by the specific choice of fermion-to-qubit mapping used.
The Jordan-Wigner~\cite{jordan1928jwtransform} and parity transformations~\cite{bravyi2002fermionicqc} produce qubit operators that are up to $\mathcal{O}(n)$-local. For example, using the Jordan-Wigner transformation, the two-electron operator $\crt{P} \crt{Q} \dst{S} \dst{R}$, with $R<S<Q<P$ labeling spin-orbitals, maps to a linear combination of Pauli strings of the form
\begin{equation}
\label{eq:pauli_cnot_string}
\pauli{M}_P
\left[ \prod_{j=Q+1}^{P-1} \pauli{Z}_j \right]
\pauli{M}_Q
\pauli{M}_S
\left[ \prod_{j=R+1}^{S-1} \pauli{Z}_j \right]
\pauli{M}_R
\end{equation}
where $\pauli{M} \in \{ \pauli{X}, \pauli{Y} \}$. The quantum circuit implementing the exponential of such a Pauli string is shown in Figure \ref{fig:cnot-ladder}.

Conversely, the Bravyi-Kitaev transformation \cite{bravyi2002fermionicqc} produces $\mathcal{O}(\log n)$-local qubit operators. The Jordan-Wigner and Bravyi-Kitaev transformations have been compared for the hydrogen molecule in the minimal basis~\cite{seeley2012} and for larger molecules~\cite{tranter2018}. 
For small systems, e.g. the hydrogen molecule, although the Bravyi-Kitaev transformation may lead to lower total gate counts, the number of two-qubit gates may be higher \cite{seeley2012}. 
Note that, although the operators produced by the Bravyi-Kitaev transformation are local in an abstract quantum circuit, a physical layout on a quantum computer without all-to-all connectivity requires a substantial overhead of SWAP operations for routing.

Large numbers of two-qubit gates and high two-qubit gate depths lead to rapid accumulation of errors on pre-fault-tolerant devices. This accumulation of errors is another obstacle to electronic structure simulations with the VQE method and the UCC ansatz on present-day quantum devices.

\paragraph{Economization of quantum circuits}

We note that several approaches, of different natures, have been proposed to lower the gate count of VQE simulations based on the UCC Ansatz.
\begin{itemize}
\item In terms of circuit transpilation, some studies suggest canceling out CNOT chains between subsequent excitation operations through circuit transformations~\cite{gujarati2023quantum}. The $2n$  CNOT gates contributing a depth of $2n$ in the ``V" structure of CNOT chains in Fig.~\ref{fig:cnot-ladder} may be optimized in circuit depth to $2\log{n}$ using a balanced-tree form \cite{cowtan2020phasegadget}, albeit with the same two-qubit gate count. One may also use dynamic circuits with $\mathcal{O}(n)$ ancilla qubits to reduce the overall depth to $\mathcal{O}(1)$ \cite{moflic2024constantdepthimplementationpauli}. Furthermore, fermionic SWAP networks \cite{kivliochan2018fermionicswapnetwork} have also been proposed for systems with spatially local interactions for shorter circuit depths, scaling as $\mathcal{O}(\sqrt{d})$.
\item In terms of fermion-to-qubit mapping, the Bravyi-Kitaev Superfast (BKSF) \cite{setia2018bksf} mapping can significantly improve on the locality of qubit operators for problems with local fermionic interactions, scaling as $\mathcal{O}(d)$ where $d$ is the degree of interaction.
\item 
In terms of Ansatz design, one may choose to retain only the most relevant excitations through an iterative selection procedure, e.g., the unitary selective coupled cluster (USCC) method ~\cite{fedorov2022unitary}. 
\end{itemize}

\begin{figure*}[ht!]
    \centering
    \includegraphics[width=.9\textwidth]{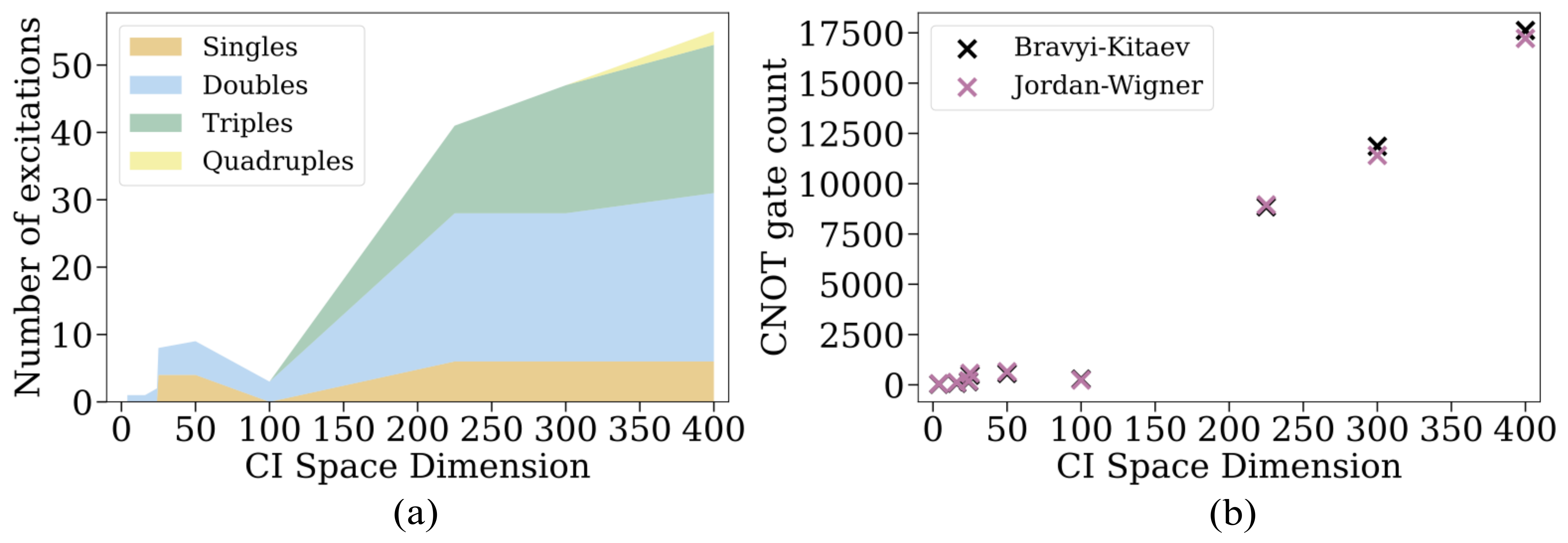}
\caption{\label{fig:uscc_resources} (a) Minimum number of USCC excitations (singles, doubles, triples, and quadruples) required to converge within 1 m$E_h$ of reference CCSD(T) energy as a function of CI space dimension across the W4-11 dataset. (b) Minimum CNOT gate counts to converge within 1 m$E_h$ of reference CCSD(T) energy for various CI space dimensions across the dataset.}
\end{figure*}

We provide resource estimates for species in the W4-11 dataset using up to 5 spatial orbitals in a minimal basis set and a frozen-core approximation. The USCC ansatz was iteratively grown until convergence to within 1 m$E_h$ of the reference CCSD(T) energy on a classical simulator, as shown in Fig.~\ref{fig:uscc_resources}. Resource estimates were obtained by decomposing the final converged ansatz circuits and transpiling them using Qiskit's preset pass managers at optimization levels 0-3 for the \device{rensselaer} backend and picking circuits with the lowest two-qubit gate depths. The results indicate that, with the preset circuit optimization, the transpiled two-qubit gate depths grow rapidly with CI space dimension. Due to the presence of noise on the current quantum hardware, these depth requirements place meaningful simulations of the W4-11 dataset using the USCC ansatz beyond the capabilities of near-term devices, even for the small systems in our test set.

These resource estimates motivate the exploration of alternative economizations of the UCC ansatz that reduce gate counts and circuit depths while preserving accuracy. The local unitary cluster Jastrow (LUCJ) ansatz, which we review in Section~\ref{sec:LUCJ}, represents one such hardware-efficient approach and forms a key ingredient of this study.

\subsection{Local Unitary Cluster Jastrow Ansatz}
\label{sec:LUCJ}

While the UCC and USCC ansatz are theoretically well established, their hardware implementation is hindered by the high gate count encountered in circuit transpilation. For platforms based on superconducting qubits, the hardware has a limited qubit connectivity, which means SWAP gates are needed to realize interactions between a generic pair of orbitals. Physical qubits also have limited coherence time -- typically on the order of 100 microseconds for superconducting qubits -- which limits the duration of circuits that one can reliably run. The unitary cluster Jastrow (UCJ) ansatz~\cite{matsuzawa2020jastrow} and local UCJ (LUCJ) ansatz were proposed to address those challenges, respectively, by introducing a truncation in the Jastrow expansion and restricting the ansatz to include only local interactions.

The UCJ ansatz consists has the following form
\begin{equation}
\label{eq:UCC}
\ansatz{U}_{\mathrm{UCJ}}=\prod_{\mu=1}^L 
e^{ \operator{K}_\mu}
e^{i\operator{J}_\mu}
e^{-\operator{K}_\mu}
\end{equation}
where
\begin{equation}
\operator{K}_\mu
=
\sum_{pq,\sigma} K^\mu_{pq} \, 
\crt{p\sigma} \dst{q\sigma}
\;,\;
\operator{J}_\mu=\sum_{pq,\sigma\tau} J_{pq,\sigma\tau}^\mu \, 
\crt{p\sigma} \dst{p\sigma} \, \crt{q\tau} \dst{q\tau}
\;,
\end{equation}
are one-body operators. $\exp(\operator{K}_\mu)$ and $\exp(\operator{J}_\mu)$ represent orbital rotations and diagonal Coulomb interactions respectively, and $pq$ label spatial orbitals while $\sigma\tau$ label spin polarizations. The form of the ansatz can be derived as a low-rank decomposition of the $t_2$ amplitudes of the unitary coupled cluster with singles and doubles (UCCSD)~\cite{motta2021low}, where the number $L$ of terms in the product corresponds to the number of terms in the decomposition. With large enough $L$ and suitable parameters $K^\mu_{pq}$ and $J_{pq,\sigma\tau}^\mu$, UCJ reproduces the UCCSD wavefunction. However, in UCJ literature~\cite{matsuzawa2020jastrow}, these parameters are independent of the underlying set of $t_2$ amplitudes.

\begin{figure*}[ht!]
    \centering
    \includegraphics[width=0.9\textwidth]{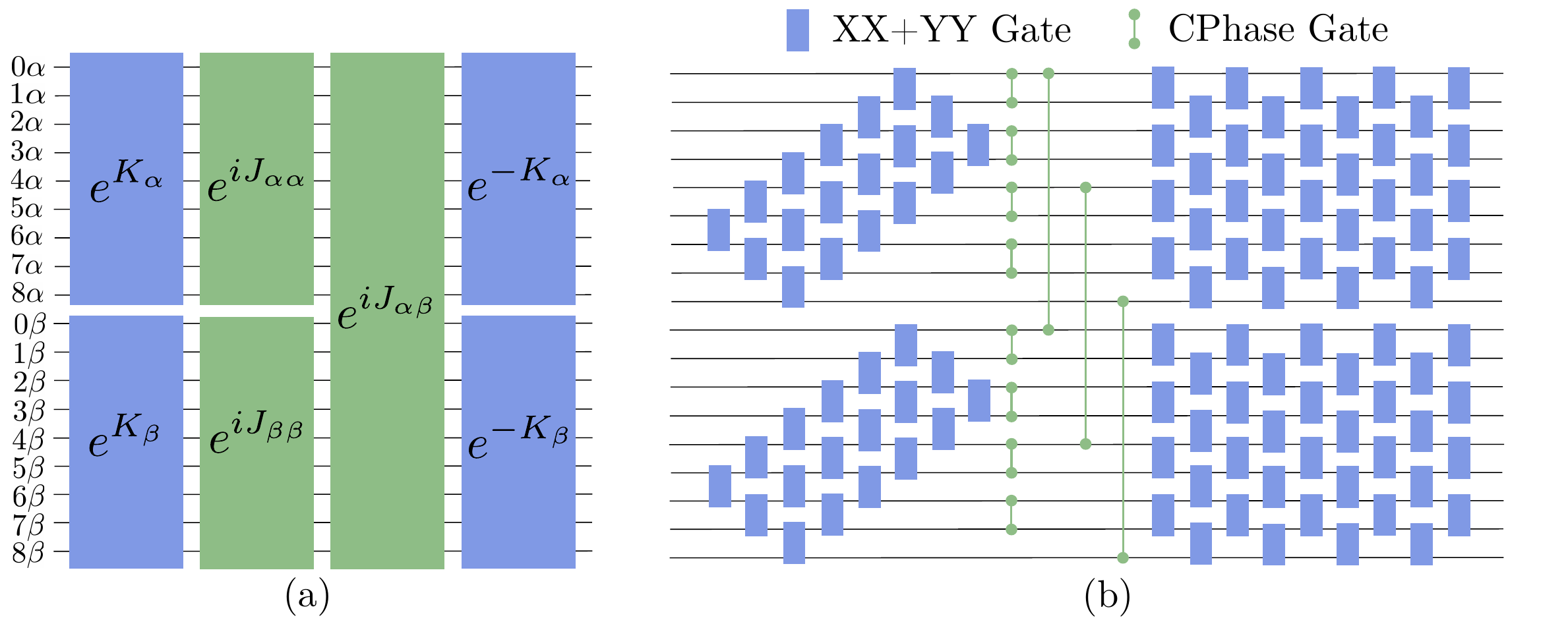}
\caption{\label{fig:lucj} (a) Structure of a single layer of the UCJ or LUCJ ansatz, with green and blue blocks denoting orbital rotations and diagonal Coulomb interactions, respectively (b) Decomposition of a single layer of the LUCJ ansatz into a quantum circuit of XX+YY gates implementing orbital rotations (blue rectangles) and ZZ gates implementing diagonal Coulomb interactions on qubits that are physically adjacent or connected by a single ancilla in a device with heavy-hex connectivity (green symbols). }
\label{fig:ucjlucj}
\end{figure*}

The unitary operations in Eq.~\eqref{eq:UCC} can be decomposed into quantum gates on universal quantum computers. This involves mapping creation and annihilation operators onto linear combinations of Pauli operators via, e.g., the Jordan-Wigner (JW) transformation, and then decomposing each exponential in Eq.~\eqref{eq:UCC} into a product of Pauli operators implemented by the single- and two-qubit quantum gates available on the hardware. In the JW representation, each orbital rotation $\exp(\pm \operator{K}_\mu)$ in the UCJ ansatz can be expressed as a circuit of Givens rotations~\cite{jiang2018quantum}, which can be implemented by $O(N_q^2)$ two-qubit exponentials of Pauli $\pauli{X} \otimes \pauli{X}$ and $\pauli{Y} \otimes \pauli{Y}$ gates acting on adjacent qubits (blue blocks in Fig. \ref{fig:ucjlucj}) with depth $O(N_q)$. The exponential of the Jastrow operator $\exp(i\operator{J}_\mu)$ is mapped to a product of $O(N_q^2)$ two-qubit $\exp(\pauli{Z} \otimes \pauli{Z})$ gates acting on pairs $(p\sigma,q\tau)$ of spin-orbitals. As a result, implementing the Jastrow operator requires all-to-all qubit connectivity or the use of a SWAP network. 

The local UCJ ansatz (LUCJ)~\cite{motta2023bridging} introduces a modification to the UCJ ansatz by excluding $J_{pq,\sigma\tau}^\mu$ terms leading to excessively expensive SWAP networks. In this study, we included only the following terms in $\operator{J}_\mu$:
\begin{itemize}
\item $J_{pq,\alpha\alpha}^\mu$ with $q=p+1$
\item $J_{pq,\beta\beta}^\mu$ with $q=p+1$
\item $J_{pq,\alpha\beta}^\mu$ with $q=p$ and $p \,  \mathrm{mod} \, 4 = 0$
\end{itemize}
which requires $O(N_q)$ gates acting on adjacent qubits within the same spin sector (green blocks in Fig. \ref{fig:ucjlucj}) and qubits in opposite spin sectors that are connected via at most one ancilla qubit on a device with heavy-hex connectivity (green blocks in Fig. \ref{fig:ucjlucj}), to minimize the number of SWAP gates required, resulting in depth $O(1)$~\cite{motta2023bridging}. To ensure the circuit duration is within the qubit coherence budget, we restrict our consideration to $L=1$. The resulting circuit sizes and depth for the molecules in the dataset are summarized in Fig.~\ref{fig:2}(a). The parameters in the LUCJ ansatz are defined performing a double-factorized decomposition~\cite{robledo2025chemistry} of the $t_2$ amplitude tensors of a classical coupled-cluster calculation carried out with the \software{PySCF} software~\cite{sun2018pyscf,sun2020recent}, and then truncated to the local form in Eq.~\eqref{eq:UCC} and to a single layer, $L=1$. These truncations lead to a reduction in circuit sizes to run on the hardware at the cost of introducing truncation errors. Further optimization for the circuit parameters is an open question but beyond the scope of this benchmark \cite{Shirakawa2025closedloop}. For open-shell systems, we restrict the $t_2$ amplitude tensors to a subset of double excitations, specifically, excitations of 
\begin{itemize}
\item opposite-spin electrons from doubly-occupied to virtual orbitals
\item same-spin electrons from doubly-occupied to singly-occupied and virtual orbitals
\end{itemize}
LUCJ circuits were constructed and parametrized using the \software{ffsim} software package \cite{ffsim}. 

\subsection{Sample-Based Quantum Diagonalization}
\label{sec:sqd}

The recently-proposed Sample-Based Quantum Diagonalization (SQD) method has lead to several recent studies on computing ground and excited states, isomers, and various other problems in quantum chemistry~\cite{robledo2025chemistry,kaliakin2025accurate,shajan2025toward,kaliakin2025implicit,bazayeva2025quantum,danilov2025enhancing,liepuoniute2025quantum,duriez2025computing,barroca2025surface,smith2025quantum,barison2025quantum}. It has established itself as a practical method for electronic structure simulations on present-day quantum devices, making the systematic assessment of its accuracy and precision over a standardized database of use cases a timely and compelling research goal.

The central assumption underlying SQD is that the ground-state wavefunction of $(N_\alpha,N_\beta)$ electrons in $M$ spatial orbitals, although a linear combination of up to $D=\binom{M}{N_\alpha} \binom{M}{N_\beta}$ electronic configurations, can be accurately approximated with a linear combination of $d \ll D$ electronic configurations~\cite{bender1969studies,ivanic2001identification}.
Such an assumption is common to the well-established classical selected configuration methods~\cite{stampfuss2005improved,roth2009importance,evangelista2014adaptive,tubman2016deterministic,holmes2016heat,schriber2016communication,liu2016ici}.

SQD uses a quantum circuit to sample important configurations -- unlike classical selected configuration methods, which instead perform an iterative search in the configuration space -- and a classical computer to solve for the lowest-energy wavefunction in the subspace of the Hilbert space spanned by the sampled configurations. While this joint use of classical and quantum computers is common to the quantum selected configuration interaction (QSCI) method~\cite{kanno2023quantum}, the latter does not specify a procedure to construct and parametrize the quantum circuits used for sampling, nor to mitigate errors affecting sampled configurations. The SQD method proposes techniques to overcome both obstacles.

\paragraph{Definition and parametrization of quantum circuits}

The study by Robledo et al~\cite{robledo2025chemistry} proposes to sample configurations from a variational ansatz, i.e. a family of parametrized quantum circuits designed to approximate electronic eigenfunctions. Specifically, it employs an LUCJ circuit with heavy-hex connectivity and a single layer, $L=1$, as described in Section~\ref{sec:LUCJ}. This choice leads to circuits with linear depth and quadratic number of gates accompanied by modest prefactors -- for example the largest simulation in the present work has qubit count $N_{\mathrm{qubits}} = 54$, two-qubit gate depth $d_{\mathrm{two-qubit}} = 122 \simeq 2.2 \cdot N_{\mathrm{qubits}}$ and two-qubit gate count $N_{\mathrm{two-qubit}} = 2640 \simeq 0.91 \cdot N_{\mathrm{qubits}}^2$ -- that are compatible with the coherence times and error rates of present-day devices like \device{rennselaer} for up to $M \sim 20-30$ spatial orbitals.
This choice is heuristic and primarily motivated by execution on present-day hardware, as LUCJ circuits require more than $L=1$ layers to accurately approximate electronic eigenfunctions~\cite{motta2023bridging}.

Following Robledo et al~\cite{robledo2025chemistry}, to parametrize LUCJ circuits, we do not employ an optimization procedure, but a classical surrogate. More specifically, we perform a CCSD calculation, extract $t_1$ and $t_2$ amplitudes from it, and use them to define the operators $\operator{K}_\mu$ and $\operator{J}_\mu$ in Eq.~\eqref{eq:UCC} (see ``Initialization of LUCJ parameters'' in Ref.~\cite{robledo2025chemistry}).

We remark that the nature of optimal wavefunctions for configuration sampling is not yet understood nor established. For example, alternative classical surrogation procedures to parametrize LUCJ circuits constitute an active research area, and multiple steps of time evolution have been proposed as an alternative to variational ansatze for configuration sampling~\cite{yu2025quantum,piccinelli2025quantum}. Consequently, systematic benchmarks like the present work are valuable tools to establish any improvements of SQD on solid numerical grounds.

\paragraph{Error mitigation by configuration recovery}

Although a physically motivated circuit ansatz like UCCSD or LUCJ conserves the number of particles with spin $\sigma$, in the presence of noise, particle-number conservation is typically violated. This symmetry breaking leads to substantial errors in the estimation of expectation values, or to a large number of useless configurations (i.e. orthogonal to the ground-space) in SQD calculations. To eliminate this source of error, SQD employs a ``self-consistent configuration recovery'' operation~\cite{robledo2025chemistry}, briefly described below.

The number of $d$ configurations and $K$ batches is defined by the user. In the first iteration (labeled by $i=0$), $K$ batches of $d$ configurations are sampled and all samples with incorrect particle numbers in the alpha- and beta-spin sectors are discarded and the Hamiltonian is diagonalized in the subspace spanned by the remaining configurations (details of the diagonalization are reported at the end of this Section). Upon diagonalization, one produces an approximation $n^{(0)}_{p\sigma}$ to the diagonal of the
unknown ground-state one-body density matrix,
\begin{equation}
n^{(gs)}_{p\sigma} = \langle \Psi_{gs} | \crt{p\sigma} \dst{p\sigma} | \Psi_{gs} \rangle \;.
\end{equation}
At a given iteration (labeled by $i>0$), for each configuration parametrized by a bitstring ${\bf{x}}$ with incorrect number of spin-$\sigma$ electrons, $\sum_{p} x_{p\sigma} \neq N^{(gs)}_{\sigma}$, one defines a probability distribution proportional to $|x_{p\sigma} - n^{(i-1)}_{p\sigma}|$ and uses it to flip entries of the bitstring ${\bf{x}}$ until the number of spin-$\sigma$ electrons assumes the target value $N^{(gs)}_{\sigma}$.
As a result, the original set of configurations is transformed into a different set of configurations, $\rm \chi_R$, with correct particle numbers. This set of configurations is sampled $K$ times to construct $K$ batches, each containing up to $d$ configurations. The Hamiltonian is diagonalized in the subspace spanned by configurations in each batch, producing a collection of wavefunctions $| \Psi^{(i)}_{\method{SQD},k} \rangle$ with $k=1 \dots K$.
With this information, one can update the approximation to $n^{(gs)}_{p\sigma}$ as
\begin{equation}
\label{eq:occ_number}
n^{(i)}_{p\sigma} = \frac{1}{K} \sum_{k=1}^K 
\langle \Psi^{(i)}_{\method{SQD},k} | \crt{p\sigma} \dst{p\sigma} | \Psi^{(i)}_{\method{SQD},k} \rangle \;.
\end{equation}
This procedure is repeated until a maximum number of iterations is reached, or the energy $E_{k} = \min_i \langle \Psi^{(i)}_{\method{SQD},k} | \operator{H} | \Psi^{(i)}_{\method{SQD},k} \rangle$ has converged within a user-defined tolerance.
In the present work, we only modify self-consistent configuration recovery by ensuring that the Hartree Fock configuration is included in each batch, so that $\langle \Psi^{(i)}_{\method{SQD},k} | \operator{H} | \Psi^{(i)}_{\method{SQD},k} \rangle \leq E_{\mathrm{HF}}$. This is necessary because, due to the size $d$ of the batches under consideration and/or the strength of the quantum noise, the Hartree Fock configuration may not be present in one or more batches, leading to high energies and/or convergence issues in the diagonalization of the projected Hamiltonians.

The resulting energy depends on various factors, including the nature of the probability distribution from which configurations are sampled, and the number $K$ and the size $d$ of the batches.
For many of the active spaces considered in this work, it is possible to increase $d$ within the available classical computational resources until SQD produces energies of FCI-like quality -- not by effectively selecting important configurations, but by brute-force.
Such a possibility is incompatible with the goal of this study, which is to assess the impact of the approximations in SQD vis-a-vis commonly used classical electronic structure methods including configuration interaction singles and doubles or CISD, coupled cluster singles and doubles or CCSD, and CCSD with perturbative triples or CCSD(T). Therefore, we limit the size of the SQD subspaces to
\begin{equation}
\label{eq:zeta_values}
d = \zeta \, N_{\method{CCSD}}
\;,
\end{equation}
where $\zeta \in \{0.25,0.50,1.00,2.00,4.00\}$ and, denoting $V_\sigma = M - N_\sigma$,
\begin{equation}
\label{eq:julia}
N_{\method{CCSD}} = 
\left\{
\begin{array}{ll}
1+\sum_\sigma N_\sigma V_\sigma+\frac{N_\sigma (N_\sigma-1)}{2} \frac{(V_\sigma) (V_\sigma-1)}{2}+\prod_\sigma N_\sigma V_\sigma & \mbox{\, if $N_\uparrow \neq N_\downarrow$} \\
1+N_\alpha V_\alpha+\sum_{i<j,a<b} \delta_{(ai) \leq (bj)} & \mbox{\, if $N_\uparrow = N_\downarrow$} \\
\end{array}
\right.
\end{equation}
is the number of CCSD parameters. 

\section{Energy-Variance Analysis}
\label{Appendix:EnergyVarianceAnalysis}

The energy-variance extrapolation procedure is performed to improve the SQD energy estimates beyond the raw SQD accuracy, as illustrated in Fig.~\ref{fig:4}. In this section, we provide the motivation and the technical details of the extrapolation procedure. The SQD method produces wavefunction $\Psi_{\method{SQD},k}$ by diagonalizing the projected Hamiltonian spanned by the selected configurations, however there is no guarantee that the output wavefunction is a good estimate to the eigenstates of the original Hamiltonian. In other words, we expect
\begin{equation}
\label{eq:ev}
\begin{split}
E_k &= \langle \Psi_{\method{SQD},k}|H|\Psi_{\method{SQD},k}\rangle > E_{\mathrm{gs}} \;, \\
V_k &= \langle\Psi_{\method{SQD},k}|H^2|\Psi_{\method{SQD},k}\rangle - \langle \Psi_{\method{SQD},k}|H|\Psi_{\method{SQD},k}\rangle^2 > 0 \;,
\end{split}
\end{equation}
where $E_{\mathrm{gs}}$ denotes the ground-state energy. A part of this work is to establish how well the SQD wavefunction, solved at various projected spaces, approximates the true ground state and its energy, and how much one can extend the accuracy of SQD with constraints on classical computational resource. To address these questions, in addition to SQD calculations, we carry out energy-variance extrapolations to examine the convergence of the SQD energy as a function of the subspace dimension and use this information to extrapolate the SQD energy toward the full-CI limit.

The energy-variance analysis conducted in this study was introduced for shell model calculations using the Lanczos diagonalization method~\cite{mizusaki2003precise} and has been adopted in SQD calculations~\cite{robledo2025chemistry}. At each subspace dimension, and for each batch in configuration recovery, the SQD procedure outputs an estimate of the ground-state wavefunction, $\Psi_{\method{SQD},k}$, from which we obtain the energy and variance in Eq.~\eqref{eq:ev}. 

When $\Psi_{\method{SQD},k}$ is sufficiently close to the ground state, the energy-variance pairs $\big( V_k, E_k \big)$ are expected to form a line in the $xy$ plane with a slope independent of the subspace dimension $\zeta$. In the limit where the subspace dimension approaches the dimension of the full Hilbert space, the SQD is expected to output the exact ground-state wavefunction and the corresponding variance is expected to approach zero. When the energy-variance pairs respect a linear relation, we can use a standard linear fit to obtain an extrapolated energy at the zero-variance limit,
$E_k \simeq m \, V_k+q$, and use the intercept of the extrapolation, $q$, to estimate the ground-state energy.
Important observations are that the result of the energy-variance extrapolation:
\begin{enumerate}
\item is generally more accurate than the SQD data
\item may undershoot the ground-state energy, i.e. individual SQD energies are variational, but their extrapolation is not
\item is affected by a statistical uncertainty, and characterizing the size of such a statistical uncertainty is necessary to determine the precision -- and therefore the practical usefulness -- of the extrapolation procedure
\item the energy-variance pairs may not follow a linear relation, and in particular cluster around the graphs of multiple lines (see Fig.~\ref{fig:fig_c5}, this situation is encountered more frequently across the dataset for smaller subspace sizes).
\end{enumerate}
To alleviate these issues, we propose and test two techniques for energy-variance extrapolation, the generalized eigenvalue extrapolation (GEV) and linear mixture model (LMM) fitting, discussed in Sections~\ref{sec:GEV} and \ref{sec:LMM} respectively.

\subsection{Generalized Eigenvalue Extrapolation (GEV)} 
\label{sec:GEV}

The purpose of the GEV technique is to improve the quality of SQD data prior to performing an energy-variance extrapolation, specifically
producing pairs $\big( V_k, E_k \big)$ with lower energy and variance, as defined in Eq.~\eqref{eq:ev}, than raw SQD data.

For each $\zeta$ in Eq.~\eqref{eq:zeta_values}, we consider the set of SQD vectors $\Psi_{\method{SQD},k}$ with dimension $d = \zeta \, (N_{occ} N_{vir})^2/4$, label them as $\psi_\mu$ to avoid clutter -- recall that $\zeta$ labels various subspace dimensions and $k=1\dots K$ labels different batches at each subspace dimension, see e.g. Eq.~\eqref{eq:occ_number} -- and use them to form the linear combination
\begin{equation}
\label{eq:gev_state}
| \psi_{\mathrm{GEV}}^{(\zeta)} \rangle = \sum_\mu c_\mu | \psi_\mu \rangle
\end{equation}
where the coefficients are the solution of the generalized eigenvector equation
\begin{equation}
\label{eq:gev_equation}
\myarray{H} \myarray{c} = \myarray{S} \myarray{c} \myarray{E}
\;,\;
\myarray{H}_{\mu\nu} = \langle \psi_\mu | \operator{H} | \psi_\nu \rangle
\;,\;
\myarray{S}_{\mu\nu} = \langle \psi_\mu | \psi_\nu \rangle
\;,
\end{equation}
with the lowest eigenvalue $\myarray{E}$.

Since the overlap matrix $\myarray{S}$ may be ill-conditioned, due to large overlaps among the states $\psi_\mu$, we regularize Eq.~\eqref{eq:gev_equation} using L\"owdin regularization: We choose a unitary gauge $\myarray{U}$ in which $\myarray{S}$ becomes diagonal, i.e., $\myarray{U}^\dagger \myarray{S} \myarray{U} = \mathsf{diag}(\myarray{s})$. Performing the substitution $\myarray{c} = \myarray{U} \tilde{\myarray{c}}$ and multiplying both sides of Eq.~\eqref{eq:gev_equation} from the left by $\myarray{U}^\dagger$ then yields
\begin{equation}
\label{eq:gev_equation_2}
\myarray{U}^\dagger \myarray{H} \myarray{U} \, \tilde{\myarray{c}}
= 
\myarray{U}^\dagger \myarray{S} \myarray{U} \, \tilde{\myarray{c}}
\, \myarray{E}
=
\mathsf{diag}(\myarray{s}) \, 
\tilde{\myarray{c}}
\, \myarray{E}.
\end{equation}
By truncating the eigenvalues $\myarray{s}$ below a user-defined threshold, in this case, $\varepsilon=10^{-5}$, we obtain a basis set defined by $\tilde {\myarray{c}}$ which yields a well-conditioned overlap matrix $\tilde{\myarray{S}}$. We numerically ensure the condition $|\kappa(\tilde{\myarray{S}}) - 1|<10^{-8}$. Repeating this procedure for all values of $\zeta$ yields a set of 5 wavefunctions, one per value of $\zeta$ in Eq.~\eqref{eq:zeta_values}, that can be used to perform an energy-variance extrapolation.

For our energy-variance extrapolation scheme, we utilize an ordinary least squares (OLS) approach. Furthermore, due to the diverse sizes of problem sizes in the dataset, some calculations may yield energy-variance pairs with variance close to or equal to zero (variance below $10^{-5}$), or the OLS fit may produce a negative slope. In these cases, the minimum energy among the computed SQD energies at various subspace dimensions is chosen instead of an extrapolated energy.

\subsection{Linear Mixture Model (LMM)}
\label{sec:LMM}
\begin{figure*}[t!]
\centering
\includegraphics[width=\textwidth]{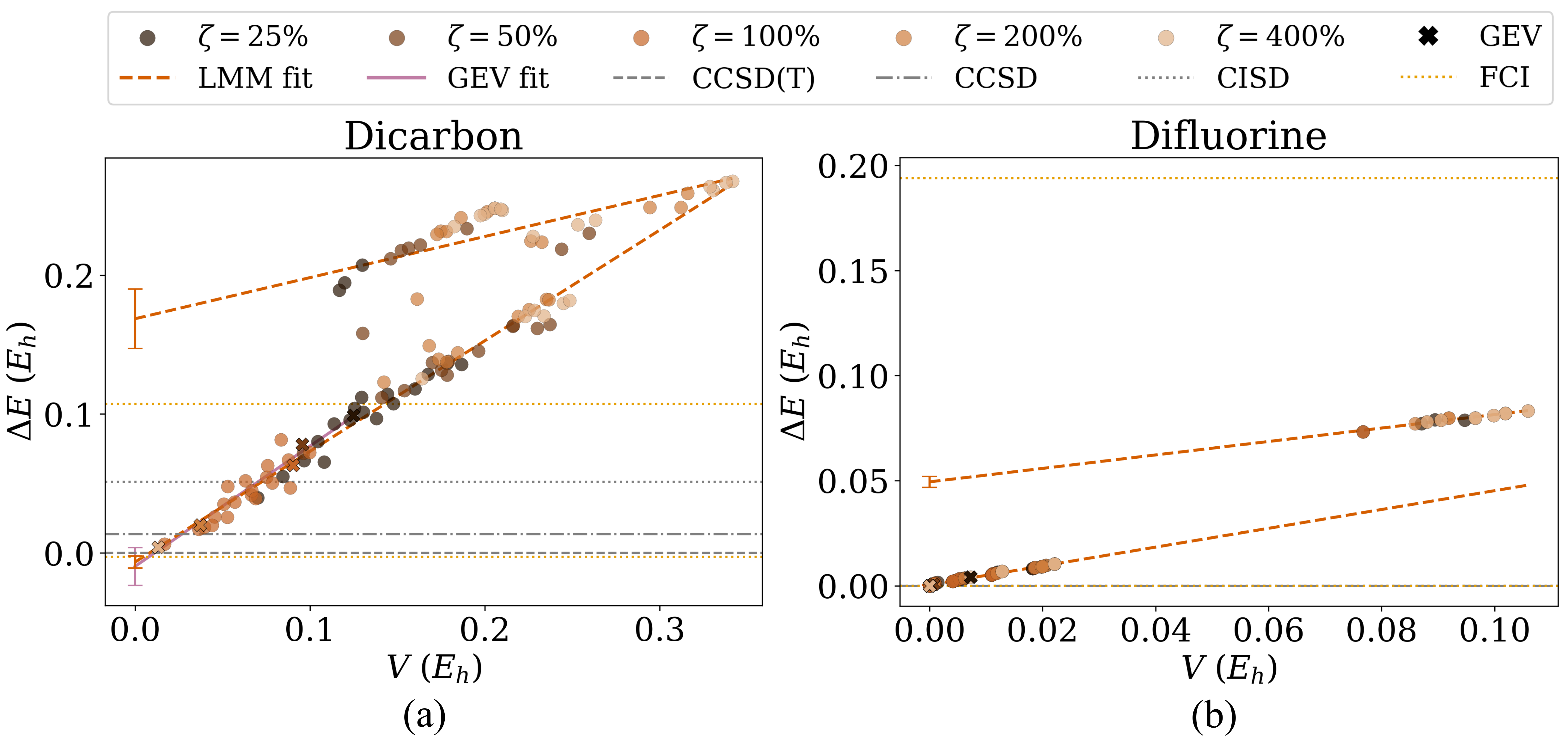}
\caption{Energy variance analysis for dicarbon (a) and difluorine (b). Variance-energy pairs of SQD wavefunctions and GEV wavefunctions (linear combinations of SQD wavefunctions) are shown as brown circles and crosses, respectively. Horizontal lines correspond to approximate ground-state energies (CISD dotted, CCSD dot-dashed, CCSD(T) dashed) and singlet excited-states from FCI (dotted orange lines). 
}
\label{fig:fig_c5}
\end{figure*}

Since energy-variance pairs in our dataset may form multiple clusters for extrapolation, as shown in Fig.~\ref{fig:fig_c5}, we employ a mixture of linear regressions, which we refer to as the linear mixture model (LMM), to fit the clustered energy-variance pairs. This approach is justified because the SQD wavefunctions at each subspace dimension can be linear combinations of multiple eigenstates of the original Hamiltonian, leading to multiple independent linear trends in the energy-variance relationships.

To fit the model, we first determine the number of clusters $N_c$ through visual inspection of the data structure. While visual inspection provides a starting point for clustering, we acknowledge that this method may introduce subjectivity and could potentially miss subtle cluster structures in the data. We then use spectral clustering with the RBF kernel (as implemented in \software{scikit\text{-}learn}) to partition the data into $N_c$ initial subsets, with the kernel bandwidth parameter $\gamma$ set to $1/d_{\text{med}}^2$, where $d_{\text{med}}$ is the median of all pairwise Euclidean distances between the energy-variance pairs $(V_k, E_k)$. Using these initial assignments, we apply an iterative expectation-maximization (E-M) algorithm to refine the clustering and fit $N_c$ separate linear regression models. The algorithm alternates between: (1) the M-step, where we fit an ordinary least squares regression to each cluster, and (2) the E-step, where we reassign each point to the cluster whose model minimizes the squared prediction error. The algorithm terminates when cluster assignments converge, the clusters collapse to fewer than $N_c$ clusters, or the maximum iteration limit (1000) is reached. We enforce a minimum cluster size of 5 points to ensure statistical validity.
The quality of each fit is assessed using adjusted $R^2$ values, which account for model complexity. For all molecules studied, the adjusted $R^2$ values exceed 0.91, indicating that the variance-energy pairs are well explained by the mixture of linear regressions. For energy extrapolation, we identify the linear model with the lowest intercept as the ground state energy estimate and report the 95\% confidence interval of the extrapolated energy. 

For the molecules amenable to full configuration interaction (FCI) calculations in \software{PySCF}, we compare the extrapolated excited state energies with FCI solutions with matching spin quantum numbers. While some SQD excited state energies coincide with the FCI solutions, as shown in Fig.~\ref{fig:4}(a), this correspondence does not always hold, as seen in Fig.~\ref{fig:fig_c5}(b). 

The statistical uncertainties on the extrapolated energies from both the GEV and the LMM methods are reported as 95\% confidence intervals on the intercept of the linear regression line, computed using the \code{OLSResults.conf_int()} method of the fitted OLS regression object from the \software{statsmodels} Python package. Table~\ref{tab:incompatible_molecules} lists molecules for which the extrapolated energies are statistically incompatible with the CCSD(T) reference values, along with their corresponding confidence intervals. Identifying these molecules reveals the practical limitations of each extrapolation technique and informs us which systems may require larger subspace dimensions or alternative approaches for achieving reliable energy estimates.

\begin{table}[ht!]
    \centering
    \footnotesize
\begin{tabular}{lrr|lrr||lrr}
    \toprule
    \multicolumn{6}{c||}{LMM} & \multicolumn{3}{c}{GEV}\\
    \midrule
    {} & $|\Delta E_\mathrm{ext}|$ & $\method{CI}(\Delta E)$ & {} & $|\Delta E_\mathrm{ext}|$ & $\method{CI}(\Delta E)$ & {} & $|\Delta E_\mathrm{ext}|$ & $\method{CI}(\Delta E)$\\
    \midrule
    NH                   &  0.0006 &  0.0006 & $\cdot$OH            &  0.0013 &  0.0004 & NH                   &  0.0013 &  0.0000 \\
    CHO$\cdot$           &  0.0031 &  0.0030 & c-N$_2$H$_2$         &  0.0019 &  0.0021 & $\cdot$OH            &  0.0006 &  0.0010 \\
    CH$_3$COOH           &  0.0424 &  0.0134 & C$_2$H$_4$O$_2$      &  0.0117 &  0.0038 & CHO$\cdot$           &  0.0027 &  0.0047 \\
    N$_2$O               &  0.0328 &  0.0081 & S$_2$                &  0.0038 &  0.0015 & c-N$_2$H$_2$         &  0.0031 &  0.0047 \\
    H$_2$CO              &  0.0045 &  0.0020 & c-HCOH               &  0.0035 &  0.0020 & CH$_3$COOH           &  0.0204 &  0.0301 \\
    $\cdot$NH$_2$        &  0.0060 &  0.0034 & CH$_3$F              &  0.0016 &  0.0010 & C$_2$H$_4$O$_2$      &  0.0107 &  0.0152 \\
    HOCN                 &  0.0060 &  0.0042 & C$_2$H$_2$O$_2$      &  0.0302 &  0.0077 & N$_2$O               &  0.0470 &  0.0923 \\
    C$_2$H$_2$O          &  0.0175 &  0.0058 & P$_2$                &  0.0055 &  0.0012 & S$_2$                &  0.0027 &  0.0007 \\
    P$_4$                &  0.0141 &  0.0036 & $\cdot$SSH           &  0.0013 &  0.0010 & H$_2$CO              &  0.0031 &  0.0035 \\
    t-N$_2$H$_2$         &  0.0009 &  0.0021 & O$_2$                &  0.0048 &  0.0030 & c-HCOH               &  0.0026 &  0.0030 \\
    N$_2$                &  0.0047 &  0.0032 & CH$_2$NH$\cdot$      &  0.0035 &  0.0019 & $\cdot$NH$_2$        &  0.0045 &  0.0063 \\
    B$_2$                &  0.0127 &  0.0110 & CH$_3$CHO            &  0.0267 &  0.0044 & CH$_3$F              &  0.0010 &  0.0019 \\
    Al                   &  0.0007 &  0.0001 & AlCl                 &  0.0016 &  0.0008 & HOCN                 &  0.0038 &  0.0071 \\
    AlCl$_3$             &  0.0015 &  0.0005 & AlF$_3$              &  0.0018 &  0.0015 & C$_2$H$_2$O$_2$      &  0.0242 &  0.0234 \\
    AlH                  &  0.0022 &  0.0010 & AlH$_3$              &  0.0009 &  0.0004 & C$_2$H$_2$O          &  0.0090 &  0.0127 \\
    C$_3$H$_4$           &  0.0448 &  0.0356 & B$_2$H$_6$           &  0.0021 &  0.0018 & P$_2$                &  0.0041 &  0.0026 \\
    Be                   &  0.0008 &  0.0003 & Be$_2$               &  0.0081 &  0.0045 & P$_4$                &  0.0222 &  0.0254 \\
    BeCl$_2$             &  0.0023 &  0.0007 & BeF$_2$              &  0.0055 &  0.0022 & $\cdot$SSH           &  0.0009 &  0.0011 \\
    BF                   &  0.0030 &  0.0021 & BF$_3$               &  0.0047 &  0.0015 & t-N$_2$H$_2$         &  0.0037 &  0.0059 \\
    BH                   &  0.0041 &  0.0021 & BN$_3$               &  0.0090 &  0.0026 & O$_2$                &  0.0031 &  0.0033 \\
    C$_2$                &  0.0065 &  0.0043 & C$_2$H$_3$F          &  0.0214 &  0.0072 & N$_2$                &  0.0028 &  0.0033 \\
    C$_2$H$_4$           &  0.0063 &  0.0025 & CF$_2$               &  0.0043 &  0.0022 & CH$_2$NH$\cdot$      &  0.0044 &  0.0043 \\
    CF$_4$               &  0.0043 &  0.0016 & CH                   &  0.0054 &  0.0021 & B$_2$                &  0.0131 &  0.0195 \\
    $^3$CH$_2$           &  0.0023 &  0.0004 & CH$_2$CH$\cdot$      &  0.0056 &  0.0066 & c-HONO               &  0.0040 &  0.0061 \\
    CH$_2$NH$_2$$\cdot$  &  0.0058 &  0.0023 & CH$_3$$\cdot$        &  0.0021 &  0.0005 & C$_2$H$_2$           &  0.0057 &  0.0098 \\
    CH$_3$NH$_2$         &  0.0020 &  0.0015 & Cl$_2$               &  0.0003 &  0.0001 & C$_2$H$_6$           &  0.0062 &  0.0105 \\
    ClF                  &  0.0006 &  0.0003 & ClO$\cdot$           &  0.0020 &  0.0008 & CCH$\cdot$           &  0.0115 &  0.0153 \\
    CN$\cdot$            &  0.0189 &  0.0042 & CO                   &  0.0024 &  0.0022 & CH$_4$               &  0.0011 &  0.0022 \\
    CS                   &  0.0025 &  0.0018 & CS$_2$               &  0.0061 &  0.0038 & ClCN                 &  0.0075 &  0.0094 \\
    C$_2$H$_5$OH         &  0.0062 &  0.0029 & F$_2$                &  0.0005 &  0.0001 & HCOF                 &  0.0034 &  0.0038 \\
    FOOF                 &  0.0169 &  0.0059 & HCOOH                &  0.0067 &  0.0044 & OF$\cdot$            &  0.0107 &  0.0419 \\
    H$_2$CN$\cdot$       &  0.0094 &  0.0026 & HCl                  &  0.0006 &  0.0002 & C$_3$H$_6$           &  0.0146 &  0.0246 \\
    HCN                  &  0.0028 &  0.0028 & HF                   &  0.0012 &  0.0004 & SO$_3$               &  0.0151 &  0.0248 \\
    HNNN                 &  0.0080 &  0.0066 & HOCl                 &  0.0018 &  0.0011 & t-HONO               &  0.0097 &  0.0144 \\
    HOF                  &  0.0028 &  0.0014 & HOO$\cdot$           &  0.0042 &  0.0018 &                      &         &         \\
    HS$\cdot$            &  0.0004 &  0.0003 & N$_2$H$_4$           &  0.0017 &  0.0014 &                      &         &         \\
    NCCN                 &  0.0330 &  0.0151 & NH$_2$$\cdot$        &  0.0023 &  0.0006 &                      &         &         \\
    NO$\cdot$            &  0.0032 &  0.0030 & NO$_2$$\cdot$        &  0.0139 &  0.0062 &                      &         &         \\
    O$_3$                &  0.0036 &  0.0035 & OClO$\cdot$          &  0.0097 &  0.0051 &                      &         &         \\
    OCS                  &  0.0099 &  0.0030 & C$_2$H$_4$O          &  0.0083 &  0.0052 &                      &         &         \\
    C$_2$H$_2$O          &  0.0099 &  0.0055 & PH$_3$               &  0.0008 &  0.0006 &                      &         &         \\
    C$_3$H$_8$           &  0.0149 &  0.0052 & S$_2$O               &  0.0039 &  0.0039 &                      &         &         \\
    S$_3$                &  0.0029 &  0.0015 & S$_4$                &  0.0082 &  0.0033 &                      &         &         \\
    Si$_2$H$_6$          &  0.0056 &  0.0017 & SiF                  &  0.0034 &  0.0017 &                      &         &         \\
    SiF$_4$              &  0.0065 &  0.0014 & SiH                  &  0.0027 &  0.0011 &                      &         &         \\
    SiH$_3$F             &  0.0023 &  0.0009 & SiH$_4$              &  0.0019 &  0.0004 &                      &         &         \\
    SO                   &  0.0117 &  0.0026 & SO$_2$               &  0.0100 &  0.0024 &                      &         &         \\
    \bottomrule
\end{tabular}
\caption{List of molecules for which the extrapolated energy using the LMM method (left 6 columns) or the GEV method (right 3 columns) are statistically incompatible with CCSD(T), i.e., the CCSD(T) energy falls outside the 95\% confidence interval of the extrapolated energy. $|\Delta E_\mathrm{ext}|$ represents the absolute energy difference between the extrapolated value and CCSD(T). Values are rounded to 4 decimals. The abbreviation ``$\method{CI}(\Delta E)$'' denotes the confidence interval. Common incompatible molecules appear at the top of each column.}
\label{tab:incompatible_molecules}
\end{table}

\section{Data Collection}

The overall strategy for the calculations performed in this work involved, for each molecule in the W4-11 database, initial pre-processing by the classical quantum chemistry code \software{PySCF} on conventional computers, to generate:
\begin{enumerate}
\item optimized restricted (closed- or open-shell) Hartree-Fock orbitals (MOs) at STO-6G level of theory
\item matrix elements of the Hamiltonian Eq.~\eqref{eq:elecHam} in an active space spanned by non-core MOs (core orbitals, i.e., Li-F[1s] and Na-Cl[1s,2s,2p], were subjected to the standard frozen-core approximation)
\item restricted (for closed-shell singlet species) and unrestricted (otherwise) MP2, CISD, and CCSD calculations, using the frozen-core approximation. In the case of CCSD calculations, we stored $t_1$ and $t_2$ coefficients
\end{enumerate}
The choice of the minimal STO-6G basis is motivated by the fact that only a few molecular orbitals can be described on present-day devices for all the species in the W4-11 database, due to the available qubit number and error rates.

Following classical preprocessing, quantum circuits were constructed by the \software{ffsim} library as detailed in Section~\ref{sec:LUCJ}, i.e. performing a low-rank decomposition of the $t_2$ coefficients accompanied by truncation of terms to produce an LUCJ wavefunction with $L=1$ layers and compatible with heavy-hex qubit connectivity.

\begin{figure*}[ht!]
    \centering
    \includegraphics[width=.5 \textwidth]{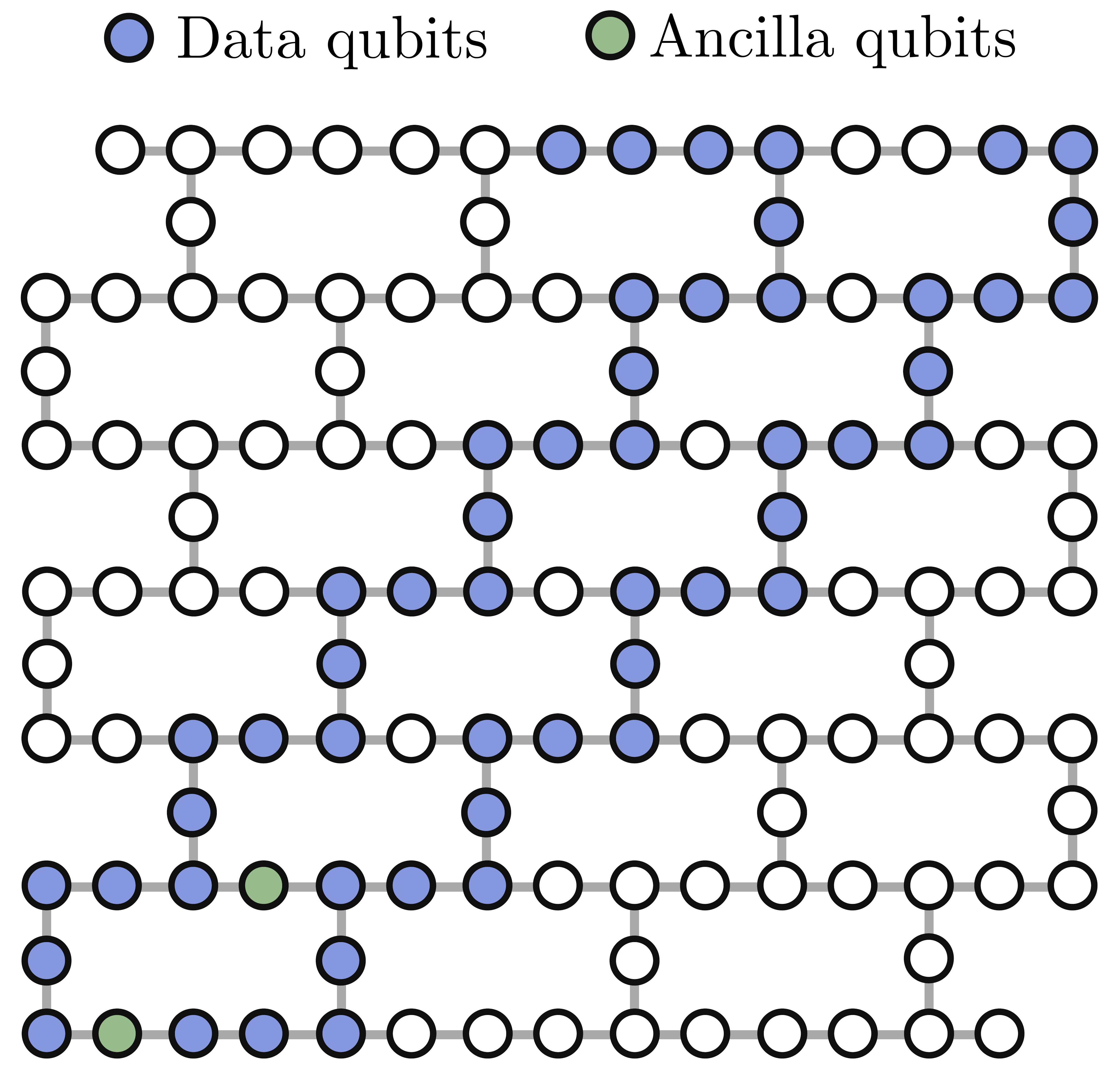}
\caption{Schematic representation of \device{rennselaer} (with circles representing qubits and horizontal/vertical lines representing connections between them) and qubit layout used for the largest simulation, cyanogen, in this study (with blue circles representing qubits encoding occupation numbers of $\alpha$ and $\beta$ spin-orbitals and green circles representing ancillae).}
\label{fig:layout}
\end{figure*}

Quantum simulations were performed on IBM's 127-qubit superconducting processor \device{rennselaer}, based on the Eagle architecture, sketched in Fig.~\ref{fig:layout}. Groups of best-performing qubits were user-selected based on monitoring average readout, measurement, and gate errors. Prior to execution, quantum circuits were transpiled using IBM's open-source Python library for quantum computing, \software{Qiskit}. The transpilation used optimization level 3 and a user-defined qubit layout, exemplified in Fig.~\ref{fig:layout} for the largest species in the database (CNNC, with 31 MOs). For each circuit, we collected $N_s = 1000000$ measurement outcomes (termed ``shots'' in quantum computing literature).

We used dynamical decoupling (DD)~\cite{lidar2014review,ezzell2023dynamical} to mitigate errors arising from quantum gates. We used the implementation of DD available in the \software{Runtime} library of \software{Qiskit}, through the Sampler primitive. DD is implemented by applying sequences of mutually-canceling pulses to idle qubits, to protect them from decoherence caused by low-frequency system-environment coupling. Here, we applied: no error suppression, DD-$\pauli{X}\pauli{X}$ (two pulses as in Ramsey echo experiments), DD-$\pauli{X}\pauli{Y}_4$ (four pulses), DD-$\pauli{X}_+\pauli{X}_-$ (two pulses).

The execution of a quantum circuit returns (i) a set of observed bitstrings ${\bf{x}}_\ell$ of length $2M$ (where $M$ is the number of spatial orbitals) and (ii) for each bitstring, the number $f_\ell$ of times the bitstring was observed, so that $N_s = \sum_\ell f_\ell$.

\subsection{Analysis of quantum samples}

As described in Section~\ref{sec:sqd}, in the standard Jordan-Wigner representation, a bitstring ${\bf{x}}$ labels a Slater determinant with $N_\sigma({\bf{x}}) = \sum_p x_{p\sigma}$ spin-$\sigma$ electrons. In other words, the total particle number is the Hamming weight of ${\bf{x}}$ and, with qubits ordered as in Fig.~\ref{fig:ucjlucj}, the number of spin-$\alpha$ (spin-$\beta$) electrons is the Hamming weight of the first (second) half of the bitstring.

\begin{figure*}[ht!]
    \centering
    \includegraphics[width=0.6\textwidth]{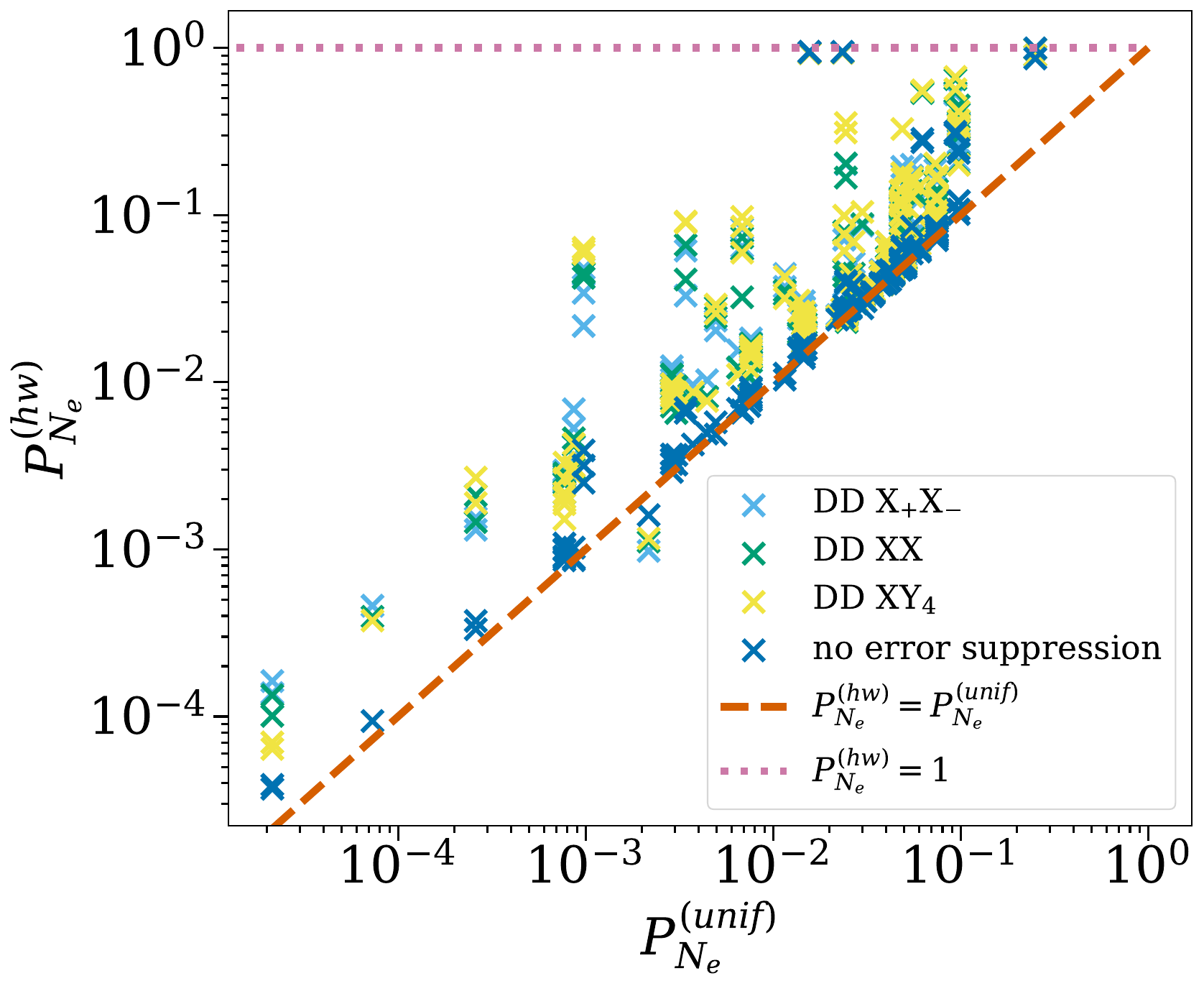}
\caption{Fraction $P^{(hw)}_{N_e}$ of bitstrings with correct particle number from simulations on \device{rensselaer} versus probability $P^{(unif)}_{N_e}$ that a uniformly distributed bitstring has correct particle number, for all the simulations in this study (i.e. all species in the W4-11 database at STO-6G level of theory with frozen-core) and various error suppression techniques (colored symbols). Simulations that do not break particle number conservation have $P^{(hw)}_{N_e}=1$ (dotted purple line), and simulations affected by strong depolarizing noise have $P^{(hw)}_{N_e}=P^{(unif)}_{N_e}$ (dashed orange line).
}
\label{fig:nelec_conservation} 
\end{figure*}

As claimed in Section~\ref{sec:sqd}, due to device noise, particle number conservation is violated. The severity of this phenomenon is assessed in Fig.~\ref{fig:nelec_conservation}, where we show, for each species in the W4-11 database, the fraction of bitstrings with correct particle number,
\begin{equation}
P^{(hw)}_{N_e} = \frac{ \sum_\ell f_\ell \, \delta_{ \sum_\sigma N_\sigma({\bf{x}}_\ell) = \sum_\sigma N_\sigma^{(gs)} } }{N_s}
\;,
\end{equation}
as a function of the ratio between the dimension of the FCI space and the dimension of the Fock space,
\begin{equation}
P^{(unif)}_{N_e} = \frac{ \binom{M}{N_\alpha} \binom{M}{N_\beta}  }{2^{2M}} 
\;.
\end{equation}
The latter quantity is the probability that a bitstring sampled from the uniform distribution has correct particle number,
i.e. in the presence of an infinitely strong global depolarizing noise channel $P^{(hw)}_{N_e} = P^{(unif)}_{N_e}$ (although a uniform distribution may be observed for other physical reasons). Therefore, values of $P^{(hw)}_{N_e} \simeq P^{(unif)}_{N_e}$ may be interpreted as arising from the presence of intense depolarizing noise.
As seen in Fig.~\ref{fig:nelec_conservation}, for some simulations we observe $P^{(hw)}_{N_e} < P^{(unif)}_{N_e}$, which cannot arise from a global depolarizing noise channel, but rather from other noise channels affecting real quantum devices (e.g. qubit relaxation, which tends to return bitstrings with low Hamming weight and lead to $P^{(hw)}_{N_e} \simeq 0$).

Fig.~\ref{fig:nelec_conservation} also allows us to compare the performance of various error suppression methods, by considering the fraction of bitstrings with correct particle number produced by each. As seen, DD tends to return higher values of $P^{(hw)}_{N_e}$ compared to no error suppression, with the DD-$\pauli{X}\pauli{Y}_4$ sequence providing slightly higher values.

\begin{figure*}[ht!]
    \centering
    \includegraphics[width=0.6\textwidth]{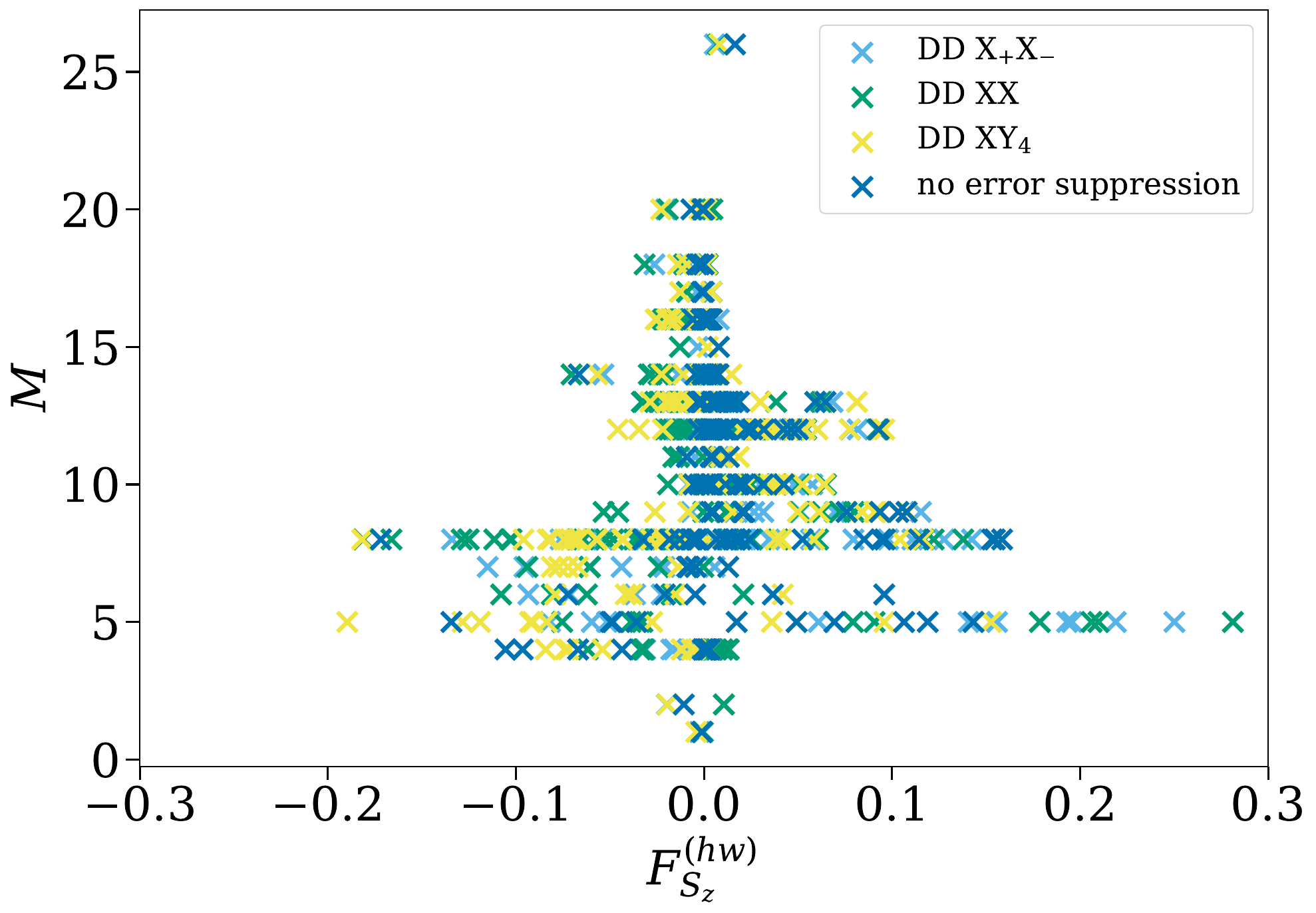}
\caption{Difference between the percentage of bitstrings with the correct number of $\alpha$ and $\beta$ electrons ($x$ axis) for different numbers of active-space orbitals ($y$ axis), using various error suppression techniques (colored symbols).}
\label{fig:sz_conservation}
\end{figure*}

In a typical electronic structure (though not, e.g., in the presence of spin-orbit coupling), it is important to conserve the number of spin-$\alpha$ and spin-$\beta$ electrons individually. In Fig.~\ref{fig:sz_conservation}, we show the difference between the percentage of bitstrings with correct number of $\alpha$ and $\beta$ electrons,
\begin{equation}
F^{(hw)}_{S_z} = \frac{ \sum_\ell f_\ell \, \delta_{ N_\alpha({\bf{x}}_\ell) = N_\alpha^{(gs)} } }{N_s}
-
\frac{ \sum_\ell f_\ell \, \delta_{ N_\beta({\bf{x}}_\ell) = N_\beta^{(gs)} } }{N_s}
\;,
\end{equation}
using different error mitigation techniques.
The experimental results indicate that the DD-$\pauli{X}\pauli{Y}_4$ sequence provides the best tradeoff on our hardware for two criteria. In the absence of noise,
$F^{(hw)}_{S_z} = 0$. DD tends to return lower values of $F^{(hw)}_{S_z}$ and a more symmetric distribution, compared to no error suppression, with the DD-$\pauli{X}\pauli{Y}_4$ sequence providing slightly higher values.

\newpage
\section{Results}
\label{sec3}
We here present an accuracy benchmark study of the SQD method applied to the W4-11 dataset~\cite{karton2011w4}, which comprises 154 molecular species and 745 thermochemical reactions: 124 total atomization energies (TAE), 83 bond dissociation energies (BDE), 20 isomerization energies (ISO), 505 heavy atom transfer (HAT) and 13 nucleophilic substitution reactions (SN). We systematically compare the performance of SQD with MP2, CISD and CCSD, using CCSD(T) as a reference benchmark for accuracy. All numerical simulations are conducted on \device{rensselaer}, a 127-qubit IBM superconducting quantum processor (IBM Eagle), and \texttt{AiMOS}, an eight petaflop IBM POWER9-equipped supercomputer. On the software side, we employed \software{PySCF}~\cite{sun2018pyscf,sun2020recent}, \software{Qiskit}~\cite{qiskit2024}, and \software{ffsim}~\cite{ffsim}. 
All molecules are discretized using a minimal basis set (STO-6G) in conjunction with a frozen-core approximation since the basis set does not include core-valence correlation.

\subsection{Ground-State Simulations}
\label{sec:ResultsGS}

\begin{figure*}[ht!]
\begin{subfigure}[t]{0.49\textwidth}
    \centering
    \includegraphics[width=\textwidth]{SN_docs/Graphics/GS_abs_violin_inset.png}
    \caption{}
\end{subfigure}%
\hfill
\begin{subfigure}[t]{0.49\textwidth}
    \centering
    \includegraphics[width=\textwidth]{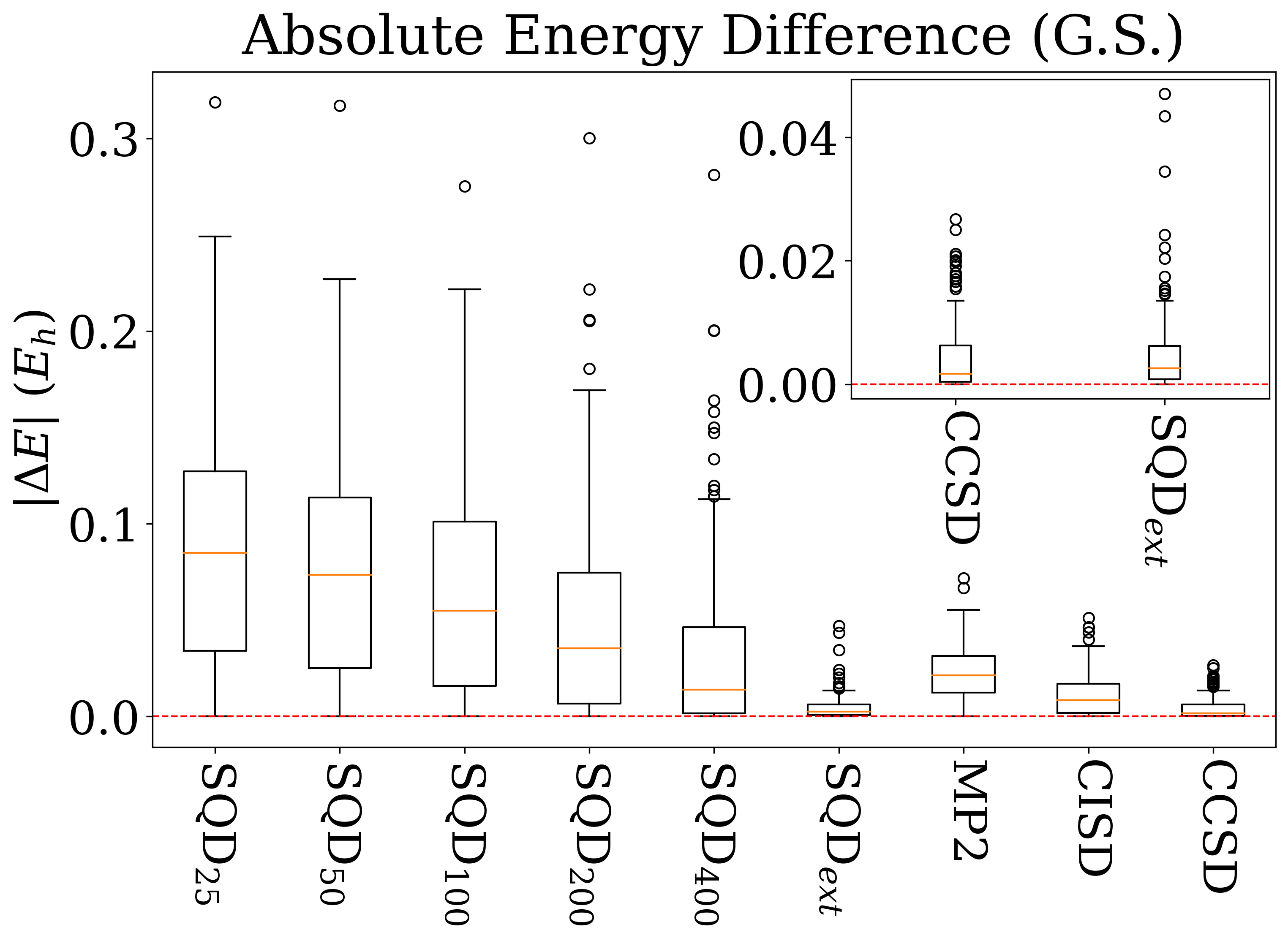}
    \caption{}
\end{subfigure}
\caption{\label{fig:APPendixQSDGS} Comparing the Sample‑based Quantum Diagonalization (SQD) algorithm across the W4‑11 thermochemistry suite with the classical 2nd order Møller-Plesset (MP2), configuration interaction singles and doubles (CISD), and coupled cluster singles and doubles (CCSD) algorithms. (a) Violin plots showing the distribution of absolute ground state energy errors for different SQD subspace sizes (25\%, 50\%, 10\%, 200\%, 400\%) and their GEV extrapolated limit, alongside CISD and CCSD. The insert shows a more detailed comparison of CCSD and SQD$_{ext}$. (b) Shows the corresponding box plots. 
}
\end{figure*}

We begin our study by assessing the accuracy of the SQD method for ground-state energy simulations. Specifically, we evaluate its performance using the absolute energy deviation
\begin{equation}
|\Delta E| = |E_X - E_{\rm CCSD(T)}|,
\end{equation}
where $X$ corresponds to MP2, CISD, CCSD, or SQD, and CCSD(T) serves as the reference benchmark. As CCSD(T) is widely regarded as the most accurate single-reference method available for the W4-11 dataset, it provides a reliable benchmark to evaluate the accuracy of approximate methods. The SQD algorithm as provided per~\cite{robledo2025chemistry} can be viewed as a quantum-enhanced selected CI method constructed on top of CCSD reference amplitudes (see Section~\ref{sec2}). We therefore report SQD results across a range of active space sizes used in the diagonalization step. These active space sizes are defined relative to the number of degrees of freedom encoded by the CCSD amplitudes, ensuring a consistent and meaningful comparison to classical methods, in particular, CCSD. Additionally, we use the proposed GEV extrapolation strategy for SQD energies (see Section~\ref{sec:GEV}).
To provide insight into the asymptotic behavior of the method as the active space is enlarged.\\

Figure~\ref{fig:APPendixQSDGS} illustrates the statistical distribution of ground-state energy errors, as defined in Eq.~\eqref{eq:EnergyDiviation}. As expected, the accuracy of the SQD procedure improves with increasing active space size. Nonetheless, in the absence of extrapolation, the method exhibits limited accuracy when compared to established classical methods. Notably, even when the active space is expanded to include significantly more degrees of freedom than those used by CCSD, the unextrapolated SQD results remain less accurate than all classical methods. For SQD$_{25}$, SQD$_{50}$, SQD$_{100}$, and SQD$_{200}$, the relative ground-state energy errors consistently exceed those obtained from CCSD. For SQD$_{\rm 400}$, the SQD error is reduced below that of CCSD, see Table~\ref{tab:combined_ene_diff}. Similarly, we find that the relative ground-state energy errors of SQD$_{25}$, SQD$_{50}$, SQD$_{100}$, and SQD$_{200}$ exceed those obtained from CISD. For SQD$_{\rm 400}$, we again observe that the SQD error is reduced below that of CISD, see Table~\ref{tab:combined_ene_diff}.

\pagebreak
\begin{table}[ht!]
    \centering
\begin{tabular}{lrrr||lrrr}
\toprule
{} & $\Delta$CCSD & $\Delta$SQD$_{\rm 400}$ & $\Delta$SQD$_{\rm ext}$ 
   & {} & $\Delta$CISD & $\Delta$SQD$_{\rm 400}$ & $\Delta$SQD$_{\rm ext}$ \\
\midrule
BH	&	0	&	0	&	0	&	C$_2$	&	0.0512	&	-0.0096	&	0.0063	\\
H$_2$S	&	0	&	-0.0001	&	0	&	BN	&	0.0438	&	-0.0069	&	0.0115	\\
AlCl	&	0.0006	&	-0.0003	&	0.0003	&	O$_3$	&	0.0365	&	-0.0075	&	0.0243	\\
H$_2$O	&	0.0001	&	-0.0004	&	-0.0001	&	SO$_2$	&	0.0293	&	-0.0061	&	0.0159	\\
HOCl	&	0.0003	&	-0.0004	&	-0.0001	&	CS$_2$	&	0.0266	&	-0.008	&	0.0249	\\
AlF	&	0.0051	&	-0.0008	&	0.0009	&	FOOF	&	0.0261	&	-0.0091	&	0.0245	\\
SiO	&	0.0176	&	-0.0008	&	0.0038	&	S$_2$O	&	0.0251	&	-0.0045	&	0.0126	\\
NH$_3$	&	0.0001	&	-0.0008	&	0.0001	&	S$_3$	&	0.0207	&	-0.0024	&	0.0113	\\
BH$_3$	&	0.0002	&	-0.0009	&	0	&	SiO	&	0.0201	&	-0.0008	&	0.0038	\\
BF	&	0.0068	&	-0.0011	&	0.0019	&	P$_2$	&	0.015	&	-0.0041	&	-0.0028	\\
AlH$_3$	&	0.0004	&	-0.0011	&	0.0001	&	CS	&	0.0141	&	-0.0021	&	0.0017	\\
Cl$_2$O	&	0.0009	&	-0.0012	&	0.0003	&	HNC	&	0.0124	&	-0.0056	&	0.0086	\\
HOF	&	0.0003	&	-0.0012	&	0	&	HCN	&	0.0124	&	-0.0085	&	0.0051	\\
CO	&	0.0074	&	-0.0017	&	0.0023	&	CO	&	0.0119	&	-0.0017	&	0.0023	\\
CS	&	0.0086	&	-0.0021	&	0.0017	&	N$_2$	&	0.0105	&	-0.0029	&	0.0008	\\
S$_3$	&	0.0114	&	-0.0024	&	0.0113	&	CH$_2$C	&	0.0104	&	-0.0039	&	0.0097	\\
N$_2$	&	0.0018	&	-0.0029	&	0.0008	&	H$_2$CO	&	0.0085	&	-0.0031	&	0.0073	\\
F$_2$O	&	0.0014	&	-0.0036	&	0.0007	&	HNO	&	0.0079	&	-0.0036	&	0.0032	\\
S$_2$O	&	0.0154	&	-0.0045	&	0.0126	&	BF	&	0.0078	&	-0.0011	&	0.0019	\\
SO$_2$	&	0.0198	&	-0.0061	&	0.0159	&	Be$_2$	&	0.0077	&	-0.0087	&	0.0044	\\
BN	&	0.0207	&	-0.0069	&	0.0115	&	F$_2$O	&	0.0064	&	-0.0036	&	0.0007	\\
C$_2$	&	0.0135	&	-0.0096	&	0.0064	&	AlF	&	0.0051	&	-0.0008	&	0.0009	\\
	&		&		&		&	HOOH	&	0.0043	&	-0.0015	&	0.0025	\\
	&		&		&		&	Cl$_2$O	&	0.0039	&	-0.0012	&	0.0003	\\
	&		&		&		&	NH$_2$Cl	&	0.0029	&	-0.0009	&	0.0023	\\
	&		&		&		&	SiH$_4$	&	0.0027	&	-0.0012	&	0.0014	\\
	&		&		&		&	CH$_4$	&	0.0026	&	-0.0011	&	0.0015	\\
	&		&		&		&	CH$_2$-sing	&	0.0026	&	0.0003	&	-0.0002	\\
	&		&		&		&	HOF	&	0.0024	&	-0.0012	&	0	\\
	&		&		&		&	PH$_3$	&	0.002	&	-0.0009	&	0.0002	\\
	&		&		&		&	BH	&	0.0017	&	0	&	0	\\
	&		&		&		&	HOCl	&	0.0017	&	-0.0004	&	-0.0001	\\
	&		&		&		&	NH$_3$	&	0.0016	&	-0.0008	&	0.0001	\\
	&		&		&		&	AlH$_3$	&	0.0015	&	-0.0011	&	0.0001	\\
	&		&		&		&	BH$_3$	&	0.0012	&	-0.0009	&	0	\\
	&		&		&		&	AlH	&	0.0009	&	-0.0003	&	0	\\
	&		&		&		&	AlCl	&	0.0009	&	-0.0003	&	0.0003	\\
	&		&		&		&	H$_2$O	&	0.0007	&	-0.0004	&	-0.0001	\\
	&		&		&		&	H$_2$S	&	0.0007	&	-0.0001	&	0	\\
	&		&		&		&	Be	&	0	&	-0.0009	&	0	\\
	&		&		&		&	HCl	&	0	&	0	&	0	\\
	&		&		&		&	HF	&	0	&	0	&	0	\\
\bottomrule
\end{tabular}
\caption{Comparison of molecules for which SQD$_{\rm 400}$ outperforms CCSD (left) and CISD (right). Values are rounded to 4 decimals, deviations below .1m$E_h$ are denoted zero.}
\label{tab:combined_ene_diff}
\end{table}

\pagebreak
\subsection{Thermochemical Reactions}
\label{sec:hermo-Chemical Reactions}
The W4-11 dataset includes diverse bonding motifs with varying covalent and ionic character. The success of CCSD(T) on this dataset is largely due to error cancellation in energy differences. While SQD, when constrained to a moderate number of configurations, does not achieve the same level of accuracy as CCSD or MP2 for ground-state energies, its behavior in thermochemical energy differences remains to be evaluated. We examine whether SQD exhibits similar error cancellation in this context.

\begin{figure*}[ht!]
\centering
\begin{subfigure}[t]{0.49\textwidth}
    \centering
    \includegraphics[width=\textwidth]{SN_docs/Graphics/Bar_ave_set1_insert.png}
    \caption{}
\end{subfigure}%
\hspace{4mm}
\begin{subfigure}[t]{0.36\textwidth}
    \centering
    \includegraphics[width=\textwidth]{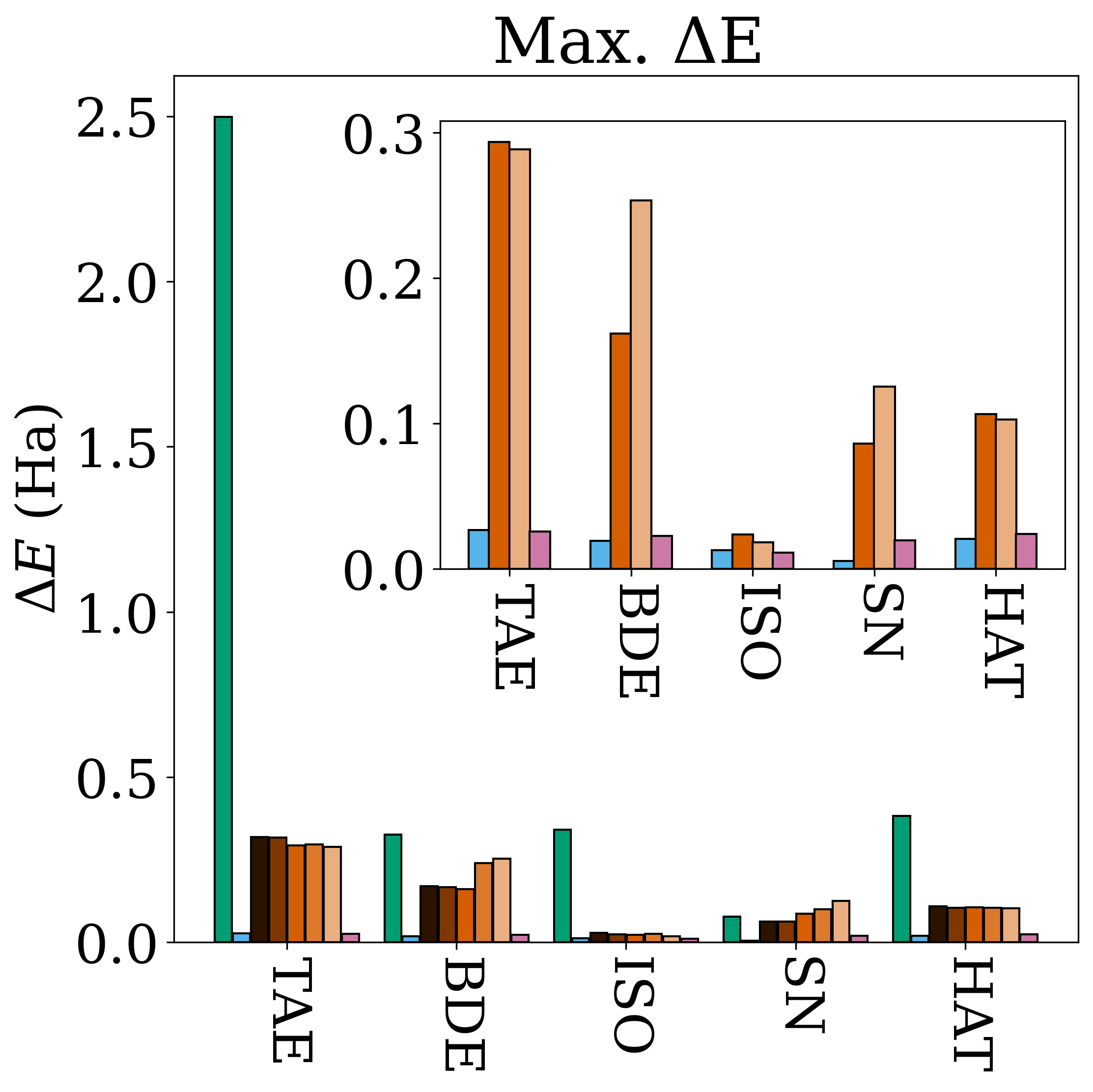}
    \caption{}
\end{subfigure}
\caption{\label{fig:AllReaction_ave} Side-by-side comparison of the averaged absolute errors observed across various thermochemical reaction types: total atomization energy (TAE), bond dissociation energy (BDE), isomerization energy (ISO), heavy atom transfer (HAT), and nucleophilic substitution (SN). Panels (a)--(c) show the results under different energy scales and method selections to enable a more detailed evaluation of the relative performance of the computational approaches.}
\end{figure*}

We begin by analyzing the absolute energy error averaged across the respective reactions, denoted as Ave.~$\Delta$E. Figure~\ref{fig:AllReaction_ave} presents a comparison between classical simulation results obtained using ROHF, MP2, CISD, and CCSD, and those from the SQD method across various active subspace sizes, including extrapolated estimates. The results show that without extrapolation techniques, the errors of SQD across the W4-11 dataset are substantially larger than those of CCSD, even when significantly larger active subspaces are used.\\

To gain deeper insight, we now examine the error statistics for individual categories of thermochemical reactions. For clarity, we organize the discussion by reaction type. 

\pagebreak
\subsubsection{Total Atomization Energies}
\label{sec:TAE}

\begin{figure*}[ht!]
\centering
\includegraphics[width = .7\textwidth]{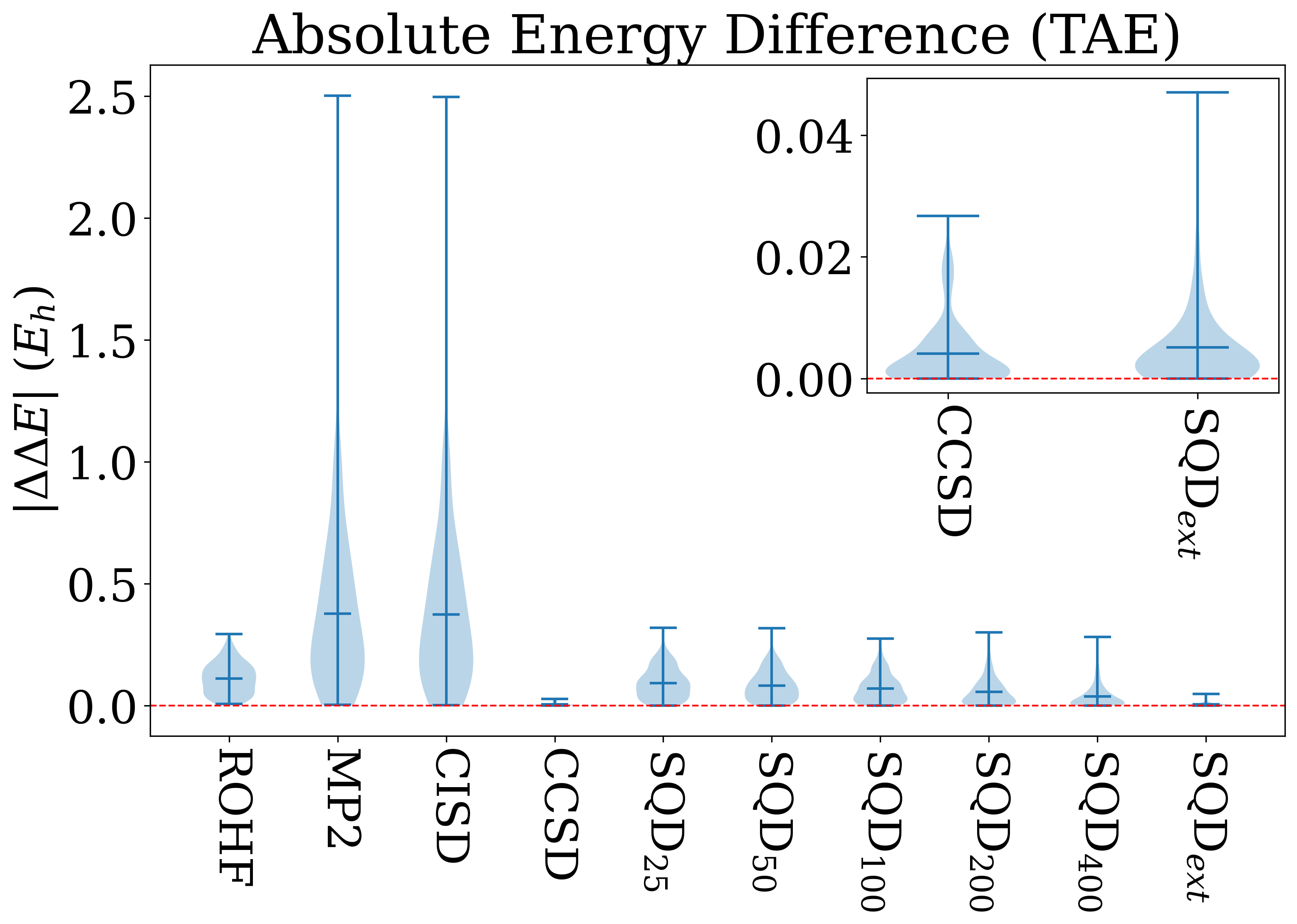}
\caption{\label{fig:TAE} Show a side-by-side comparison of violin plots of the error distribution for total atomization energies.}
\end{figure*}

The W4-11 dataset includes 124 total atomization processes, for which we analyze the statistical behavior of relative errors using CCSD(T) as the reference benchmark. As illustrated in Figure~\ref{fig:TAE}, the violin plot reveals a pronounced degradation in accuracy for SQD in the absence of extrapolation, highlighting its limited reliability under constrained resource settings.  To better quantify this observation, we compare the statistical error profiles of CCSD, SQD$_{\rm 400}$, and SQD$_{\rm ext}$ in terms of key descriptive statistics: median, first quartile ($Q_1$), third quartile ($Q_3$), interquartile range (IQR), and the min and max whiskers defined by
\begin{equation}
\label{eq:min_max_whisk}
{\rm min}_W = Q_1 - \frac{3}{2} (Q_3-Q_1)
\qquad {\rm and} \qquad
{\rm max}_W = Q_3+\frac{3}{2} (Q_3-Q_1)
\end{equation}

The results, summarized in Table~\ref{tab:Stat_profil_TAE}, show that the median errors of CCSD and SQD$_{\rm ext}$ are very close (0.0017 vs.~0.0024 $E_h$), indicating that extrapolation can indeed recover the accuracy of high-level classical methods. Both methods also share identical lower quartiles up to 4 decimals (0.0006 $E_h$), with CCSD exhibiting a slightly larger interquartile range (0.0051 $E_h$ vs.~0.0047 $E_h$ for SQD$_{\rm ext}$). In contrast, SQD$_{\rm 400}$ shows a markedly larger median error (0.0187 $E_h$) and a substantially wider statistical spread (IQR = 0.0520 $E_h$), exceeding CCSD by more than an order of magnitude. The maximum and minimum whiskers further confirm this trend: SQD$_{\rm 400}$ spans nearly an order of magnitude broader error range compared to CCSD and SQD$_{\rm ext}$. Together, these results demonstrate that while SQD$_{\rm ext}$ achieves error profiles comparable to CCSD, SQD$_{\rm 400}$ suffers from both reduced accuracy and significantly inflated variability.

\begin{table}[ht!]
    \centering
    \begin{tabular}{r|cccccc}
    \toprule
    Method      &  median & 1$^{\rm st}$ quartile & 3$^{\rm rd}$ quartile & IQR & max$_W$ & min$_W$\\
    \midrule
CCSD        &   0.0017 & 0.0006 & 0.0056 & 0.0051 & 0.0133 & -0.0071\\
SQD$_{\rm ext}$  &   0.0027 & 0.0009 & 0.0061 & 0.0052 & 0.0139 & -0.0069\\
SQD$_{\rm 400}$ &   0.0203 & 0.0028 & 0.0503 & 0.0475 & 0.1216 & -0.0686\\
\bottomrule
    \end{tabular}
    \caption{Statistical error profiles for TAEs of CCSD, SQD$_{\rm 400}$, and SQD$_{\rm ext}$ in terms of median, first quartile ($Q_1$), third quartile ($Q_3$), interquartile range (IQR), and the min and max whiskers}
    \label{tab:Stat_profil_TAE}
\end{table}

Based on the minimum and maximum whiskers derived from the interquartile range, we identify statistical outliers in Figure~\ref{fig:TAE} for CCSD, SQD$_{\rm 400}$, and SQD$_{\rm ext}$, see Table~\ref{tab:TAE_outliers}. The distribution of outliers has little overlap between classical and quantum approaches. For CCSD, the flagged reactions are predominantly small inorganic species such as BeF$_2$ or SiO, together with a handful of triatomics and sulfur oxides, as well as nitrogen oxides and chalcogenides, e.g. NO$_2$ or N$_2$O. In contrast, SQD$_{\rm ext}$ identifies a mixture of small unsaturated or oxygenated organic molecules such as C$_3$H$_4$ or C$_3$H$_6$, and compact, strongly bound inorganics (C$_2$F$_2$ or SO$_3$). SQD$_{\rm 400}$ further amplifies the organic bias, with large errors for e.g. C$_3$H$_4$ or C$_3$H$_8$, and the oxygenated systems e.g. C$_2$H$_4$O$_2$, along with the strongly multibonded C$_2$N$_2$.\\

Notably, there is no consistent set of reactions flagged across all three methods, underscoring that the error mechanisms differ significantly between CCSD and SQD. Classical CCSD appears to struggle with highly electronegative or multivalent fragments involving multiple bonds, whereas the SQD method shows systematic deficiencies for polyatomic organic molecules and conjugated or multiply bonded systems such as NCCN and glyoxal. The fact that extrapolation substantially reduces the spread relative to SQD$_{\rm 400}$ but does not eliminate these organic and multibonded outliers suggests that the SQD framework faces inherent challenges in capturing the correlation balance required for multi-center bonding and delocalization effects.

\begin{table}[h!]
    \centering
    \begin{tabular}{c|c|c|c|c|c}
    \toprule
    \multicolumn{2}{c|}{CCSD} & \multicolumn{2}{|c|}{SQD$_{\rm ext}$} & \multicolumn{2}{|c}{SQD$_{\rm 400}$} \\
    \midrule
    TAE Reaction & $\Delta$E & TAE Reaction & $\Delta$E & TAE Reaction & $\Delta$E\\
    \midrule
BeF$_2$$\rightarrow$Be+2F	&	0.0135	&	C$_3$H$_4$$\rightarrow$3C+4H	&	0.0344			&	C$_3$H$_4$$\rightarrow$3C+4H	&	0.1472	\\	
SiO$\rightarrow$Si+O	&	0.0176	&	C$_3$H$_6$$\rightarrow$3C+6H	&	0.0146			&	CH$_3$C$\equiv$CH$\rightarrow$3C+4H	&	0.1335	\\		
OCS$\rightarrow$O+C+S	&	0.0181	&	C$_2$F$_2$$\rightarrow$2C+2F	&	0.0156			&	C$_2$H$_6$O$\rightarrow$2C+6H+O	&	0.158	\\
HCNO$\rightarrow$C+O+N+H	&	0.0166	&	C$_2$H$_2$O$\rightarrow$2C+2H+O	&	0.0174	&	C$_3$H$_6$$\rightarrow$3C+6H	&	0.1639	\\		
CO$_2$$\rightarrow$C+2O	&	0.0211	&	C$_2$N$_2$$\rightarrow$2C+2N	&	0.0435			&	C$_3$H$_8$$\rightarrow$3C+8H	&	0.2003	\\		
HNCO$\rightarrow$C+O+N+H	&	0.017	&	C$_2$H$_2$O$_2$$\rightarrow$2C+2H+2O	&	0.0242	&	C$_2$N$_2$$\rightarrow$2C+2N	&	0.2812	\\	
S$_2$O$\rightarrow$2S+O	&	0.0154	&	C$_2$H$_4$O$_2$$\rightarrow$2C+4H+2O	&	0.0204	&	C$_2$H$_2$O$_2$$\rightarrow$2C+2H+2O	&	0.1501	\\
NO$_2$$\rightarrow$N+2O	&	0.02	&	CN$\rightarrow$C+N	&	0.0147			&	C$_2$H$_4$O$_2$$\rightarrow$2C+4H+2O	&	0.2001	\\
SO$_2$$\rightarrow$S+2O	&	0.0198	&	N$_2$O$\rightarrow$2N+O	&	0.0471	&&\\												
N$_2$O$\rightarrow$2N+O	&	0.0192	&	P$_4$$\rightarrow$4P	&	0.0221			&&\\												
CS$_2$$\rightarrow$C+2S	&	0.0159	&	SO$_3$$\rightarrow$S+3O	&	0.0152	&&\\												
SO$_3$$\rightarrow$S+3O	&	0.0267	&	&	&	&	\\																
    \bottomrule
    \end{tabular}
    \caption{Statistical outliers in TAE predictions for CCSD, SQD$_{\rm 400}$, and SQD$_{\rm ext}$, as identified based on interquartile range whisker criteria.}
    \label{tab:TAE_outliers}
\end{table}

\pagebreak
\subsubsection{Bond Dissociation Energies}
\label{sec:BDE}

\begin{figure*}[ht!]
\centering
\includegraphics[width = .7\textwidth]{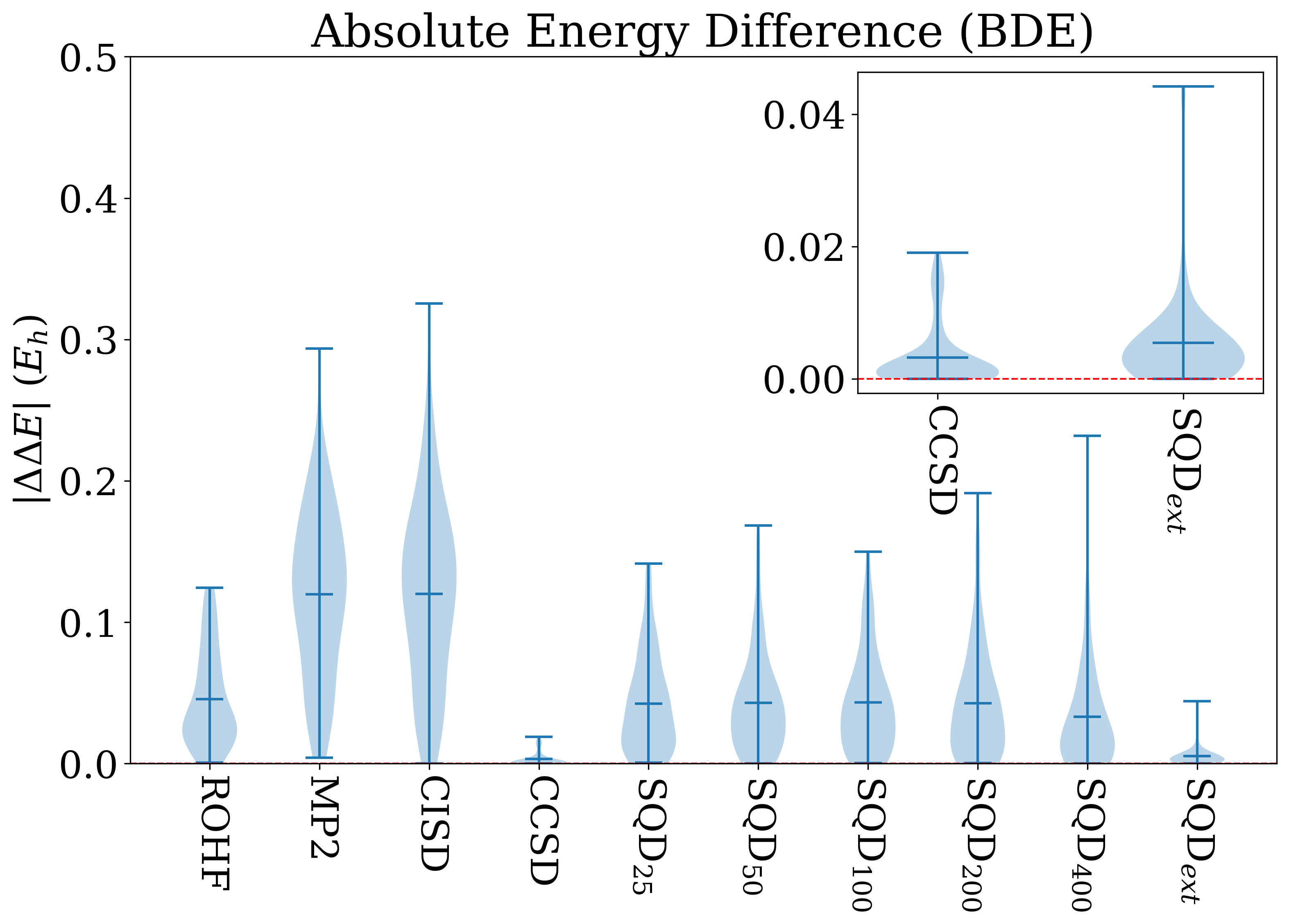}
\caption{\label{fig:BDE} Show a side-by-side comparison of violin plots of the error distribution for bond dissociation energies.}
\end{figure*}

We now turn to an analogous analysis for the BDEs. The W4-11 dataset comprises 83 bond dissociation processes, for which we evaluate the statistical distribution of relative errors using CCSD(T) as the reference. As shown in Figure~\ref{fig:BDE}, the violin plots reveal substantial deviations in predictive accuracy across the various methods considered. Notably, the error distributions exhibit significant skewness due to the presence of outliers. To enable a more meaningful comparison, particularly among CCSD, SQD$_{\rm 400}$, and SQD$_{\rm ext}$, we compute statistical descriptors following the same procedure used in the analysis of total atomization energies (TAEs).\\

The results summarized in Table~\ref{tab:Stat_profil_BDE} show that CCSD achieves the lowest median error (0.0013 $E_h$), with an interquartile range (0.0027 $E_h$) that is nearly half that of SQD$_{\rm ext}$ and more than an order of magnitude smaller than that of SQD$_{\rm 400}$. SQD$_{\rm ext}$ recovers accuracy close to CCSD, with a modest increase in both median error (0.0035 $E_h$) and spread (IQR = 0.0048 $E_h$), demonstrating that extrapolation substantially improves the reliability of the quantum approach. In contrast, SQD$_{\rm 400}$ exhibits markedly larger errors, with a median (0.0216 $E_h$) an order of magnitude higher than CCSD and a statistical spread (IQR = 0.0418 $E_h$) exceeding CCSD by more than a factor of 15. The extreme whisker values for SQD$_{\rm 400}$ further highlight its variability, indicating that a fixed resource budget without extrapolation is insufficient for consistently accurate bond dissociation energies.

\begin{table}[ht!]
    \centering
    \begin{tabular}{r|cccccc}
    \toprule
    Method      &  median & 1$^{\rm st}$ quartile & 3$^{\rm rd}$ quartile & IQR & max$_W$ & min$_W$\\
    \midrule
CCSD            &   0.0013 & 0.0004 & 0.0031 & 0.0027 & 0.0071 & -0.0036 \\
SQD$_{\rm ext}$ &   0.0035 & 0.0013 & 0.0062 & 0.0048 & 0.0134 & -0.0059 \\
SQD$_{\rm 400}$ &   0.0216 & 0.0061 & 0.0479 & 0.0418 & 0.1106 & -0.0566 \\
\bottomrule
    \end{tabular}
    \caption{Statistical error profiles for BDEs of CCSD, SQD$_{\rm 400}$, and SQD$_{\rm ext}$ in terms of median, first quartile ($Q_1$), third quartile ($Q_3$), interquartile range (IQR), and the min and max whiskers}
    \label{tab:Stat_profil_BDE}
\end{table}

\newpage

Based on the minimum and maximum whiskers computed from the interquartile range, we identify statistical outliers in Figure~\ref{fig:BDE} for CCSD, SQD$_{\rm 400}$, and SQD$_{\rm ext}$; the corresponding reactions are listed in Table~\ref{tab:BDE_outliers}. As before, we find no common set of outliers across all three methods, indicating that the dominant error mechanisms again differ between CCSD and SQD. For CCSD, the flagged reactions are dominated by small heteroatomic species, together with several isomerization-prone systems (HCNO, HNNN, HNCO). These molecules involve multiple bonds to oxygen or sulfur and often feature significant near-degeneracy effects, making them challenging for single-reference CCSD. In contrast, SQD$_{\rm ext}$ highlights a largely different set of systems, dominated by the two dissociation channels of N$_2$O, along with allene, glyoxal, cyanogen, and acetic acid. These reactions involve delocalized $\pi$-bonding, multi-center electronic structures, or electronic reorganizations upon bond cleavage, which appear more difficult for the extrapolated SQD scheme to capture accurately. SQD$_{\rm 400}$, operating under fixed quantum resources, exhibits a still more distinct outlier profile. The large deviations in cyanogen, allene, P$_4$, and N$_2$O reflect the increased difficulty of treating multibonded or electronically flexible systems with restricted quantum resources. The broader error distribution and larger whiskers of SQD$_{\rm 400}$ further emphasize its sensitivity to such electronically complex fragments. 

\begin{table}[h!]
    \centering
    \begin{tabular}{c|c|c|c|c|c}
    \toprule
    \multicolumn{2}{c|}{CCSD} &
    \multicolumn{2}{|c|}{SQD$_{\rm ext}$} &
    \multicolumn{2}{|c}{SQD$_{\rm 400}$} \\
    \midrule
    BDE Reaction & $\Delta$E &
    BDE Reaction & $\Delta$E &
    BDE Reaction & $\Delta$E \\
    \midrule
SO$_2$$\rightarrow$SO+O        & 0.0190 &
N$_2$O$\rightarrow$N$_2$+O                & 0.0442 &
C$_2$N$_2$$\rightarrow$CN+CN              & 0.2319 \\
N$_2$O$\rightarrow$N$_2$+O     & 0.0174 &
N$_2$O$\rightarrow$NO+N                   & 0.0429 &
C$_3$H$_4$$\rightarrow$CH$_2$C+CH$_2$     & 0.1319 \\
NO$_2$$\rightarrow$NO+O        & 0.0166 &
C$_3$H$_4$$\rightarrow$CH$_2$C+CH$_2$     & 0.0322 &
P$_4$$\rightarrow$P$_2$+P$_2$             & 0.1231 \\
N$_2$O$\rightarrow$NO+N        & 0.0157 &
C$_2$H$_2$O$_2$$\rightarrow$HCO+HCO       & 0.0189 &
N$_2$O$\rightarrow$N$_2$+O                & 0.1119 \\
S$_2$O$\rightarrow$S$_2$+O     & 0.0149 &
C$_2$N$_2$$\rightarrow$CN+CN              & 0.0141 &
                                           &        \\
S$_2$O$\rightarrow$S+SO        & 0.0146 &
P$_4$$\rightarrow$P$_2$+P$_2$             & 0.0139 &
                                           &        \\
CO$_2$$\rightarrow$CO+O        & 0.0137 &
                              &        &
                                           &        \\
t-HONO$\rightarrow$H+NO$_2$    & 0.0133 &
                              &        &
                                           &        \\
HCNO$\rightarrow$CH+NO         & 0.0131 &
                              &        &
                                           &        \\
HNNN$\rightarrow$N$_2$+NH      & 0.0107 &
                              &        &
                                           &        \\
HNCO$\rightarrow$NH+CO         & 0.0096 &
                              &        &
                                           &        \\
CS$_2$$\rightarrow$CS+S        & 0.0073 &
                              &        &
                                           &        \\
    \bottomrule
    \end{tabular}
    \caption{Statistical outliers in BDE predictions for CCSD, SQD$_{\rm 400}$, and SQD$_{\rm ext}$, as identified based on interquartile range whisker criteria.}
    \label{tab:BDE_outliers}
\end{table}

\newpage
\subsubsection{Isomerization Energies}
\label{sec:ISO}

\begin{figure*}[ht!]
\centering
\includegraphics[width = .7\textwidth]{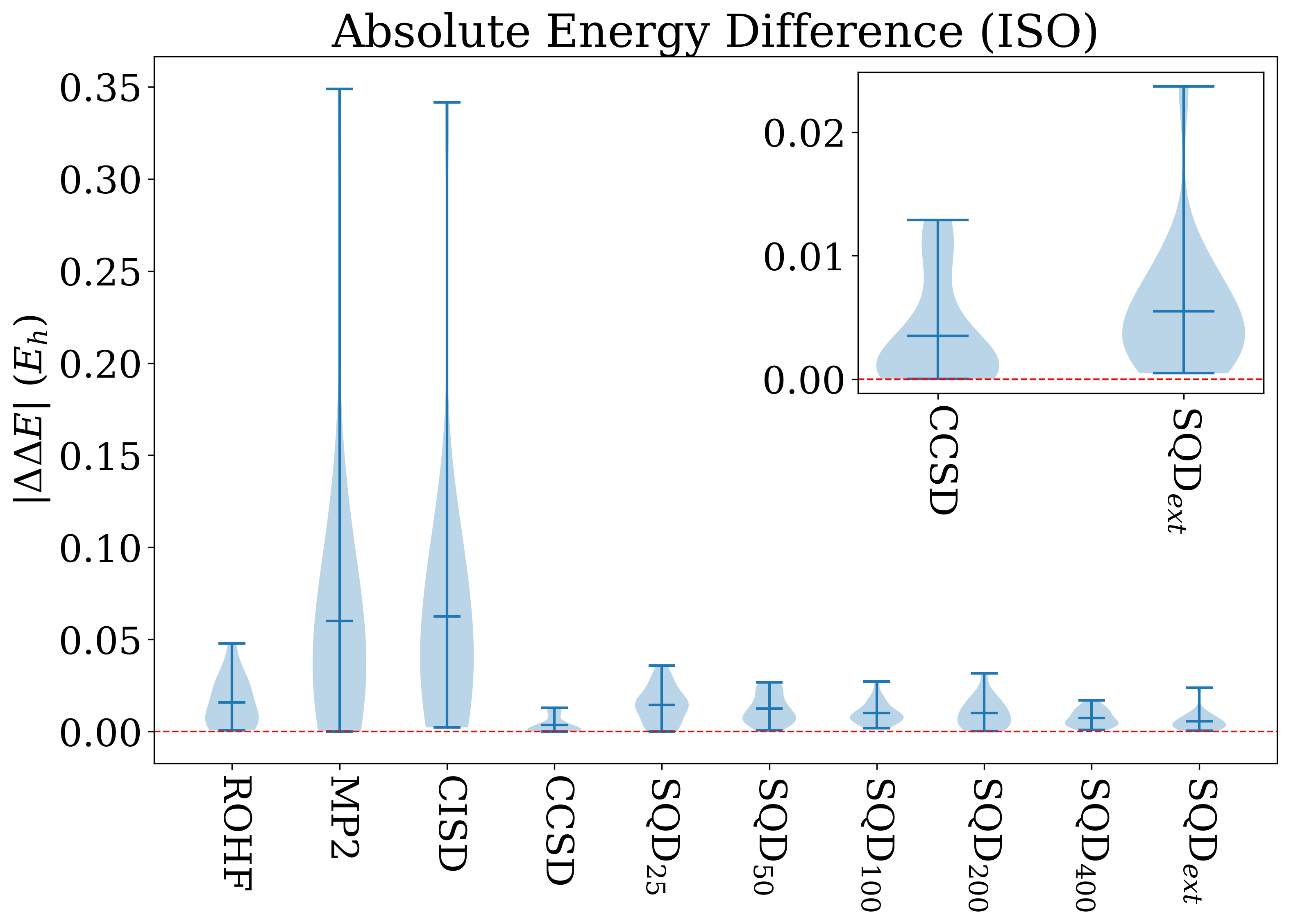}
\caption{\label{fig:ISO} Show a side-by-side comparison of violin plots of the error distribution for isomerization energies.}
\end{figure*}

We now turn to an analogous analysis for the ISOs. The W4-11 dataset contains 20 isomerization reactions. As shown in Figure~\ref{fig:ISO}, the violin plots highlight marked variations in error distributions across the methods investigated. Compared to bond dissociation processes, the deviations are generally more symmetric but still feature notable outliers that broaden the tails of the distributions. To ensure a consistent basis of comparison, we again evaluate statistical descriptors using CCSD(T) as the reference, following the same procedure applied in the TAE and BDE analyses. Particular attention is given to the relative performance of CCSD, SQD$_{\rm 400}$, and SQD$_{\rm ext}$, which display discernible differences in their treatment of subtle correlation effects central to isomerization energetics.\\

The results are summarized in Table~\ref{tab:Stat_profil_ISO}. CCSD achieves the lowest median deviation (0.0381 $E_h$) with a relatively compact interquartile range (0.1112 $E_h$), although its distribution still includes a notable negative outlier. SQD$_{\rm ext}$ shows a median error more than double in size (0.0842 $E_h$) as well as a larger spread (IQR = 0.2744 $E_h$). SQD$_{\rm 400}$ yields substantially inflated errors, with both the median (0.1182 $E_h$) and interquartile range (0.4551 $E_h$) far exceeding those of the other methods. These results highlight that while CCSD remains the most accurate approach overall, SQD$_{\rm ext}$ introduces an increase in error, whereas SQD$_{\rm 400}$ shows markedly larger deviations, indicating that restricted-resource SQD calculations struggle to capture the delicate energetic balance of isomerization reactions.

\pagebreak
\begin{table}[ht!]
    \centering
    \begin{tabular}{r|cccccc}
    \toprule
    Method      &  median & 1$^{\rm st}$ quartile & 3$^{\rm rd}$ quartile & IQR & max$_W$ & min$_W$\\
    \midrule
CCSD             &   0.0381 & 0.0133 & 0.1244 & 0.1112 & 0.2911 & -0.1535\\
SQD$_{\rm ext}$  &   0.0842 & 0.0481 & 0.3225 & 0.2744 & 0.7341 & -0.3635\\
SQD$_{\rm 400}$  &   0.1182 & 0.0616 & 0.5168 & 0.4551 & 1.1994 & -0.6210\\
\bottomrule
    \end{tabular}
    \caption{Statistical error profiles for ISOs of CCSD, SQD$_{\rm 400}$, and SQD$_{\rm ext}$ in terms of median, first quartile ($Q_1$), third quartile ($Q_3$), interquartile range (IQR), and the min and max whiskers}
    \label{tab:Stat_profil_ISO}
\end{table}

Based on the minimum and maximum whiskers computed from the interquartile range, we identify statistical outliers in Figure~\ref{fig:ISO} for CCSD, SQD$_{\rm 400}$, and SQD$_{\rm ext}$; the corresponding reactions are listed in Table~\ref{tab:ISO_outliers}. The largest deviations in isomerization energies arise from chemically labile rearrangements and strongly multi-reference pathways. For CCSD, the dominant outliers are the tautomerizations HNCO$\rightarrow$HOCN and HCNO$\rightarrow$HONC, both exceeding 2.6~$E_h$, reflecting the inherent limitations of single-reference coupled-cluster theory for proton-transfer and bond-reordering processes. SQD$_{\rm ext}$ highlights a related set of electronically flexible species, including the proton shift in HNCO and the rearrangement t-HOOO$\rightarrow$c-HOOO, as well as the allene–propyne isomerization, all of which feature delicate balances between competing bonding motifs. SQD$_{\rm 400}$, constrained by its reduced quantum resources, identifies an even broader range of systems, with large errors for t-HONO$\rightarrow$c-HONO, HNCO$\rightarrow$HOCN, HCNO$\rightarrow$HONC, and t-HOOO$\rightarrow$c-HOOO, spanning roughly 1-4~$E_h$. Taken together, these results indicate that all methods (classical and quantum) struggle with isomerizations involving proton transfers, oxygenated radical rearrangements, and near-degenerate bonding topologies. While CCSD exhibits large individual deviations, SQD methods show greater variability in which systems are flagged, reflecting their sensitivity to resource limitations and the structure of the extrapolation protocol. The persistence of these challenging reactions across methods underscores the intrinsic complexity of strongly multi-reference isomerization pathways, where subtle electron correlation and open-shell effects govern the energetics.

\begin{table}[h!]
    \centering
    \begin{tabular}{c|c|c|c|c|c}
    \toprule
    \multicolumn{2}{c|}{CCSD} &
    \multicolumn{2}{|c|}{SQD$_{\rm ext}$} &
    \multicolumn{2}{|c}{SQD$_{\rm 400}$} \\
    \midrule
    ISO Reaction & $\Delta$E &
    ISO Reaction & $\Delta$E &
    ISO Reaction & $\Delta$E \\
    \midrule
HNCO$\rightarrow$HOCN      & 2.6026 &
t-HONO$\rightarrow$c-HONO   & 1.8385 &
t-HONO$\rightarrow$c-HONO   & 3.6663 \\
HCNO$\rightarrow$HONC      & 2.7343 &
HNCO$\rightarrow$HOCN        & 1.9704 &
HNCO$\rightarrow$HOCN        & 1.4666 \\
                           &        &
t-HOOO$\rightarrow$c-HOOO    & 1.4811 &
HCNO$\rightarrow$HONC        & 2.4510 \\
                           &        &
H3C-C$\equiv$CH$\rightarrow$H2C=C=CH2   & 1.1027 &
t-HOOO$\rightarrow$c-HOOO    & 2.8647 \\
    \bottomrule
    \end{tabular}
    \caption{Statistical outliers in ISO predictions for CCSD, SQD$_{\rm 400}$, and SQD$_{\rm ext}$, as identified based on interquartile range whisker criteria.}
    \label{tab:ISO_outliers}
\end{table}

\pagebreak
\subsubsection{Nucleophilic Substitution}
\label{sec:SN}

\begin{figure*}[ht!]
\centering
\includegraphics[width = .7\textwidth]{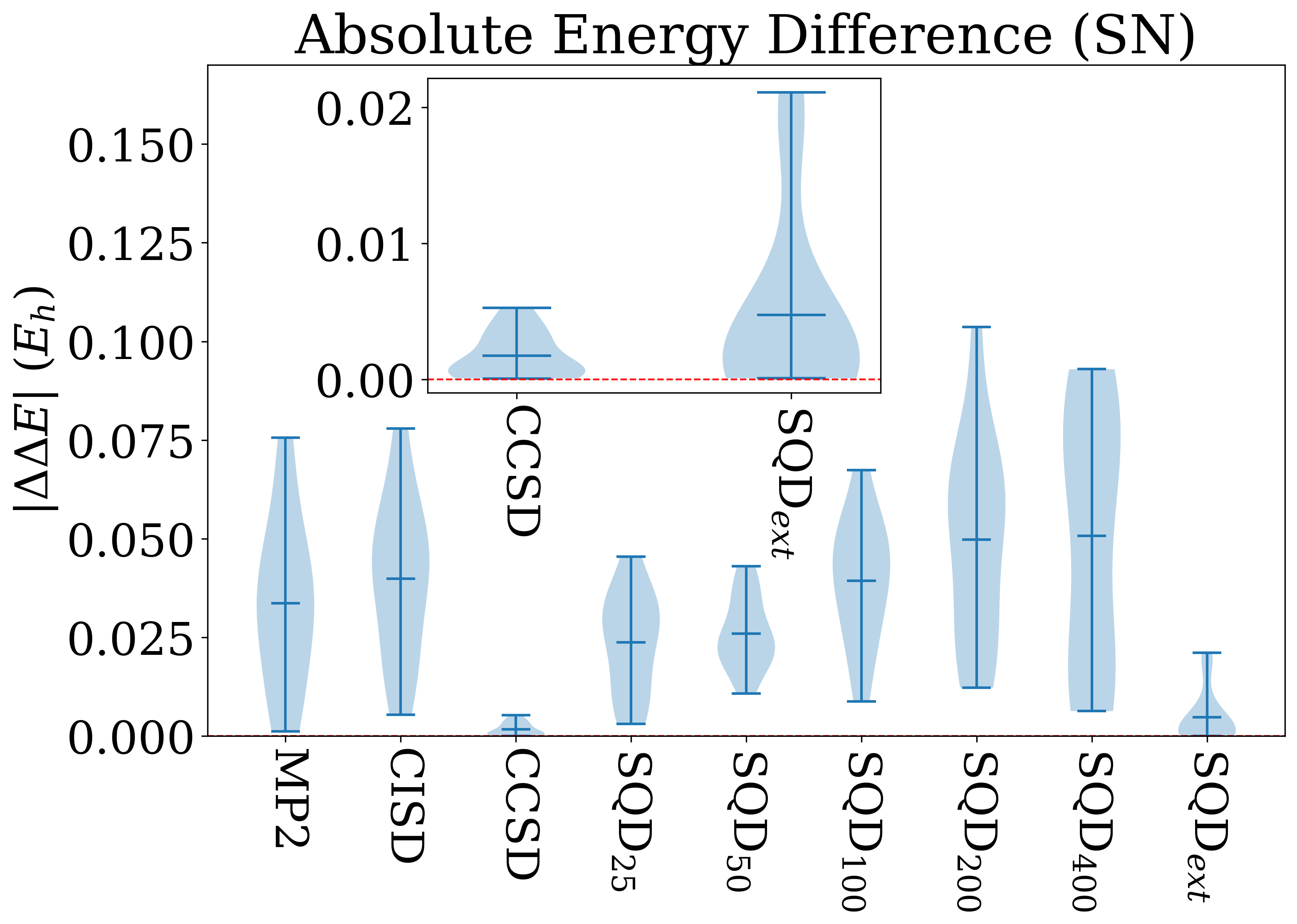}
\caption{\label{fig:SN} Show a side-by-side comparison of violin plots of the error distribution for nucleophilic substitution energies.}
\end{figure*}

We now turn to an analogous analysis for the nucleophilic substitution reactions. The W4-11 dataset contains 13 such processes. As shown in Figure~\ref{fig:ISO}, the violin plots show the large variations in error distributions across the methods investigated. In particular, we observe the counterintuitive behavior in SQD that, as the computational resources increase, the results deteriorate, indicating an error propagation in SN reactions. To ensure a consistent basis of comparison, we again evaluate statistical descriptors using CCSD(T) as the reference, following the same procedure applied in the TAE and BDE analyses. Particular attention is given to the relative performance of CCSD, SQD$_{\rm 400}$, and SQD$_{\rm ext}$, which display discernible differences in their treatment of the subtle correlation effects central to substitution energetics.\\

The results are summarized in Table~\ref{tab:Stat_profil_SN}. CCSD achieves the lowest median deviation (0.0010~$E_h$) with a very compact interquartile range (0.0023~$E_h$), indicating excellent consistency and minimal statistical spread. SQD$_{\rm ext}$ shows a larger median error (0.0016~$E_h$) and a broader IQR (0.0047~$E_h$), but still remains in close agreement with CCSD, demonstrating that extrapolation largely yields comparable accuracy. In contrast, SQD$_{\rm 400}$ exhibits a substantially inflated median deviation (0.0521~$E_h$) and wide interquartile range (0.0622~$E_h$), together with large whisker values that highlight the reduced reliability. Overall, CCSD provides the most accurate and consistent description of SN energetics, SQD$_{\rm ext}$ offers a stable quantum alternative, and SQD$_{\rm 400}$ struggles to capture the delicate balance of charge transfer and correlation in nucleophilic substitution reactions.

\begin{table}[ht!]
    \centering
    \begin{tabular}{r|cccccc}
    \toprule
    Method      &  median & 1$^{\rm st}$ quartile & 3$^{\rm rd}$ quartile & IQR & max$_W$ & min$_W$\\
    \midrule
CCSD             &  0.0010 & 0.0004 & 0.0028 & 0.0023 & 0.0063 & -0.0031\\
SQD$_{\rm ext}$  &  0.0016 & 0.0009 & 0.0056 & 0.0047 & 0.0126 & -0.0062\\
SQD$_{\rm 400}$  &  0.0521 & 0.0181 & 0.0802 & 0.0622 & 0.1735 & -0.0752\\
\bottomrule
    \end{tabular}
    \caption{Statistical error profiles for SNs of CCSD, SQD$_{\rm 400}$, and SQD$_{\rm ext}$ in terms of median, first quartile ($Q_1$), third quartile ($Q_3$), interquartile range (IQR), and the min and max whiskers}
    \label{tab:Stat_profil_SN}
\end{table}

Based on the minimum and maximum whiskers computed from the interquartile range, we identify statistical outliers in Figure~\ref{fig:SN} for CCSD, SQD$_{\rm 400}$, and SQD$_{\rm ext}$. The only reactions classified as outliers occur for SQD$_{\rm ext}$, namely C$_2$H$_5$F+CH$_3$$\rightarrow$F+C$_3$H$_8$ and HCOF+HCO$\rightarrow$F+C$_2$H$_2$O$_2$, with deviations of 0.0211~$E_h$ and 0.0182~$E_h$, respectively. No outliers are detected for CCSD, reflecting its consistently compact error distribution. Although SQD$_{\rm 400}$ does not produce individual reactions beyond the whisker threshold, its large median shift and broad interquartile range indicate substantial statistical spread. Taken together, these results show that while CCSD and SQD$_{\rm ext}$ provide robust and reliable predictions for substitution energetics, SQD$_{\rm 400}$ remains significantly less stable, with wide variability across the dataset even in the absence of formally classified outliers.

\begin{table}[h!]
    \centering
    \begin{tabular}{c|c|c|c|c|c}
    \toprule
    \multicolumn{2}{c|}{CCSD} &
    \multicolumn{2}{|c|}{SQD$_{\rm ext}$} &
    \multicolumn{2}{|c}{SQD$_{\rm 400}$} \\
    \midrule
    SN Reaction & $\Delta$E &
    SN Reaction & $\Delta$E &
    SN Reaction & $\Delta$E \\
    \midrule
--- & --- &
C$_2$H$_5$F+CH$_3$$\rightarrow$F+C$_3$H$_8$ & 0.0211 &
--- & --- \\
--- & --- &
HCOF+HCO$\rightarrow$F+C$_2$H$_2$O$_2$ & 0.0182 &
--- & --- \\
    \bottomrule
    \end{tabular}
    \caption{Statistical outliers in SN predictions for CCSD, SQD$_{\rm ext}$, and SQD$_{\rm 400}$ based on interquartile range whisker criteria.}
    \label{tab:SN_outliers}
\end{table}

\newpage
\subsubsection{Heavy Atom Transfer}
\label{sec:HAT}

\begin{figure*}[ht!]
\centering
\includegraphics[width = .7\textwidth]{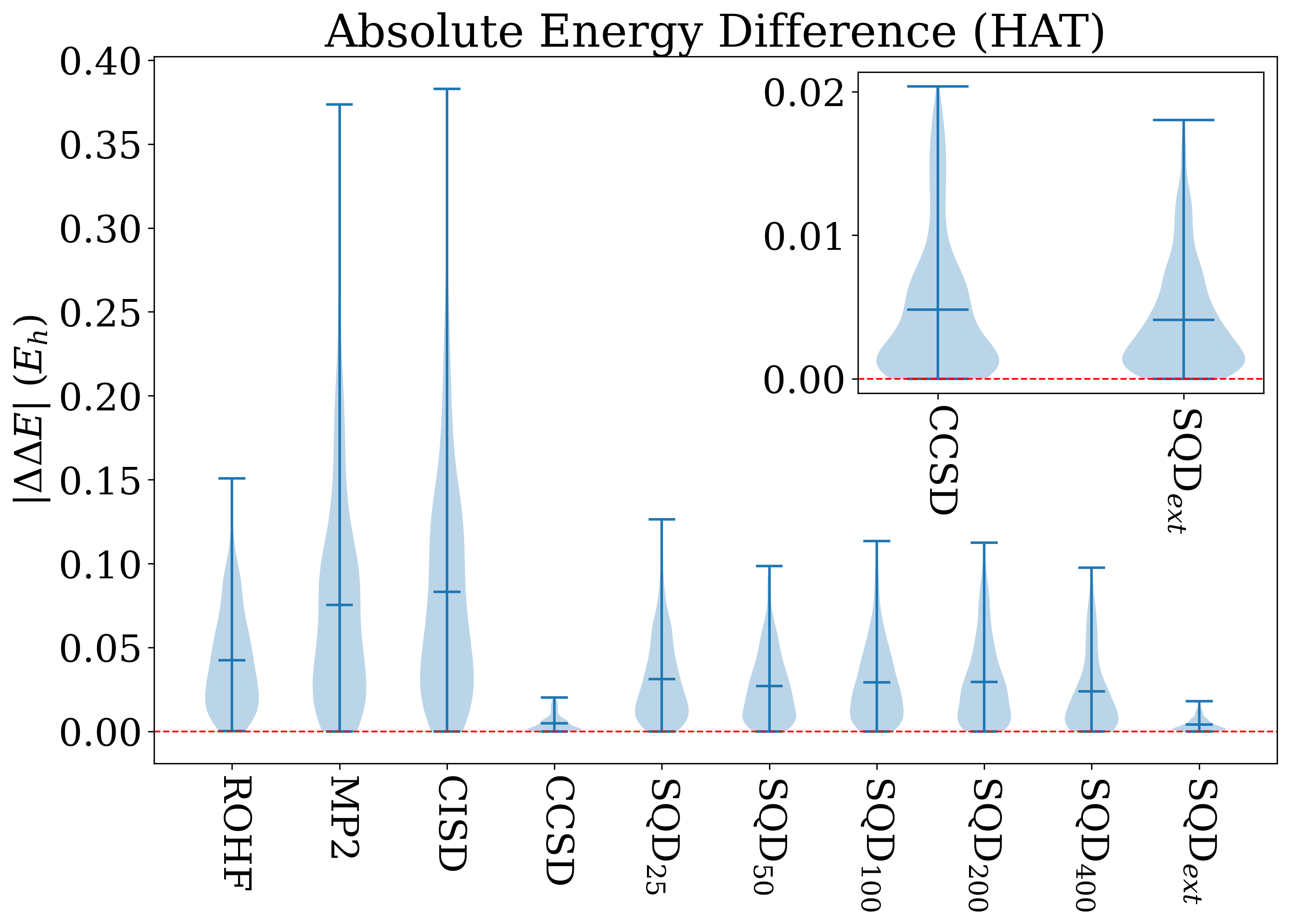}
\caption{\label{fig:HAT} Show a side-by-side comparison of violin plots of the error distribution for heavy atom transfer energies.}
\end{figure*}

We now turn to an analogous analysis for the HAT reactions. The W4-11 dataset contains 505 such processes. As shown in Figure~\ref{fig:HAT}, the violin plots reveal notable variations in predictive accuracy across the methods investigated. MP2 and CISD display wide error distributions with long whiskers, highlighting their limitations for these reactions. CCSD, in contrast, yields substantially smaller deviations. For the SQD hierarchy, the whiskers remain of comparable size across different resource levels, but the mean errors show a gradual decline and the violin plots reveal increasing concentration of values, indicating that additional resources primarily reduce the spread around low-error predictions rather than tightening the extreme outliers. The inset emphasizes that both CCSD and SQD$_{\rm ext}$ achieve very high accuracy, with deviations well below chemical significance. We again evaluate statistical descriptors using CCSD(T) as the reference, following the same procedure as above.\\

The results are summarized in Table~\ref{tab:Stat_profil_HAT}. CCSD achieves a low median deviation (0.0032~$E_h$) together with a narrow interquartile range (0.0058~$E_h$), indicating a high level of accuracy and consistency. SQD$_{\rm ext}$ performs comparably, with a nearly identical median error (0.0030~$E_h$) and a slightly smaller spread (IQR = 0.0048~$E_h$), confirming that extrapolation effectively recovers CCSD-level performance for heavy-atom transfer energetics. In contrast, SQD$_{\rm 400}$ exhibits a median deviation (0.0164~$E_h$) several times larger than CCSD, and its interquartile range (0.0277~$E_h$) is correspondingly broader. The whiskers of SQD$_{\rm 400}$, extending up to 0.0758~$E_h$ and down to $-0.0352$~$E_h$, further highlight its pronounced statistical variability, demonstrating that SQD$_{\rm 400}$ remains substantially less stable under fixed resource constraints.

\begin{table}[ht!]
    \centering
    \begin{tabular}{r|cccccc}
    \toprule
    Method      &  median & 1$^{\rm st}$ quartile & 3$^{\rm rd}$ quartile & IQR & max$_W$ & min$_W$\\
    \midrule
CCSD             &   0.0032 & 0.0010 & 0.0068 & 0.0058 & 0.0154 & -0.0077\\
SQD$_{\rm ext}$  &   0.0030 & 0.0011 & 0.0059 & 0.0048 & 0.0131 & -0.0061\\
SQD$_{\rm 400}$  &   0.0164 & 0.0064 & 0.0342 & 0.0277 & 0.0758 & -0.0352\\
\bottomrule
    \end{tabular}
    \caption{Statistical error profiles for HATs of CCSD, SQD$_{\rm 400}$, and SQD$_{\rm ext}$ in terms of median, first quartile ($Q_1$), third quartile ($Q_3$), interquartile range (IQR), and the min and max whiskers}
    \label{tab:Stat_profil_HAT}
\end{table}

Based on the minimum and maximum whiskers computed from the interquartile range, we identify statistical outliers in Figure~\ref{fig:HAT} for CCSD, SQD$_{\rm 400}$, and SQD$_{\rm ext}$; the corresponding reactions are listed in Table~\ref{tab:HAT_outliers}. The set of HAT outliers is extensive and strongly method-dependent. CCSD flags a broad variety of small-radical reactions, dominated by H-atom abstractions from NO$_2$, CO$_2$, SO$_2$, SiO, and related species, all with deviations of about 0.015–0.020 $E_h$. SQD$_{\rm ext}$ highlights a somewhat different subset, most prominently hydrogen and heteroatom additions to glyoxal (C$_2$H$_2$O$2$), as well as sulfur- and oxygen-centered transformations, with errors typically in the 0.015–0.025 $E_h$ range. By contrast, SQD$_{\rm 400}$ produces a cluster of substantially larger outliers, all involving heavy-atom transfer from carbonyl and oxygenated species such as ketene (C$_2$H$_2$O), glyoxal (C$_2$H$_2$O$_2$), acetaldehyde (C$_2$H$_4$O), and related O/S heteroatom systems, with deviations ranging from 0.08 $E_h$ to more than 0.10 $E_h$.\\

Taken together, these results indicate that while all methods face challenges for radical-heavy atom transfer chemistry, the dominant sources of error differ. CCSD primarily struggles with small radicals and multi-reference open-shell fragments, but errors remain modest in magnitude. SQD$_{\rm ext}$ narrows the spread relative to SQD$_{\rm 400}$, but still reveals systematic deficiencies in describing delocalized oxygenated intermediates. SQD$_{\rm 400}$, under fixed resource constraints, performs worst, with broad statistical spread and consistently large outliers in reactions involving conjugated or multi-bonded heavy-atom systems. This underscores that heavy-atom transfer remains one of the most demanding classes of reactions, requiring high-level correlation treatments or effective extrapolation strategies to achieve reliable accuracy.

\begin{table}[h!]
\rotatebox{270}{%
    \centering
    \begin{tabular}{c|c|c|c|c|c}
    \toprule
    \multicolumn{2}{c|}{CCSD} &
    \multicolumn{2}{|c|}{SQD$_{\rm ext}$} &
    \multicolumn{2}{|c}{SQD$_{\rm 400}$} \\
    \midrule
    HAT Reaction & $\Delta$E &
    HAT Reaction & $\Delta$E &
    HAT Reaction & $\Delta$E \\
    \midrule
H+NO$_2$$\rightarrow$OH+NO              & 0.0166 &
H+C$_2$H$_2$O$_2$$\rightarrow$CH+HCOOH & 0.0180 &
H+NO$_2$$\rightarrow$OH+NO            & 0.0781 \\
H+NO$_2$$\rightarrow$NH+O$_2$             & 0.0193 &
H+CN$\rightarrow$CH+N                 & 0.0147 &
H+C$_2$H$_2$O$_2$$\rightarrow$CH+HCOOH & 0.0827 \\
H+HNCO$\rightarrow$OH+HCN                 & 0.0157 &
NH+C$\rightarrow$CN+H                  & 0.0161 &
H+C$_2$H$_4$O$\rightarrow$CH+CH$_3$OH  & 0.0968 \\
H+OCS$\rightarrow$CH+SO                   & 0.0173 &
H+C$_2$F$_2$$\rightarrow$CH+CF$_2$      & 0.0132 &
OH+C$_2$H$_4$$\rightarrow$H+C$_2$H$_4$ & 0.0796 \\
H+CO$_2$$\rightarrow$CH+O$_2$             & 0.0204 &
N+CH$_2$O$\rightarrow$CN+H$_2$O        & 0.0133 &
H+C$_3$H$_4$$\rightarrow$CH+C$_2$H$_4$ & 0.0916 \\
H+SiO$\rightarrow$SiH+O                   & 0.0175 &
N+C$_2$H$_4$O$\rightarrow$CN+CH$_3$OH   & 0.0149 &
H+C$_3$H$_6$$\rightarrow$CH+C$_2$H$_6$ & 0.0783 \\
H+SiO$\rightarrow$OH+Si                   & 0.0176 &
N+CH$_2$ (triplet)$\rightarrow$CN+H$_2$ & 0.0163 &
C+NO$_2$$\rightarrow$CO+NO            & 0.0767 \\
H+SO$_2$$\rightarrow$OH+SO                & 0.0190 &
O+C$_2$H$_2$O$\rightarrow$O$_2$+C$_2$H$_2$ & 0.0148 &
C+C$_2$H$_4$O$\rightarrow$CO+C$_2$H$_4$ & 0.0782 \\
H+SO$_2$$\rightarrow$HS+O$_2$             & 0.0191 &
O+C$_2$H$_2$O$_2$$\rightarrow$CO+HCOOH & 0.0163 &
O+C$_2$H$_2$O$_2$$\rightarrow$CO+HCOOH & 0.0829 \\
C+NO$_2$$\rightarrow$CN+O$_2$             & 0.0169 &
F+C$_2$H$_2$O$_2$$\rightarrow$CF+HCOOH & 0.0163 &
O+C$_2$H$_4$O$\rightarrow$CO+CH$_3$OH  & 0.0970 \\
N+NO$_2$$\rightarrow$N$_2$+O$_2$          & 0.0175 &
S+C$_2$H$_2$O$_2$$\rightarrow$CS+HCOOH & 0.0160 &
O+C$_3$H$_4$$\rightarrow$CO+C$_2$H$_4$ & 0.0918 \\
N+CO$_2$$\rightarrow$CN+O$_2$             & 0.0180 &
OH+C$_2$H$_2$O$_2$$\rightarrow$HCO+HCOOH & 0.0148 &
O+C$_3$H$_6$$\rightarrow$CO+C$_2$H$_6$  & 0.0785 \\
N+SO$_2$$\rightarrow$NO+SO                & 0.0156 &
                                          &&
F+C$_2$H$_4$O$\rightarrow$CF+CH$_3$OH  & 0.0847 \\
O+NO$_2$$\rightarrow$O$_2$+NO             & 0.0158 &
                                          &&
F+C$_3$H$_4$$\rightarrow$CF+C$_2$H$_4$ & 0.0795 \\
O+SiO$\rightarrow$O$_2$+Si                & 0.0168 &
                                          &&
S+C$_2$H$_2$O$_2$$\rightarrow$CS+HCOOH & 0.0834 \\
O+SO$_2$$\rightarrow$O$_2$+SO             & 0.0183 &
                                          &&
S+C$_2$H$_4$O$\rightarrow$CS+CH$_3$OH  & 0.0975 \\
S+NO$_2$$\rightarrow$SO+NO                & 0.0158 &
                                          &&
S+C$_3$H$_4$$\rightarrow$CS+C$_2$H$_4$ & 0.0924 \\
S+SiO$\rightarrow$SO+Si                   & 0.0168 &
                                          &&
S+C$_3$H$_6$$\rightarrow$CS+C$_2$H$_6$ & 0.0791 \\
S+SO$_2$$\rightarrow$SO+SO                & 0.0183 &
                                          &
                                          &        \\
S+SO$_2$$\rightarrow$S$_2$+O$_2$          & 0.0186 &
                                          &
                                          &        \\
Cl+NO$_2$$\rightarrow$ClO+NO              & 0.0160 &
                                          &
                                          &        \\
Cl+SiO$\rightarrow$ClO+Si                 & 0.0170 &
                                          &
                                          &        \\
Cl+SO$_2$$\rightarrow$ClO+SO              & 0.0185 &
                                          &
                                          &        \\
OH+NO$_2$$\rightarrow$HNO+O$_2$           & 0.0177 &
                                          &
                                          &        \\
OH+CO$_2$$\rightarrow$HCO+O$_2$           & 0.0159 &
                                          &
                                          &        \\
OH+SiO$\rightarrow$HOO+Si                 & 0.0163 &
                                          &
                                          &        \\
OH+SO$_2$$\rightarrow$HOO+SO              & 0.0178 &
                                          &
                                          &        \\
    \bottomrule
    \end{tabular}
    }
    \caption{Statistical outliers in HAT predictions for CCSD, SQD$_{\rm 400}$, and SQD$_{\rm ext}$, based on interquartile range whisker criteria.}
    \label{tab:HAT_outliers}
\end{table}

\end{appendices}
\end{refsegment}

\printbibliography[segment=2,heading=bibintoc,title={References (Appendix)}]

@string{ACP="Atm. Chem. Phys"}

@string{ACR="Acc. Chem. Res"}

@string{Annals="Ann. Phys"}

@string{ARCBE="Annu. Rev. Chem. Biomol. Eng"}

@string{ARCC="Annu. Rep. Comput. Chem"}

@string{ChemRev="Chem. Rev"}

@string{CMP="Comm. Math. Phys"}

@string{CPL="Chem. Phys. Lett"}

@string{CP="Comm. Phys"}

@string{CPC="Chem. Phys. Chem"}

@string{CS="Chem. Sci"}

@string{CSR="Chem. Soc. Rev"}

@string{EJMC="Eur. J. Med. Chem"}

@string{EPTCS="Electron. Proc. Theor. Comp. Sci"}

@string{FGCS="Fut. Gen. Comput. Sys"}

@string{IJTP="Int. J. Theor. Phys"}

@string{JACS="J. Am. Chem. Soc"}

@string{JBIC="J. Biol. Inorg. Chem"}

@string{JCC="J. Comp. Chem"}

@string{JCP="J. Chem. Phys"}

@string{JCTC="J. Chem. Theory Comput"}

@string{JFC="J. Fluor. Chem"}

@string{JOC="J. Org. Chem"}

@string{JPCA="J. Phys. Chem. A"}

@string{JPCB="J. Phys. Chem B"}

@string{JPCL="J. Phys. Chem. Lett"}

@string{JPCRD="J. Phys. Chem. Ref. Data"}

@string{JPD="J. Phys. D"}

@string{JPPC="J. Photochem. Photobiol. C"}

@string{JPOC="J. Phys. Org. Chem"}

@string{MolInf="Mol. Inf"}

@string{MolPhys="Mol. Phys"}

@string{NatComm="Nat. Comm"}

@string{NJP="New J. Phys"}

@string{NPJQI="npj Quantum Inf"}

@string{NRP="Nat. Rev. Phys"}

@string{PCCP="Phys. Chem. Chem. Phys"}

@string{PRAppl="Phys. Rev. Appl"}

@string{PRA="Phys. Rev. A"}

@string{PRC="Phys. Rev. C"}

@string{PRL="Phys. Rev. Lett"}

@string{PRX="Phys. Rev. X"}

@string{QIC="Quantum Info. Comput"}

@string{QICC="Quantum Info. Comput. Chem"}

@string{QST="Quant. Sci. Tech"}

@string{RMP="Rev. Mod. Phys"}

@string{RSA="Proc. R. Soc. Lond. A"}

@string{SciAdv="Sci. Adv"}

@string{TCA="Theor. Chem. Acc"}

@string{WIRES="WIREs Comput. Mol. Sci"}

@article{barison2022quantum,
  title={Quantum simulations of molecular systems with intrinsic atomic orbitals},
  author={Barison, Stefano and Galli, Davide E and Motta, Mario},
  journal=PRA,
  volume={106},
  number={2},
  pages={022404},
  year={2022},
  publisher={APS}
}

@article{liu2022quantum,
  title={Quantum algorithms for electronic structures: basis sets and boundary conditions},
  author={Liu, Jie and Fan, Yi and Li, Zhenyu and Yang, Jinlong},
  journal=CSR,
  volume={51},
  number={8},
  pages={3263--3279},
  year={2022},
  publisher={Royal Society of Chemistry}
}

@article{kwon2023adaptive,
  title={Adaptive basis sets for practical quantum computing},
  author={Kwon, Hyuk-Yong and Curtin, Gregory M and Morrow, Zachary and Kelley, CT and Jakubikova, Elena},
  journal={International Journal of Quantum Chemistry},
  volume={123},
  number={14},
  pages={e27123},
  year={2023},
  publisher={Wiley Online Library}
}

@article{feniou2025real,
  title={Real-Space Chemistry on Quantum Computers: A Fault-Tolerant Algorithm with Adaptive Grids and Transcorrelated Extension},
  author={Feniou, C{\'e}sar and Cherfan, Christopher and Zylberman, Julien and Claudon, Baptiste and Piquemal, Jean-Philip and Giner, Emmanuel},
  journal={arXiv:2507.20583},
  year={2025}
}

@article{faulstich2022discontinuous,
  title={Discontinuous Galerkin method with Voronoi partitioning for quantum simulation of chemistry},
  author={Faulstich, Fabian M and Wu, Xiaojie and Lin, Lin},
  journal={Research in the Mathematical Sciences},
  volume={9},
  number={4},
  pages={68},
  year={2022},
  publisher={Springer}
}

@article{mcclean2020discontinuous,
  title={Discontinuous Galerkin discretization for quantum simulation of chemistry},
  author={McClean, Jarrod R and Faulstich, Fabian M and Zhu, Qinyi and O’Gorman, Bryan and Qiu, Yiheng and White, Steven R and Babbush, Ryan and Lin, Lin},
  journal={New Journal of Physics},
  volume={22},
  number={9},
  pages={093015},
  year={2020},
  publisher={IOP Publishing}
}

@article{georges2025quantum,
  title={Quantum simulations of chemistry in first quantization with any basis set},
  author={Georges, Timothy N and Bothe, Marius and S{\"u}nderhauf, Christoph and Berntson, Bjorn K and Izs{\'a}k, R{\'o}bert and Ivanov, Aleksei V},
  journal={npj Quantum Information},
  volume={11},
  number={1},
  pages={55},
  year={2025},
  publisher={Nature Publishing Group UK London}
}

@article{berry2024quantum,
  title={Quantum simulation of realistic materials in first quantization using non-local pseudopotentials},
  author={Berry, Dominic W and Rubin, Nicholas C and Elnabawy, Ahmed O and Ahlers, Gabriele and DePrince III, A Eugene and Lee, Joonho and Gogolin, Christian and Babbush, Ryan},
  journal={npj Quantum Information},
  volume={10},
  number={1},
  pages={130},
  year={2024},
  publisher={Nature Publishing Group UK London}
}

@article{babbush2019quantum,
  title={Quantum simulation of chemistry with sublinear scaling in basis size},
  author={Babbush, Ryan and Berry, Dominic W and McClean, Jarrod R and Neven, Hartmut},
  journal={npj Quantum Information},
  volume={5},
  number={1},
  pages={92},
  year={2019},
  publisher={Nature Publishing Group UK London}
}

@article{childs2022quantum,
  title={Quantum simulation of real-space dynamics},
  author={Childs, Andrew M and Leng, Jiaqi and Li, Tongyang and Liu, Jin-Peng and Zhang, Chenyi},
  journal={Quantum},
  volume={6},
  pages={860},
  year={2022},
  publisher={Verein zur F{\"o}rderung des Open Access Publizierens in den Quantenwissenschaften}
}

@article{su2021fault,
  title={Fault-tolerant quantum simulations of chemistry in first quantization},
  author={Su, Yuan and Berry, Dominic W and Wiebe, Nathan and Rubin, Nicholas and Babbush, Ryan},
  journal={PRX Quantum},
  volume={2},
  number={4},
  pages={040332},
  year={2021},
  publisher={APS}
}

@article{chan2023grid,
  title={Grid-based methods for chemistry simulations on a quantum computer},
  author={Chan, Hans Hon Sang and Meister, Richard and Jones, Tyson and Tew, David P and Benjamin, Simon C},
  journal={Science Advances},
  volume={9},
  number={9},
  pages={eabo7484},
  year={2023},
  publisher={American Association for the Advancement of Science}
}

@article{kivlichan2017bounding,
  title={Bounding the costs of quantum simulation of many-body physics in real space},
  author={Kivlichan, Ian D and Wiebe, Nathan and Babbush, Ryan and Aspuru-Guzik, Al{\'a}n},
  journal={Journal of Physics A: Mathematical and Theoretical},
  volume={50},
  number={30},
  pages={305301},
  year={2017},
  publisher={IOP Publishing}
}

@article{kassal2008polynomial,
  title={Polynomial-time quantum algorithm for the simulation of chemical dynamics},
  author={Kassal, Ivan and Jordan, Stephen P and Love, Peter J and Mohseni, Masoud and Aspuru-Guzik, Al{\'a}n},
  journal={Proceedings of the National Academy of Sciences},
  volume={105},
  number={48},
  pages={18681--18686},
  year={2008},
  publisher={National Academy of Sciences}
}

@article{berry2019qubitization,
  title={Qubitization of arbitrary basis quantum chemistry leveraging sparsity and low rank factorization},
  author={Berry, Dominic W and Gidney, Craig and Motta, Mario and McClean, Jarrod R and Babbush, Ryan},
  journal={Quantum},
  volume={3},
  pages={208},
  year={2019},
  publisher={Verein zur F{\"o}rderung des Open Access Publizierens in den Quantenwissenschaften}
}

@article{motta2021low,
  title={Low rank representations for quantum simulation of electronic structure},
  author={Motta, Mario and Ye, Erika and McClean, Jarrod R and Li, Zhendong and Minnich, Austin J and Babbush, Ryan and Chan, Garnet Kin-Lic},
  journal={npj Quantum Information},
  volume={7},
  number={1},
  pages={83},
  year={2021},
  publisher={Nature Publishing Group UK London}
}

@online{w411Challenge,
  title = {W4 Challenge},
  url = {https://w4dataset.foci.rpi.edu/}
}

@article{dirac1928quantum,
  title   = {The quantum theory of the electron},
  author  = {Dirac, Paul Adrien Maurice},
  journal = RSA,
  volume  = {117},
  number  = {778},
  pages   = {610--624},
  year    = {1928},
  url     = {https://doi.org/10.1098/rspa.1928.0023}
}

@article{vogiatzis2017pushing,
  author  = {Vogiatzis, Konstantinos D. and Ma, Dongxia and Olsen, Jeppe and Gagliardi, Laura and de Jong, Wibe A.},
  title   = {Pushing configuration-interaction to the limit: towards massively parallel MCSCF calculations},
  journal = JCP,
  volume  = {147},
  number  = {18},
  pages   = {184111},
  year    = {2017},
  url     = {https://doi.org/10.1063/1.4989858}
}

@article{gao2024distributed,
  author  = {Gao, Hong and Imamura, Satoshi and Kasagi, Akihiko and Yoshida, Eiji},
  title   = {Distributed implementation of full configuration interaction for one trillion determinants},
  journal = JCTC,
  volume  = {20},
  number  = {3},
  pages   = {1185-1192},
  year    = {2024},
  url     = {https://doi.org/10.1021/acs.jctc.3c01190}
}

@article{leblanc2015solutions,
  title   = {Solutions of the two-dimensional Hubbard model: benchmarks and results from a wide range of numerical algorithms},
  author  = {LeBlanc, James P. F. and Antipov, Andrey E and Becca, Federico and Bulik, Ireneusz W and Chan, Garnet Kin-Lic and Chung, Chia-Min and Deng, Youjin and Ferrero, Michel and Henderson, Thomas M and Jim{\'e}nez-Hoyos, Carlos A and others},
  journal = PRX,
  volume  = {5},
  number  = {4},
  pages   = {041041},
  year    = {2015},
  url     = {https://doi.org/10.1103/PhysRevX.5.041041}
}

@article{zheng2017stripe,
  title   = {Stripe order in the underdoped region of the two-dimensional Hubbard model},
  author  = {Zheng, Bo-Xiao and Chung, Chia-Min and Corboz, Philippe and Ehlers, Georg and Qin, Ming-Pu and Noack, Reinhard M and Shi, Hao and White, Steven R and Zhang, Shiwei and Chan, Garnet Kin-Lic},
  journal = {Science},
  volume  = {358},
  number  = {6367},
  pages   = {1155--1160},
  year    = {2017},
  url     = {https://www.science.org/doi/10.1126/science.aam7127}
}

@article{motta2017towards,
  title   = {Towards the solution of the many-electron problem in real materials: equation of state of the hydrogen chain with state-of-the-art many-body methods},
  author  = {Motta, Mario and Ceperley, David M and Chan, Garnet Kin-Lic and Gomez, John A and Gull, Emanuel and Guo, Sheng and Jim{\'e}nez-Hoyos, Carlos A and Lan, Tran Nguyen and Li, Jia and Ma, Fengjie and others},
  journal = PRX,
  volume  = {7},
  number  = {3},
  pages   = {031059},
  year    = {2017},
  url     = {https://doi.org/10.1103/PhysRevX.7.031059}
}

@article{eriksen2020ground,
  title   = {The ground-state electronic energy of benzene},
  author  = {Eriksen, Janus J and Anderson, Tyler A and Deustua, J Emiliano and Ghanem, Khaldoon and Hait, Diptarka and Hoffmann, Mark R and Lee, Seunghoon and Levine, Daniel S and Magoulas, Ilias and Shen, Jun and others},
  journal = JPCL, 
  volume  = {11},
  number  = {20},
  pages   = {8922--8929},
  year    = {2020},
  url     = {https://doi.org/10.1021/acs.jpclett.0c02621}
}

@article{williams2020direct,
  title   = {Direct comparison of many-body methods for realistic electronic Hamiltonians},
  author  = {Williams, Kiel T and Yao, Yuan and Li, Jia and Chen, Li and Shi, Hao and Motta, Mario and Niu, Chunyao and Ray, Ushnish and Guo, Sheng and Anderson, Robert J and others},
  journal = PRX,
  volume  = {10},
  number  = {1},
  pages   = {011041},
  year    = {2020},
  url     = {https://doi.org/10.1103/PhysRevX.10.011041}
}

@article{alexeev2024quantum,
  title   = {Quantum-centric supercomputing for materials science: A perspective on challenges and future directions},
  author  = {Alexeev, Yuri and Amsler, Maximilian and Barroca, Marco Antonio and Bassini, Sanzio and Battelle, Torey and Camps, Daan and Casanova, David and Choi, Young Jay and Chong, Frederic T and Chung, Charles and others},
  journal = FGCS,
  volume  = {160},
  pages   = {666--710},
  year    = {2024},
  url     = {https://doi.org/10.1016/j.future.2024.04.060}
}

@Article{feynman1982simulating,
  author  = {Feynman, Richard P.},
  title   = {Simulating physics with computers},
  journal = IJTP,
  year    = {1982},
  volume  = {21},
  number  = {6},
  pages   = {467-488},
  url     = {https://doi.org/10.1007/BF02650179}
}

@article{georgescu2014quantum,
  title   = {Quantum simulation},
  author  = {Georgescu, Iulia M and Ashhab, Sahel and Nori, Franco},
  journal = RMP,
  volume  = {86},
  number  = {1},
  pages   = {153},
  year    = {2014},
  url     = {https://doi.org/10.1103/RevModPhys.86.153}
}

@article{cao2019quantum,
  title   = {Quantum chemistry in the age of quantum computing},
  author  = {Cao, Yudong and Romero, Jonathan and Olson, Jonathan P and Degroote, Matthias and Johnson, Peter D and Kieferov{\'a}, M{\'a}ria and Kivlichan, Ian D and Menke, Tim and Peropadre, Borja and Sawaya, Nicolas PD and others},
  journal = ChemRev,
  volume  = {119},
  number  = {19},
  pages   = {10856--10915},
  year    = {2019},
  url     = {https://doi.org/10.1021/acs.chemrev.8b00803}
}

@article{cerezo2020variational,
  title   = {Variational quantum algorithms},
  author  = {Cerezo, Marco and Arrasmith, Andrew and Babbush, Ryan and Benjamin, Simon C and Endo, Suguru and Fujii, Keisuke and McClean, Jarrod R and Mitarai, Kosuke and Yuan, Xiao and Cincio, Lukasz and others},
  journal = NRP,
  pages   = {1--20},
  year    = {2021},
  url     = {https://doi.org/10.1038/s42254-021-00348-9}
}

@article{mcardle2020quantum,
  title   = {Quantum computational chemistry},
  author  = {McArdle, Sam and Endo, Suguru and Aspuru-Guzik, Al\'an and Benjamin, Simon C. and Yuan, Xiao},
  journal = RMP,
  volume  = {92},
  issue   = {1},
  pages   = {015003},
  year    = {2020},
  url     = {https://doi.org/10.1103/RevModPhys.92.015003}
}

@article{bauer2020quantum,
  title    = {Quantum algorithms for quantum chemistry and quantum materials science},
  author   = {Bauer, Bela and Bravyi, Sergey and Motta, Mario and Kin-Lic Chan, Garnet},
  journal  = ChemRev,
  volume   = {120},
  number   = {22},
  pages    = {12685--12717},
  year     = {2020},
  url      = {https://doi.org/10.1021/acs.chemrev.9b00829}
}

@article{eisert2020quantum,
  title={Quantum certification and benchmarking},
  author={Eisert, Jens and Hangleiter, Dominik and Walk, Nathan and Roth, Ingo and Markham, Damian and Parekh, Rhea and Chabaud, Ulysse and Kashefi, Elham},
  journal=NRP,
  volume={2},
  number={7},
  pages={382--390},
  year={2020},
  publisher={Nature Publishing Group UK London},
  url={https://doi.org/10.1038/s42254-020-0186-4}
}

@article{proctor2025benchmarking,
  title={Benchmarking quantum computers},
  author={Proctor, Timothy and Young, Kevin and Baczewski, Andrew D and Blume-Kohout, Robin},
  journal=NRP,
  volume={7},
  number={2},
  pages={105--118},
  year={2025},
  publisher={Nature Publishing Group UK London},
  url={https://doi.org/10.1038/s42254-024-00796-z}
}

@article{karton2011w4,
  title   = {W4-11: A high-confidence benchmark dataset for computational thermochemistry derived from first-principles W4 data},
  author  = {Karton, Amir and Daon, Shauli and Martin, Jan ML},
  journal = CPL,
  volume  = {510},
  number  = {4-6},
  pages   = {165--178},
  year    = {2011},
  url     = {https://doi.org/10.1016/j.cplett.2011.05.007}
}

@article{karton2017w4,
  title   = {W4-17: A diverse and high-confidence dataset of atomization energies for benchmarking high-level electronic structure methods},
  author  = {Karton, Amir and Sylvetsky, Nitai and Martin, Jan ML},
  journal = JCC,
  volume  = {38},
  number  = {24},
  pages   = {2063--2075},
  year    = {2017},
  url     = {https://doi.org/10.1002/jcc.24854}
}

@article{raghavachari1989fifth,
  title   = {A fifth-order perturbation comparison of electron correlation theories},
  author  = {Raghavachari, Krishnan and Trucks, Gary W and Pople, John A and Head-Gordon, Martin},
  journal = CPL,
  volume  = {157},
  number  = {6},
  pages   = {479--483},
  year    = {1989},
  url     = {https://doi.org/10.1016/S0009-2614(89)87395-6}
}

@article{robledo2025chemistry,
  title   = {Chemistry beyond the scale of exact diagonalization on a quantum-centric supercomputer},
  author  = {Robledo-Moreno, Javier and Motta, Mario and Haas, Holger and Javadi-Abhari, Ali and Jurcevic, Petar and Kirby, William and Martiel, Simon and Sharma, Kunal and Sharma, Sandeep and Shirakawa, Tomonori and others},
  journal = SciAdv,
  volume  = {11},
  number  = {25},
  pages   = {eadu9991},
  year    = {2025},
  url     = {https://www.science.org/doi/10.1126/sciadv.adu9991}
}

@article{hehre1969self,
  title   = {Self-consistent molecular-orbital methods. I. Use of Gaussian expansions of Slater-type atomic orbitals},
  author  = {Hehre, Warren J and Stewart, Robert F and Pople, John A},
  journal = JCP,
  volume  = {51},
  number  = {6},
  pages   = {2657--2664},
  year    = {1969},
  url     = {https://doi.org/10.1063/1.1672392}
}

@article{tubman2016deterministic,
  title   = {A deterministic alternative to the full configuration interaction quantum Monte Carlo method},
  author  = {Tubman, Norm M and Lee, Joonho and Takeshita, Tyler Y and Head-Gordon, Martin and Whaley, K Birgitta},
  journal = JCP,
  volume  = {145},
  number  = {4},
  year    = {2016},
  url     = {https://doi.org/10.1063/1.4955109}
}

@article{holmes2016heat,
  title   = {Heat-bath configuration interaction: An efficient selected configuration interaction algorithm inspired by heat-bath sampling},
  author  = {Holmes, Adam A and Tubman, Norm M and Umrigar, Cyrus J},
  journal = JCTC,
  volume  = {12},
  number  = {8},
  pages   = {3674--3680},
  year    = {2016},
  url     = {https://pubs.acs.org/doi/10.1021/acs.jctc.6b00407}
}

@article{schriber2016communication,
  title   = {Communication: An adaptive configuration interaction approach for strongly correlated electrons with tunable accuracy},
  author  = {Schriber, Jeffrey B and Evangelista, Francesco A},
  journal = JCP,
  volume  = {144},
  number  = {16},
  year    = {2016},
  url     = {https://doi.org/10.1063/1.4948308}
}

@article{kanno2023quantum,
  title   = {Quantum-selected configuration interaction: classical diagonalization of Hamiltonians in subspaces selected by quantum computers},
  author  = {Kanno, Keita and Kohda, Masaya and Imai, Ryosuke and Koh, Sho and Mitarai, Kosuke and Mizukami, Wataru and Nakagawa, Yuya O},
  journal = {arXiv:2302.11320},
  year    = {2023},
  url     = {https://arxiv.org/abs/2302.11320}
}

@article{motta2023bridging,
  title   = {Bridging physical intuition and hardware efficiency for correlated electronic states: the local unitary cluster Jastrow ansatz for electronic structure},
  author  = {Motta, Mario and Sung, Kevin J and Whaley, K Birgitta and Head-Gordon, Martin and Shee, James},
  journal = CS,
  volume  = {14},
  number  = {40},
  pages   = {11213--11227},
  year    = {2023},
  url     = {https://doi.org/10.1039/D3SC02516K}
}

@article{kaliakin2025accurate,
  title   = {Accurate quantum-centric simulations of intermolecular interactions},
  author  = {Kaliakin, Danil and Shajan, Akhil and Liang, Fangchun and Robledo Moreno, Javier and Li, Zhen and Mitra, Abhishek and Motta, Mario and Johnson, Caleb and Saki, Abdullah Ash and Das, Susanta and others},
  journal = CP,
  volume  = {8},
  number  = {1},
  pages   = {396},
  year    = {2025},
  url     = {https://doi.org/10.1038/s42005-025-02305-9}
}

@article{shajan2025toward,
  title   = {Toward quantum-centric simulations of extended molecules: sample-based quantum diagonalization enhanced with density matrix embedding theory},
  author  = {Shajan, Akhil and Kaliakin, Danil and Mitra, Abhishek and Robledo Moreno, Javier and Li, Zhen and Motta, Mario and Johnson, Caleb and Saki, Abdullah Ash and Das, Susanta and Sitdikov, Iskandar and others},
  journal = JCTC,
  volume  = {21},
  number  = {14},
  pages   = {6801--6810},
  year    = {2025},
  url     = {https://doi.org/10.1021/acs.jctc.5c00114}
}

@article{kaliakin2025implicit,
  title   = {Implicit solvent sample-based quantum diagonalization},
  author  = {Kaliakin, Danil and Shajan, Akhil and Liang, Fangchun and Merz Jr, Kenneth M},
  journal = JPCB,
  volume  = {129},
  number  = {23},
  pages   = {5788--5796},
  year    = {2025},
  url     = {https://pubs.acs.org/doi/10.1021/acs.jpcb.5c01030}
}

@article{bazayeva2025quantum,
  title   = {Quantum-Centric Alchemical Free Energy Calculations},
  author  = {Bazayeva, Milana and Li, Zhen and Kaliakin, Danil and Liang, Fangchun and Shajan, Akhil and Das, Susanta and Merz Jr, Kenneth M},
  journal = {arXiv:2506.20825},
  year    = {2025},
  url     = {https://arxiv.org/abs/2506.20825}
}

@article{danilov2025enhancing,
  title   = {Enhancing the accuracy and efficiency of sample-based quantum diagonalization with phaseless auxiliary-field quantum Monte Carlo},
  author  = {Danilov, Don and Robledo-Moreno, Javier and Sung, Kevin J and Motta, Mario and Shee, James},
  journal = {arXiv:2503.05967},
  year    = {2025},
  url     = {https://arxiv.org/abs/2503.05967}
}

@article{liepuoniute2025quantum,
  title   = {Quantum-centric computational study of methylene singlet and triplet states},
  author  = {Liepuoniute, Ieva and Doney, Kirstin D and Robledo Moreno, Javier and Job, Joshua A and Friend, William S and Jones, Gavin O},
  journal = JCTC,
  volume  = {21},
  number  = {10},
  pages   = {5062--5070},
  year    = {2025},
  url     = {https://doi.org/10.1021/acs.jctc.5c00075}
}

@article{duriez2025computing,
  title   = {Computing band gaps of periodic materials via sample-based quantum diagonalization},
  author  = {Duriez, Alan and Carvalho, Pamela C and Barroca, Marco Antonio and Zipoli, Federico and Jaderberg, Ben and Ferreira, Rodrigo Neumann Barros and Sharma, Kunal and Mezzacapo, Antonio and Wunsch, Benjamin and Steiner, Mathias},
  journal = {arXiv:2503.10901},
  year    = {2025},
  url     = {https://arxiv.org/abs/2503.10901}
}

@article{barroca2025surface,
  title   = {Surface reaction simulations for battery materials through sample-based quantum diagonalization and local embedding},
  author  = {Barroca, Marco Antonio and Gujarati, Tanvi and Sharma, Vidushi and Ferreira, Rodrigo Neumann Barros and Na, Young-Hye and Giammona, Maxwell and Mezzacapo, Antonio and Wunsch, Benjamin and Steiner, Mathias},
  journal = {arXiv:2503.10923},
  year    = {2025},
  url     = {https://arxiv.org/abs/2503.10923}
}

@article{smith2025quantum,
  title   = {Quantum-centric simulation of hydrogen abstraction by sample-based quantum diagonalization and entanglement forging},
  author  = {Smith, Tyler and Gujarati, Tanvi P and Motta, Mario and Link, Ben and Liepuoniute, Ieva and Friedhoff, Triet and Nishimura, Hiromichi and Nguyen, Nam and Williams, Kristen S and Moreno, Javier Robledo and others},
  journal = {arXiv:2508.08229},
  year    = {2025},
  url     = {https://arxiv.org/abs/2508.08229}
}

@article{barison2025quantum,
  title   = {Quantum-centric computation of molecular excited states with extended sample-based quantum diagonalization},
  author  = {Barison, Stefano and Moreno, Javier Robledo and Motta, Mario},
  journal = QST,
  volume  = {10},
  number  = {2},
  pages   = {025034},
  year    = {2025},
  url     = {https://iopscience.iop.org/article/10.1088/2058-9565/adb781}
}

@article{scuseria1987closed,
  title   = {The closed-shell coupled cluster single and double excitation (CCSD) model for the description of electron correlation. A comparison with configuration interaction (CISD) results},
  author  = {Scuseria, Gustavo E and Scheiner, Andrew C and Lee, Timothy J and Rice, Julia E and Schaefer, Henry F},
  journal = JCP,
  volume  = {86},
  number  = {5},
  pages   = {2881--2890},
  year    = {1987},
  url     = {https://doi.org/10.1063/1.452039}
}

@article{mizusaki2003precise,
  title   = {Precise estimation of shell-model energy by second-order extrapolation method},
  author  = {Mizusaki, Takahiro and Imada, Masatoshi},
  journal = PRC,
  volume  = {67},
  number  = {4},
  pages   = {041301},
  year    = {2003},
  url     = {https://doi.org/10.1103/PhysRevC.67.041301}
}

@article{yu2025quantum,
  title   = {Quantum-centric algorithm for sample-based Krylov diagonalization},
  author  = {Yu, Jeffery and Moreno, Javier Robledo and Iosue, Joseph T and Bertels, Luke and Claudino, Daniel and Fuller, Bryce and Groszkowski, Peter and Humble, Travis S and Jurcevic, Petar and Kirby, William and others},
  journal = {arXiv:2501.09702},
  year    = {2025},
  url     = {https://arxiv.org/abs/2501.09702}
}

@article{piccinelli2025quantum,
  title   = {Quantum chemistry with provable convergence via randomized sample-based quantum diagonalization},
  author  = {Piccinelli, Samuele and Baiardi, Alberto and Rossmannek, Max and Vazquez, Almudena Carrera and Tacchino, Francesco and Mensa, Stefano and Altamura, Edoardo and Alavi, Ali and Motta, Mario and Robledo-Moreno, Javier and others},
  journal = {arXiv:2508.02578},
  year    = {2025},
  url     = {https://arxiv.org/abs/2508.02578}
}

@misc{lin2025pushing, 
  author       = "Wan-Hsuan Lin and Fangchun Liang and Mario Motta and Haimeng Zhang and Kevin J. Sung",
  title        = "Improved parameter initialization for the (local) unitary cluster Jastrow ansatz",
  howpublished = "private communication",
  year         = "2025",
}

@article{moreno2023enhancing,
  title   = {Enhancing the expressivity of variational neural, and hardware-efficient quantum states through orbital rotations},
  author  = {Moreno, Javier Robledo and Cohn, Jeffrey and Sels, Dries and Motta, Mario},
  journal = {arXiv:2302.11588},
  year    = {2023},
  url     = {https://arxiv.org/abs/2302.11588}
}

@article{shirakawa2025closed,
  title={Closed-loop calculations of electronic structure on a quantum processor and a classical supercomputer at full scale},
  author={Shirakawa, Tomonori and Robledo-Moreno, Javier and Itoko, Toshinari and Tripathi, Vinay and Ueda, Kento and Kawashima, Yukio and Broers, Lukas and Kirby, William and Pathak, Himadri and Paik, Hanhee and others},
  journal={arXiv:2511.00224},
  year={2025},
  url={https://www.arxiv.org/abs/2511.00224}
}

@article{hehre1972self,
  title   = {Self-consistent molecular orbital methods. XII. Further extensions of Gaussian-type basis sets for use in molecular orbital studies of organic molecules},
  author  = {Hehre, Warren J and Ditchfield, Robert and Pople, John A},
  journal = JCP,
  volume  = {56},
  number  = {5},
  pages   = {2257--2261},
  year    = {1972},
  url     = {https://doi.org/10.1063/1.1677527}
}

@misc{website,
  author  = {Petervary, Lukas and Raisuddin, Osama M. and Zhang, Haimeng and Motta, Mario and Faulstich, Fabian M.},
  title   = {An algorithm benchmarking tool built around the W4-11 dataset},
  year    = {2025},
  url     = {https://pypi.org/project/w4benchmark/}
}

@article{karton2025highly,
  title   = {A highly diverse and dccurate database of 3366 total atomization energies calculated at the CCSD(T)/CBS level by means of W1-F12 theory},
  author  = {Karton, Amir},
  journal = CPL,
  volume  = {868},
  pages   = {142030},
  year    = {2025},
  url     = {https://doi.org/10.1016/j.cplett.2025.142030}
}

@article{karton2022quantum,
  title   = {Quantum mechanical thermochemical predictions 100 years after the Schr{\"o}dinger equation},
  author  = {Karton, Amir},
  journal = ARCC,
  volume  = {18},
  pages   = {123--166},
  year    = {2022},
  url     = {https://doi.org/10.1016/bs.arcc.2022.09.003}
}

@incollection{feller2013improved,
  title     = {Improved accuracy benchmarks of small molecules using correlation consistent basis sets},
  author    = {Feller, David and Peterson, Kirk A and Ruscic, Branko},
  booktitle = {Thom H. Dunning, Jr. A Festschrift from Theor. Chem. Acc},
  pages     = {31--46},
  year      = {2013},
  url       = {https://doi.org/10.1007/978-3-662-47051-0_4}
}

@article{goerigk2011thorough,
  title   = {A thorough benchmark of density functional methods for general main group thermochemistry, kinetics, and noncovalent interactions},
  author  = {Goerigk, Lars and Grimme, Stefan},
  journal = PCCP,
  volume  = {13},
  number  = {14},
  pages   = {6670--6688},
  year    = {2011},
  url     = {https://doi.org/10.1039/C0CP02984J}
}

@article{mardirossian2017thirty,
  title   = {Thirty years of density functional theory in computational chemistry: an overview and extensive assessment of 200 density functionals},
  author  = {Mardirossian, Narbe and Head-Gordon, Martin},
  journal = MolPhys,
  volume  = {115},
  number  = {19},
  pages   = {2315--2372},
  year    = {2017},
  url     = {https://doi.org/10.1080/00268976.2017.1333644}
}

@article{goerigk2017look,
  title   = {A look at the density functional theory zoo with the advanced GMTKN55 database for general main group thermochemistry, kinetics and noncovalent interactions},
  author  = {Goerigk, Lars and Hansen, Andreas and Bauer, Christoph and Ehrlich, Stephan and Najibi, Asim and Grimme, Stefan},
  journal = PCCP,
  volume  = {19},
  number  = {48},
  pages   = {32184--32215},
  year    = {2017},
  url     = {https://doi.org/10.1039/C7CP04913G}
}

@article{benson1976thermochemical,
  title   = {Thermochemical kinetics: methods for the estimation of thermochemical data and rate parameters},
  author  = {Benson, Sidney William},
  journal = {Angew. Chemie},
  volume  = {89},
  pages   = {921},
  year    = {1976},
  url     = {https://doi.org/10.1002/ange.19770891237}
}

@article{gani2018understanding,
  title   = {Understanding and breaking scaling relations in single-site catalysis: methane to methanol conversion by $\mathrm{Fe^{IV}=O}$},
  author  = {Gani, Terry Z. H. and Kulik, Heather J.},
  journal = {ACS Catalysis},
  volume  = {8},
  number  = {2},
  pages   = {975--986},
  year    = {2018},
  url     = {https://pubs.acs.org/doi/10.1021/acscatal.7b03597}
}

@article{kim2019experimental,
  title   = {Experimental and theoretical insight into the soot tendencies of the methylcyclohexene isomers},
  author  = {Kim, Seonah and Fioroni, Gina M and Park, Ji-Woong and Robichaud, David J and Das, Dhrubajyoti D and John, Peter C St and Lu, Tianfeng and McEnally, Charles S and Pfefferle, Lisa D and Paton, Robert S and others},
  journal = {Proc. Comb. Ins},
  volume  = {37},
  number  = {1},
  pages   = {1083--1090},
  year    = {2019},
  url     = {https://doi.org/10.1016/j.proci.2018.06.095}
}

@article{lin2011linear,
  title   = {Linear-free energy relationships for modeling structure-reactivity trends in controlled radical polymerization},
  author  = {Lin, Ching Yeh and Marque, Sylvain RA and Matyjaszewski, Krzysztof and Coote, Michelle L},
  journal = {Macromolecules},
  volume  = {44},
  number  = {19},
  pages   = {7568--7583},
  year    = {2011},
  url     = {https://pubs.acs.org/doi/10.1021/ma2014996}
}

@article{giannetti2005thermal,
  title   = {Thermal stability and bond dissociation energy of fluorinated polymers: A critical evaluation},
  author  = {Giannetti, Enzo},
  journal = JFC,
  volume  = {126},
  number  = {4},
  pages   = {623--630},
  year    = {2005},
  url     = {https://doi.org/10.1016/j.jfluchem.2005.01.008}
}

@article{bian2016thermal,
  title   = {Thermal stability of phenolic resin: new insights based on bond dissociation energy and reactivity of functional groups},
  author  = {Bian, Cheng and Wang, Shujuan and Liu, Yuhong and Jing, Xinli},
  journal = {RSC Adv},
  volume  = {6},
  number  = {60},
  pages   = {55007--55016},
  year    = {2016},
  url     = {https://doi.org/10.1039/C6RA07597E}
}

@article{kim2011computational,
  title   = {Computational study of bond dissociation enthalpies for a large range of native and modified lignins},
  author  = {Kim, Seonah and Chmely, Stephen C and Nimlos, Mark R and Bomble, Yannick J and Foust, Thomas D and Paton, Robert S and Beckham, Gregg T},
  journal = JPCL, 
  volume  = {2},
  number  = {22},
  pages   = {2846--2852},
  year    = {2011},
  url     = {https://doi.org/10.1021/jz201182w}
}

@article{lienard2015predicting,
  title   = {Predicting drug substances autoxidation},
  author  = {Lienard, P and Gavartin, J and Boccardi, G and Meunier, M},
  journal = {Pharm. Res},
  volume  = {32},
  pages   = {300--310},
  year    = {2015},
  url     = {https://doi.org/10.1007/s11095-014-1463-7}
}

@article{drew2012impact,
  title   = {The impact of carbon--hydrogen bond dissociation energies on the prediction of the cytochrome P450 mediated major metabolic site of drug-like compounds},
  author  = {Drew, Kurt LM and Reynisson, J{\'o}hannes},
  journal = EJMC,
  volume  = {56},
  pages   = {48--55},
  year    = {2012},
  url     = {https://doi.org/10.1016/j.ejmech.2012.08.017}
}

@article{zhao2005assessment,
  title   = {Assessment of the metabolic stability of the methyl groups in heterocyclic compounds using C-H bond dissociation energies: effects of diverse aromatic groups on the stability of methyl radicals},
  author  = {Zhao, Su-Wen and Liu, Lei and Fu, Yao and Guo, Qing-Xiang},
  journal = JPOC,
  volume  = {18},
  number  = {4},
  pages   = {353--367},
  year    = {2005},
  url     = {https://doi.org/10.1002/poc.856}
}

@article{harris1997ab,
  title   = {Ab initio density functional computations of conformations and bond dissociation energies for hexahydro-1, 3, 5-trinitro-1, 3, 5-triazine},
  author  = {Harris, Nathan J and Lammertsma, Koop},
  journal = JACS,
  volume  = {119},
  number  = {28},
  pages   = {6583--6589},
  year    = {1997},
  url     = {https://doi.org/10.1021/ja970392i}
}

@article{warr2014short,
  title   = {A short review of chemical reaction database systems, computer-aided synthesis design, reaction prediction and synthetic feasibility},
  author  = {Warr, Wendy A},
  journal = MolInf,
  volume  = {33},
  number  = {6-7},
  pages   = {469--476},
  year    = {2014},
  url     = {https://doi.org/10.1002/minf.201400052}
}

@article{ahneman2018predicting,
  title   = {Predicting reaction performance in C-N cross-coupling using machine learning},
  author  = {Ahneman, Derek T and Estrada, Jes{\'u}s G and Lin, Shishi and Dreher, Spencer D and Doyle, Abigail G},
  journal = {Science},
  volume  = {360},
  number  = {6385},
  pages   = {186--190},
  year    = {2018},
  url     = {https://www.science.org/doi/10.1126/science.aar5169}
}

@article{wilcox2018stable,
  title   = {Stable radical materials for energy applications},
  author  = {Wilcox, Daniel A and Agarkar, Varad and Mukherjee, Sanjoy and Boudouris, Bryan W},
  journal = ARCBE,
  volume  = {9},
  number  = {1},
  pages   = {83--103},
  year    = {2018},
  url     = {https://doi.org/10.1146/annurev-chembioeng-060817-083945}
}

@article{bak2000accuracy,
  title   = {Accuracy of atomization energies and reaction enthalpies in standard and extrapolated electronic wave function/basis set calculations},
  author  = {Bak, Keld L and J{\o}rgensen, Poul and Olsen, Jeppe and Helgaker, Trygve and Klopper, Wim},
  journal = JCP,
  volume  = {112},
  number  = {21},
  pages   = {9229--9242},
  year    = {2000},
  url     = {https://doi.org/10.1063/1.481544}
}

@article{ruzsinszky2015insight,
  title   = {Insight into organic reactions from the direct random phase approximation and its corrections},
  author  = {Ruzsinszky, Adrienn and Zhang, Igor Ying and Scheffler, Matthias},
  journal = JCP,
  volume  = {143},
  number  = {14},
  pages   = {144115},
  year    = {2015},
  url     = {https://doi.org/10.1063/1.4932306}
}

@article{lesiuk2022quintic,
  title   = {Quintic-scaling rank-reduced coupled cluster theory with single and double excitations},
  author  = {Lesiuk, Micha{\l}},
  journal = JCP,
  volume  = {156},
  number  = {6},
  pages   = {064103},
  year    = {2022},
  url     = {https://doi.org/10.1063/5.0071916}
}

@article{dasgupta2017standard,
  title   = {Standard grids for high-precision integration of modern density functionals: SG-2 and SG-3},
  author  = {Dasgupta, Saswata and Herbert, John M},
  journal = JCC, 
  volume  = {38},
  number  = {12},
  pages   = {869--882},
  year    = {2017},
  url     = {https://doi.org/10.1002/jcc.24761},
}

@article{grimme2007compute,
  title   = {How to compute isomerization energies of organic molecules with quantum chemical methods},
  author  = {Grimme, Stefan and Steinmetz, Marc and Korth, Martin},
  journal = JOC,
  volume  = {72},
  number  = {6},
  pages   = {2118--2126},
  year    = {2007},
  url     = {https://doi.org/10.1021/jo062446p},
}

@article{wedemeyer2002proline,
  title   = {Proline cis-trans isomerization and protein folding},
  author  = {Wedemeyer, William J and Welker, Ervin and Scheraga, Harold A},
  journal = {Biochemistry},
  volume  = {41},
  number  = {50},
  pages   = {14637--14644},
  year    = {2002},
  url     = {https://doi.org/10.1021/bi020574b}
}

@article{laplaza2024overcoming,
  title   = {Overcoming the pitfalls of computing reaction selectivity from ensembles of transition states},
  author  = {Laplaza, Ruben and Wodrich, Matthew D and Corminboeuf, Clemence},
  journal = JPCL,
  volume  = {15},
  number  = {29},
  pages   = {7363--7370},
  year    = {2024},
  url     = {https://doi.org/10.1021/acs.jpclett.4c01657}
}

@article{zahrt2019prediction,
  title   = {Prediction of higher-selectivity catalysts by computer-driven workflow and machine learning},
  author  = {Zahrt, Andrew F and Henle, Jeremy J and Rose, Brennan T and Wang, Yang and Darrow, William T and Denmark, Scott E},
  journal = {Science},
  volume  = {363},
  number  = {6424},
  pages   = {eaau5631},
  year    = {2019},
  url     = {https://www.science.org/doi/10.1126/science.aau5631}
}

@article{yu2004alignment,
  title   = {Alignment modulation of azobenzene-containing liquid crystal systems by photochemical reactions},
  author  = {Yu, Yanlei and Ikeda, Tomiki},
  journal = JPPC,
  volume  = {5},
  number  = {3},
  pages   = {247--265},
  year    = {2004},
  url     = {https://doi.org/10.1016/j.jphotochemrev.2004.10.004}
}

@article{legge1992photo,
  title   = {Photo-induced phase transitions in azobenzene-doped liquid crystals},
  author  = {Legge, C. H. and Mitchell, G. R.},
  journal = JPD,
  volume  = {25},
  number  = {3},
  pages   = {492},
  year    = {1992},
  url     = {https://doi.org/10.1088/0022-3727/25/3/024},
}

@article{chen2015reactivity,
  title   = {The reactivity of the active metal oxo and hydroxo intermediates and their implications in oxidations},
  author  = {Chen, Zhuqi and Yin, Guochuan},
  journal = CSR,
  volume  = {44},
  number  = {5},
  pages   = {1083--1100},
  year    = {2015},
  url     = {https://doi.org/10.1039/C4CS00244J}
}

@article{holm1987metal,
  title   = {Metal-centered oxygen atom transfer reactions},
  author  = {Holm, RH},
  journal = ChemRev,
  volume  = {87},
  number  = {6},
  pages   = {1401--1449},
  year    = {1987},
  url     = {https://doi.org/10.1021/cr00082a005}
}

@article{masunov2016chemical,
  title   = {Chemical reaction $\mathrm{CO+OH^\bullet \to CO_2 + H^\bullet}$ autocatalyzed by carbon dioxide: quantum chemical study of the potential energy surfaces},
  author  = {Masunov, Artem E and Wait, Elizabeth and Vasu, Subith S},
  journal = JPCA,
  volume  = {120},
  number  = {30},
  pages   = {6023--6028},
  year    = {2016},
  url     = {https://doi.org/10.1021/acs.jpca.6b03242}
}

@article{harvey2007understanding,
  title   = {Understanding the kinetics of spin-forbidden chemical reactions},
  author  = {Harvey, Jeremy N},
  journal = PCCP,
  volume  = {9},
  number  = {3},
  pages   = {331--343},
  year    = {2007},
  url     = {https://doi.org/10.1039/B614390C}
}

@article{schroder2000two,
  title   = {Two-state reactivity as a new concept in organometallic chemistry},
  author  = {Schr{\"o}der, Detlef and Shaik, Sason and Schwarz, Helmut},
  journal = ACR,
  volume  = {33},
  number  = {3},
  pages   = {139--145},
  year    = {2000},
  url     = {https://doi.org/10.1021/ar990028j}
}

@article{zhao2005benchmark,
  title   = {Benchmark database of barrier heights for heavy atom transfer, nucleophilic substitution, association, and unimolecular reactions and its use to test theoretical methods},
  author  = {Zhao, Yan and Gonz{\'a}lez-Garc{\'\i}a, N{\'u}ria and Truhlar, Donald G},
  journal = JPCA,
  volume  = {109},
  number  = {9},
  pages   = {2012--2018},
  year    = {2005},
  url     = {https://pubs.acs.org/doi/10.1021/jp045141s}
}

@article{cohen2012challenges,
  title   = {Challenges for density functional theory},
  author  = {Cohen, Aron J and Mori-S{\'a}nchez, Paula and Yang, Weitao},
  journal = ChemRev,
  volume  = {112},
  number  = {1},
  pages   = {289--320},
  year    = {2012},
  url     = {https://doi.org/10.1021/cr200107z}
}

@article{semidalas2020canonical,
  title   = {Canonical and DLPNO-based G4 (MP2) XK-inspired composite wave function methods parametrized against large and chemically diverse training sets: are they more accurate and/or robust than double-hybrid DFT?},
  author  = {Semidalas, Emmanouil and Martin, Jan ML},
  journal = JCTC,
  volume  = {16},
  number  = {7},
  pages   = {4238--4255},
  year    = {2020},
  url     = {https://doi.org/10.1021/acs.jctc.0c00189}
}

@article{dzib2024enhancing,
  title   = {Enhancing Eyringpy: accurate rate constants with canonical variational transition state theory and the hindered rotor model},
  author  = {Dzib, Eugenia and Quintal, Alan and Merino, Gabriel},
  journal = JCTC,
  volume  = {20},
  number  = {22},
  pages   = {9999--10009},
  year    = {2024},
  url     = {https://pubs.acs.org/doi/10.1021/acs.jctc.4c00926}
}

@article{baulch2005evaluated,
  title   = {Evaluated kinetic data for combustion modeling: supplement II},
  author  = {Baulch, DL and Bowman, Craig T and Cobos, Carlos J and Cox, Richard Anthony and Just, Th and Kerr, JA and Pilling, MJ and Stocker, D and Troe, Juergen and Tsang, Wing and others},
  journal = JPCRD,
  volume  = {34},
  number  = {3},
  pages   = {757--1397},
  year    = {2005},
  url     = {https://doi.org/10.1063/1.1748524}
}

@article{atkinson2003atmospheric,
  title   = {Atmospheric degradation of volatile organic compounds},
  author  = {Atkinson, Roger and Arey, Janet},
  journal = ChemRev,
  volume  = {103},
  number  = {12},
  pages   = {4605--4638},
  year    = {2003},
  url     = {https://doi.org/10.1021/cr0206420}
}

@article{saunders2003protocol,
  title   = {Protocol for the development of the Master Chemical Mechanism, MCM v3 (Part A): tropospheric degradation of non-aromatic volatile organic compounds},
  author  = {Saunders, Sandra M and Jenkin, Michael E and Derwent, Richard G and Pilling, Mike J},
  journal = ACP,
  volume  = {3},
  number  = {1},
  pages   = {161--180},
  year    = {2003},
  url     = {https://doi.org/10.5194/acp-3-161-2003}
}

@article{wang2017oxygen,
  title={Oxygen activation by mononuclear nonheme iron dioxygenases involved in the degradation of aromatics},
  author={Wang, Yifan and Li, Jiasong and Liu, Aimin},
  journal=JBIC,
  volume={22},
  number={2},
  pages={395--405},
  year={2017},
  url = {https://doi.org/10.1007/s00775-017-1436-5}
}

@article{hamlin2018nucleophilic,
  title={Nucleophilic substitution (SN2): dependence on nucleophile, leaving group, central atom, substituents, and solvent},
  author={Hamlin, Trevor A and Swart, Marcel and Bickelhaupt, F Matthias},
  journal=CPC,
  volume={19},
  number={11},
  pages={1315--1330},
  year={2018},
  url={https://doi.org/10.1002/cphc.201701363}
}

@article{kubelka2017activation,
  title={Activation strain analysis of SN2 reactions at C, N, O, and F centers},
  author={Kubelka, Jan and Bickelhaupt, F Matthias},
  journal=JPCA,
  volume={121},
  number={4},
  pages={885--891},
  year={2017},
  url={https://doi.org/10.1021/acs.jpca.6b12240}
}

@article{hohenstein2012wavefunction,
  title   = {Wavefunction methods for noncovalent interactions},
  author  = {Hohenstein, Edward G and Sherrill, C David},
  journal = WIRES,
  volume  = {2},
  number  = {2},
  pages   = {304--326},
  year    = {2012},
  url     = {https://doi.org/10.1002/wcms.84}
}

@article{chabinyc1998gas,
  title   = {Gas-phase ionic reactions: dynamics and mechanism of nucleophilic displacements},
  author  = {Chabinyc, Michael L and Craig, Stephen L and Regan, Colleen K and Brauman, John I},
  journal = {Science},
  volume  = {279},
  number  = {5358},
  pages   = {1882--1886},
  year    = {1998},
  url     = {https://www.science.org/doi/10.1126/science.279.5358.1882}
}

@article{houk2008computational,
  title   = {Computational prediction of small-molecule catalysts},
  author  = {Houk, Kendall N and Cheong, Paul Ha-Yeon},
  journal = {Nature},
  volume  = {455},
  number  = {7211},
  pages   = {309--313},
  year    = {2008},
  url     = {https://doi.org/10.1038/nature07368}
}

@article{mikosch2008imaging,
  title   = {Imaging nucleophilic substitution dynamics},
  author  = {Mikosch, J and Trippel, S and Eichhorn, C and Otto, R and Lourderaj, U and Zhang, JX and Hase, WL and Weidemuller, M and Wester, R},
  journal = {Science},
  volume  = {319},
  number  = {5860},
  pages   = {183--186},
  year    = {2008},
  url     = {https://www.science.org/doi/10.1126/science.1150238}
}

@article{kitaev1995qpe,
  title   = {Quantum measurements and the Abelian stabilizer problem}, 
  author  = {A. Yu. Kitaev},
  year    = {1995},
  journal = {arXiv:quant-ph/9511026},
  url     = {https://arxiv.org/abs/quant-ph/9511026}, 
}

@article{peruzzo2014vqe,
  author  = {Peruzzo, Alberto and McClean, Jarrod and Shadbolt, Peter and Yung, Man-Hong and Zhou, Xiao-Qi and Love, Peter J. and Aspuru-Guzik, Alan and O'Brien, Jeremy L.},
  title   = {A variational eigenvalue solver on a photonic quantum processor},
  journal = NatComm,
  volume  = {5},
  number  = {1},
  url     = {http://dx.doi.org/10.1038/ncomms5213},
  year    = {2014}
}

@article{lee2023evaluating,
  title   = {Evaluating the evidence for exponential quantum advantage in ground-state quantum chemistry},
  author  = {Lee, Seunghoon and Lee, Joonho and Zhai, Huanchen and Tong, Yu and Dalzell, Alexander M and Kumar, Ashutosh and Helms, Phillip and Gray, Johnnie and Cui, Zhi-Hao and Liu, Wenyuan and others},
  journal = NatComm,
  volume  = {14},
  number  = {1},
  pages   = {1952},
  year    = {2023},
  url     = {https://doi.org/10.1038/s41467-023-37587-6}
}

@article{mcclean2016theory,
  title   = {The theory of variational hybrid quantum-classical algorithms},
  author  = {McClean, Jarrod R and Romero, Jonathan and Babbush, Ryan and Aspuru-Guzik, Al{\'a}n},
  journal = NJP,
  volume  = {18},
  number  = {2},
  pages   = {023023},
  year    = {2016},
  url     = {https://iopscience.iop.org/article/10.1088/1367-2630/18/2/023023}
}

@article{patel2025quantum,
  title   = {Quantum measurement for quantum chemistry on a quantum computer},
  author  = {Patel, Smik and Jayakumar, Praveen and Yen, Tzu-Ching and Izmaylov, Artur F},
  journal = ChemRev,
  volume  = {125},
  number  = {16},
  pages   = {7490--7524},
  year    = {2025},
  url     = {https://doi.org/10.1021/acs.chemrev.5c00055}
}

@article{cerezo2021variational,
  author  = {Cerezo,  M. and Arrasmith,  Andrew and Babbush,  Ryan and Benjamin,  Simon C. and Endo,  Suguru and Fujii,  Keisuke and McClean,  Jarrod R. and Mitarai,  Kosuke and Yuan,  Xiao and Cincio,  Lukasz and Coles,  Patrick J.},
  title   = {Variational quantum algorithms},
  journal = NRP,
  year    = {2021},
  volume  = {3},
  pages   = {625–644},
  url     = {http://dx.doi.org/10.1038/s42254-021-00348-9},
}

@article{cerezo2021local,
  author  = {Cerezo, M. and Sone, Akira and Volkoff, Tyler and Cincio, Lukasz and Coles, Patrick J.}, 
  title   = {Cost function dependent barren plateaus in shallow parametrized quantum circuits},
  volume  = {12}, 
  pages   = {1791}, 
  journal = NatComm,
  year    = {2021},
  url     = {https://doi.org/10.1038/s41467-021-21728-w}
}

@article{spall2002multivariate,
  title   = {Multivariate stochastic approximation using a simultaneous perturbation gradient approximation},
  author  = {Spall, James C},
  journal = {IEEE Trans. Autom. Control},
  volume  = {37},
  number  = {3},
  pages   = {332--341},
  year    = {2002},
  url     = {https://doi.org/10.1109/9.119632}
}

@article{suzuki1976productformulae,
	author  = {Suzuki, Masuo},
	title   = {Generalized Trotter's formula and systematic approximants of exponential operators and inner derivations with applications to many-body problems},
	journal = CMP,
	volume  = {51},
	pages   = {183--190},
	year    = {1976},
	url     = {https://doi.org/10.1007/BF01609348},
}

@article{low2017qsphamsim,
  title   = {Optimal Hamiltonian Simulation by Quantum Signal Processing},
  author  = {Low, Guang Hao and Chuang, Isaac L.},
  journal = PRL,
  volume  = {118},
  issue   = {1},
  pages   = {010501},
  year    = {2017},
  url     = {https://link.aps.org/doi/10.1103/PhysRevLett.118.010501}
}

@article{childs2012lcuhamsim,
  author  = {Childs, Andrew M. and Wiebe, Nathan},
  title   = {Hamiltonian simulation using linear combinations of unitary operations},
  year    = {2012},
  volume  = {12},
  number  = {11–12},
  journal = QIC,
  pages   = {901–924},
  url     = {https://dl.acm.org/doi/abs/10.5555/2481569.2481570}
}

@article{childs2021theory,
  title   = {Theory of Trotter error with commutator scaling},
  author  = {Childs, Andrew M. and Su, Yuan and Tran, Minh C. and Wiebe, Nathan and Zhu, Shuchen},
  journal = PRX,
  volume  = {11},
  pages   = {011020},
  year    = {2021},
  url     = {https://link.aps.org/doi/10.1103/PhysRevX.11.011020}
}

@article{jordan1928jwtransform,
  author  = {Jordan,  P. and Wigner,  E.},
  title   = {\"Uber das Paulische \"Aquivalenzverbot},
  journal = {Zeit. Phys},
  volume  = {47},
  pages   = {631–651},
  year    = {1928},
  url     = {http://dx.doi.org/10.1007/BF01331938},
}

@article{bravyi2002fermionicqc,
  author  = {Sergey B. Bravyi and Alexei Yu. Kitaev},
  title   = {Fermionic Quantum Computation},
  journal = Annals,
  volume  = {298},
  number  = {1},
  pages   = {210-226},
  year    = {2002},
  url     = {https://doi.org/10.1006/aphy.2002.6254},
}

@article{seeley2012,
   author  = {Seeley, Jacob T. and Richard, Martin J. and Love, Peter J.},
   title   = {The Bravyi-Kitaev transformation for quantum computation of electronic structure},
   volume  = {137},
   pages   = {224109},
   url     = {https://doi.org/10.1063/1.4768229},
   journal = JCP,
   year    = {2012}
}

@article{tranter2018, 
   title   = {A comparison of the Bravyi–Kitaev and Jordan–Wigner transformations for the quantum simulation of quantum chemistry}, 
   volume  = {14},
   number  = {11},
   journal = JCTC,
   author  = {Tranter, Andrew and Love, Peter J. and Mintert, Florian and Coveney, Peter V.},
   year    = {2018},
   pages   = {5617–5630},
   url     = {https://doi.org/10.1021/acs.jctc.8b00450}
}

@article{gujarati2023quantum,
  title   = {Quantum computation of reactions on surfaces using local embedding},
  author  = {Gujarati, Tanvi P and Motta, Mario and Friedhoff, Triet Nguyen and Rice, Julia E and Nguyen, Nam and Barkoutsos, Panagiotis Kl and Thompson, Richard J and Smith, Tyler and Kagele, Marna and Brei, Mark and others},
  journal = NPJQI,
  volume  = {9},
  pages   = {88},
  year    = {2023},
  url     = {https://doi.org/10.1038/s41534-023-00753-1}
}

@article{cowtan2020phasegadget,
   author  ={Cowtan, Alexander and Dilkes, Silas and Duncan, Ross and Simmons, Will and Sivarajah, Seyon},
   title   = {Phase gadget synthesis for shallow circuits},
   journal = EPTCS,
   volume  = {318},
   pages   = {213–228},
   year    = {2020},
   url     = {http://dx.doi.org/10.4204/EPTCS.318.13},
}

@article{moflic2024constantdepthimplementationpauli,
      title   = {On the Constant Depth Implementation of Pauli Exponentials}, 
      author  = {Ioana Moflic and Alexandru Paler},
      year    = {2024},
      journal = {arXiv:2408.08265},
      url     = {https://arxiv.org/abs/2408.08265}, 
}

@article{kivliochan2018fermionicswapnetwork,
  title   = {Quantum simulation of electronic structure with linear depth and connectivity},
  author  = {Kivlichan, Ian D. and McClean, Jarrod and Wiebe, Nathan and Gidney, Craig and Aspuru-Guzik, Al\'an and Chan, Garnet Kin-Lic and Babbush, Ryan},
  journal = PRL,
  volume  = {120},
  pages   = {110501},
  year    = {2018},
  url     = {https://link.aps.org/doi/10.1103/PhysRevLett.120.110501}
}

@article{setia2018bksf,
    author  = {Setia, Kanav and Whitfield, James D.},
    title   = {Bravyi-Kitaev Superfast simulation of electronic structure on a quantum computer},
    journal = JCP,
    volume  = {148},
    pages   = {164104},
    year    = {2018},
    url     = {https://doi.org/10.1063/1.5019371},
}

@article{fedorov2022unitary,
  title   = {Unitary selective coupled-cluster method},
  author  = {Fedorov, Dmitry A and Alexeev, Yuri and Gray, Stephen K and Otten, Matthew},
  journal = {Quantum},
  volume  = {6},
  pages   = {703},
  year    = {2022},
  url     = {https://doi.org/10.22331/q-2022-05-02-703}
}

@article{matsuzawa2020jastrow,
  title   = {Jastrow-type decomposition in quantum chemistry for low-depth quantum circuits},
  author  = {Matsuzawa, Yuta and Kurashige, Yuki},
  journal = JCTC,
  volume  = {16},
  number  = {2},
  pages   = {944--952},
  year    = {2020},
  url     = {https://doi.org/10.1021/acs.jctc.9b00963}
}

@article{jiang2018quantum,
  title   = {Quantum algorithms to simulate many-body physics of correlated fermions},
  author  = {Jiang, Zhang and Sung, Kevin J and Kechedzhi, Kostyantyn and Smelyanskiy, Vadim N and Boixo, Sergio},
  journal = PRAppl,
  volume  = {9},
  pages   = {044036},
  year    = {2018},
  url     = {https://doi.org/10.1103/PhysRevApplied.9.044036}
}

@article{sun2018pyscf,
  title   = {PySCF: the python-based simulations of chemistry framework},
  author  = {Sun, Qiming and Berkelbach, Timothy C and Blunt, Nick S and Booth, George H and Guo, Sheng and Li, Zhendong and Liu, Junzi and McClain, James D and Sayfutyarova, Elvira R and Sharma, Sandeep and others},
  journal = WIRES,
  volume  = {8},
  number  = {1},
  pages   = {e1340},
  year    = {2018},
  url     = {https://doi.org/10.1002/wcms.1340}
}

@article{sun2020recent,
  title   = {Recent developments in the {PySCF} program package},
  author  = {Sun, Qiming and Zhang, Xing and Banerjee, Samragni and Bao, Peng and Barbry, Marc and Blunt, Nick S and Bogdanov, Nikolay A and Booth, George H and Chen, Jia and Cui, Zhi-Hao and others},
  journal = JCP,
  volume  = {153},
  pages   = {024109},
  year    = {2020},
  url     = {https://doi.org/10.1063/5.0006074}
}

@software{ffsim,
  author = {The ffsim developers},
  title  = {ffsim: Faster simulations of fermionic quantum circuits.},
  url    = {https://github.com/qiskit-community/ffsim}
}

@article{bender1969studies,
  title   = {Studies in configuration interaction: the first-row diatomic hydrides},
  author  = {Bender, Charles F and Davidson, Ernest R},
  journal = {Phys. Rev},
  volume  = {183},
  pages   = {23},
  year    = {1969},
  url     = {https://doi.org/10.1103/PhysRev.183.23}
}

@article{ivanic2001identification,
  title   = {Identification of deadwood in configuration spaces through general direct configuration interaction},
  author  = {Ivanic, Joseph and Ruedenberg, Klaus},
  journal = TCA, 
  volume  = {106},
  pages   = {339--351},
  year    = {2001},
  url     = {https://doi.org/10.1007/s002140100285}
}

@article{stampfuss2005improved,
  title   = {Improved implementation and application of the individually selecting configuration interaction method},
  author  = {Stampfuss, P and Wenzel, W},
  journal = JCP,
  volume  = {122},
  pages   = {024110},
  year    = {2005},
  url     = {https://doi.org/10.1063/1.1829045}
}

@article{roth2009importance,
  title   = {Importance truncation for large-scale configuration interaction approaches},
  author  = {Roth, Robert},
  journal = PRC,
  volume  = {79},
  number  = {6},
  pages   = {064324},
  year    = {2009},
  url     = {https://doi.org/10.1103/PhysRevC.79.064324}
}

@article{evangelista2014adaptive,
  title   = {Adaptive multiconfigurational wave functions},
  author  = {Evangelista, Francesco A},
  journal = JCP,
  volume  = {140},
  number  = {12},
  year    = {2014},
  url     = {https://doi.org/10.1063/1.4869192}
}

@article{liu2016ici,
  title   = {iCI: Iterative CI toward full CI},
  author  = {Liu, Wenjian and Hoffmann, Mark R},
  journal = JCTC,
  volume  = {12},
  number  = {3},
  pages   = {1169--1178},
  year    = {2016},
  url     = {https://doi.org/10.1021/acs.jctc.5b01099}
}

@article{lidar2014review,
  title   = {Review of decoherence-free subspaces, noiseless subsystems, and dynamical decoupling},
  author  = {Lidar, Daniel A},
  journal = QICC,
  pages   = {295--354},
  year    = {2014},
  url     = {https://doi.org/10.1002/9781118742631.ch11}
}

@article{ezzell2023dynamical,
  title   = {Dynamical decoupling for superconducting qubits: a performance survey},
  author  = {Ezzell, Nic and Pokharel, Bibek and Tewala, Lina and Quiroz, Gregory and Lidar, Daniel A},
  journal = PRAppl,
  volume  = {20},
  pages   = {064027},
  year    = {2023},
  url     = {https://doi.org/10.1103/PhysRevApplied.20.064027}
}

@article{qiskit2024,
  title   = {Quantum computing with {Q}iskit},
  author  = {Javadi-Abhari, Ali and Treinish, Matthew and Krsulich, Kevin and Wood, Christopher J. and Lishman, Jake and Gacon, Julien and Martiel, Simon and Nation, Paul D. and Bishop, Lev S. and Cross, Andrew W. and Johnson, Blake R. and Gambetta, Jay M.},
  year    = {2024},
  journal = {arXiv:2405.08810},
  url     = {https://arxiv.org/abs/2405.08810}
}

@article{grimsley2020is,
title = {Is the Trotterized UCCSD Ansatz Chemically Well-Defined?},
author = {Grimsley, Harper R. and Claudino, Daniel and Economou, Sophia E. and Barnes, Edwin and Mayhall, Nicholas J.},
journal = {Journal of Chemical Theory and Computation},
volume = {16},
number = {1},
pages = {1-6},
year = {2020},
url = { 
        https://doi.org/10.1021/acs.jctc.9b01083
},}

@article{barkoutsos2018quantum,
  title = {Quantum algorithms for electronic structure calculations: Particle-hole Hamiltonian and optimized wave-function expansions},
  author = {Barkoutsos, Panagiotis Kl. and Gonthier, Jerome F. and Sokolov, Igor and Moll, Nikolaj and Salis, Gian and Fuhrer, Andreas and Ganzhorn, Marc and Egger, Daniel J. and Troyer, Matthias and Mezzacapo, Antonio and Filipp, Stefan and Tavernelli, Ivano},
  journal = {Phys. Rev. A},
  volume = {98},
  issue = {2},
  pages = {022322},
  year = {2018},
  url = {https://link.aps.org/doi/10.1103/PhysRevA.98.022322}
}

@article{evangelista2019exact,
    author = {Evangelista, Francesco A. and Chan, Garnet Kin-Lic and Scuseria, Gustavo E.},
    title = {Exact parameterization of fermionic wave functions via unitary coupled cluster theory},
    journal = {The Journal of Chemical Physics},
    volume = {151},
    pages = {244112},
    year = {2019},
    url = {https://doi.org/10.1063/1.5133059},
}

@article{Ding2024quantummultiple,
  doi = {10.22331/q-2024-10-02-1487},
  url = {https://doi.org/10.22331/q-2024-10-02-1487},
  title = {Quantum {M}ultiple {E}igenvalue {G}aussian filtered {S}earch: an efficient and versatile quantum phase estimation method},
  author = {Ding, Zhiyan and Li, Haoya and Lin, Lin and Ni, HongKang and Ying, Lexing and Zhang, Ruizhe},
  journal = {{Quantum}},
  issn = {2521-327X},
  publisher = {{Verein zur F{\"{o}}rderung des Open Access Publizierens in den Quantenwissenschaften}},
  volume = {8},
  pages = {1487},
  month = oct,
  year = {2024}
}

@article{Ding2023simultaneous,
  doi = {10.22331/q-2023-10-11-1136},
  url = {https://doi.org/10.22331/q-2023-10-11-1136},
  title = {Simultaneous estimation of multiple eigenvalues with short-depth quantum circuit on early fault-tolerant quantum computers},
  author = {Ding, Zhiyan and Lin, Lin},
  journal = {{Quantum}},
  issn = {2521-327X},
  publisher = {{Verein zur F{\"{o}}rderung des Open Access Publizierens in den Quantenwissenschaften}},
  volume = {7},
  pages = {1136},
  month = oct,
  year = {2023}
}

@misc{Shirakawa2025closedloop,
      title={Closed-loop calculations of electronic structure on a quantum processor and a classical supercomputer at full scale}, 
      author={Tomonori Shirakawa and Javier Robledo-Moreno and Toshinari Itoko and Vinay Tripathi and Kento Ueda and Yukio Kawashima and Lukas Broers and William Kirby and Himadri Pathak and Hanhee Paik and Miwako Tsuji and Yuetsu Kodama and Mitsuhisa Sato and Constantinos Evangelinos and Seetharami Seelam and Robert Walkup and Seiji Yunoki and Mario Motta and Petar Jurcevic and Hiroshi Horii and Antonio Mezzacapo},
      year={2025},
      eprint={2511.00224},
      archivePrefix={arXiv},
      primaryClass={quant-ph},
      url={https://arxiv.org/abs/2511.00224}, 
}

\end{document}